\documentclass[preprint,12pt]{elsarticle}

\usepackage{hyperref}
\usepackage{graphicx}
\usepackage{subcaption}
\usepackage{amssymb}
\usepackage{amsmath}
\usepackage{multirow}
\usepackage{relsize}
\usepackage[utf8]{inputenc}
\usepackage{cleveref}
\usepackage{algorithm}
\usepackage[noend]{algpseudocode}
\usepackage[section]{placeins}

\usepackage{xcolor}
\usepackage{xargs}
\usepackage[colorinlistoftodos,textsize=footnotesize]{todonotes}

\newcommandx{\unsure}[2][1=]{\todo[linecolor=red,backgroundcolor=red!25,bordercolor=red,#1]{#2}}
\newcommandx{\change}[2][1=]{\todo[linecolor=blue,backgroundcolor=blue!25,bordercolor=blue,#1]{#2}}
\newcommandx{\info}[2][1=]{\todo[linecolor=OliveGreen,backgroundcolor=OliveGreen!25,bordercolor=OliveGreen,#1]{#2}}

\newcommand{\teng}[1]{\ensuremath{\boldsymbol{#1}}}
\newcommand{\ten}[1]{\ensuremath{\mathbf{#1}}}

\usepackage{lineno}

\journal{}

\begin{document}

\begin{frontmatter}

  \title{A corrected transport-velocity formulation for fluid and structural mechanics with SPH}
  \author[IITB]{Dinesh Adepu\corref{cor1}}
  \ead{adepu.dinesh.a@gmail.com}
  \author[IITB]{Prabhu Ramachandran}
  \ead{prabhu@aero.iitb.ac.in}
\address[IITB]{Department of Aerospace Engineering, Indian Institute of
  Technology Bombay, Powai, Mumbai 400076}

\cortext[cor1]{Corresponding author}

\begin{abstract}
  Particle shifting techniques (PST) have been used to improve the accuracy of
  the Smoothed Particle Hydrodynamics (SPH) method. Shifting ensures that the
  particles are distributed homogeneously in space. This may be performed by
  moving the particles using a transport velocity. In this paper, we propose an
  extension to the class of Transport Velocity Formulation (TVF) methods. We
  derive the equations in a consistent manner and show that there are
  additional terms that significantly improve the accuracy of the method. In
  particular, we apply this to the Entropically Damped Artificial
  Compressibility SPH method. We identify the free-surface particles and their
  normals using a simple approach and thereby adapt the method for
  free-surface problems. We show how the new method can be applied to the
  problem of elastic dynamics. We consider a suite of benchmark problems
  involving both fluid and solid mechanics to demonstrate the accuracy and
  applicability of the method. The implementation is open source, and the
  manuscript is fully reproducible.
\end{abstract}


\end{frontmatter}


\section{Introduction}
\label{sec:intro}

The smoothed particle hydrodynamics (SPH) method has been widely applied since
it was originally proposed to simulate hydrodynamic problems in astrophysics
independently by \citet{lucy77}, and \citet{monaghan-gingold-stars-mnras-77}.
The method has been applied in particular to both
compressible~\cite{monaghan-review:2005}, incompressible fluid
flows~\cite{sph:fsf:monaghan-jcp94,sph:psph:cummins-rudman:jcp:1999} as well
as elastic dynamics problems~\cite{randles-1996,gray-ed-2001} in addition to a
variety of other problems\cite{rafiee:fsi-2009,khayyer-fsi-2018,
  sun2021accurate, bui2008lagrangian}.

The method is meshless and Lagrangian, and therefore particles move with the
local velocity. This motion can introduce disorder in the particles and
thereby significantly reduce the accuracy of the method.
\citet{acc_stab_xu:jcp:2009} proposed an approach to shift the particles so as
to obtain a uniform distribution of particles. This significantly improves
accuracy and the method is referred to as the Particle Shifting Technique
(PST). Many different kinds of PST methods are available in the
literature~\cite{diff_smoothing_sph:lind:jcp:2012,fickian_smoothing_sph:skillen:cmame:2013,huang_kernel_2019,ye2019sph}.
An alternative approach that ensures particle homogenization for
incompressible fluid flow was proposed as the Transport Velocity Formulation
(TVF)~\cite{Adami2013}. The method introduced an additional stress term to
account for the motion introduced by the particle shifting. The TVF produces
very accurate results but only works for internal flows.
\citet{zhang_hu_adams17} proposed the Generalized Transport Velocity
Formulation (GTVF) thereby allowing the TVF to be used for free-surface
problems as well as elastic dynamics problems. This allows for a unified
treatment of both fluids and solids. Similarly, \citet{oger_ale_sph_2016}
introduce ideas from a consistent ALE formulation for improving the accuracy
of SPH. They employ a Riemann-based formulation to solve fluid mechanics
problems and introduce particle shifting to obtain highly accurate simulations
for internal and free-surface problems may also be handled. The PST has also
been employed in the context of the $\delta$-SPH
schemes\cite{sun_consistent_2019}.

The Entropically Damped Artificial Compressibility SPH scheme
(EDAC-SPH)~\cite{edac-sph:cf:2019} introduces an evolution equation for the
pressure and significantly reduces the noise in the pressure since it features
a pressure diffusion term. The approach has a thermodynamic
justification~\cite{Clausen2013} and produces very accurate
results~\cite{edac-sph:cf:2019}. The EDAC-SPH method uses the TVF formulation
for internal flows and for free-surface flows it does not employ any form of
particle shifting.

Recently, \citet{antuono2021delta} carefully combine the ALE-SPH method of
\citet{oger_ale_sph_2016} and the consistent $\delta$-SPH formulation of
\citet{sun_consistent_2019} to improve the accuracy of the $\delta$-SPH
method. They show the importance of the additional terms to the accuracy.

With the notable exception of the GTVF scheme~\cite{zhang_hu_adams17}, most
other applications of the PST have been in the context of fluid mechanics. The
GTVF method provides a unified approach to solve both weakly-compressible
fluids as well as solids. However, the method suffers from a few issues. In
order to work for free-surface problems the method relies on using a different
background pressure for each particle and introduces a few numerical
corrections to work around issues. For example, the smoothing length of the
homogenization force is different from that used by the other equations and
this parameter is somewhat ad-hoc. For solid mechanics problems the method
uses the transport velocity of the particle rather than the true velocity in
order to compute the strain and rotation tensor. In addition there are some
terms in the governing equations that are ignored which play a major role. We
also note that the method is not robust to a change in the particle
homogenization force.

In this work we propose a scheme which we called Corrected Transport Velocity
Formulation (CTVF) that is inspired by the various recent developments but is
consistent and which works for both solid mechanics and fluid mechanics
problems. We derive the transport velocity equations afresh and note that
there are some important terms that are ignored in earlier approaches using
TVF. These terms are significant and improve the accuracy of the method.
Similar to \cite{oger_ale_sph_2016,sun_consistent_2019}, we detect the free
surface particles and compute their normals using a simpler and
computationally efficient approach which does not require the computation of
eigenvalues. This allows the method to work with free-surfaces without the
introduction of numerical parameters or a variable background pressure. We
employ the EDAC formulation and show that there are additional correction
terms in the EDAC scheme that should be introduced to improve the accuracy of
the method. Furthermore, we show how the EDAC scheme can be used in the
context of solid mechanics problems. We make use of the particle velocity
rather than the transport velocity to compute the velocity gradient, strain,
and rotation rate tensors. Our method can be used with any PST and we consider
the method of \citet{sun_consistent_2019} as well as the iterative PST of
\citet{huang_kernel_2019}. The method is also robust to the choice of the
smoothing kernel. The resulting method works for both weakly-compressible
fluids as well as solids. The new method may be thought of as an improved
extension of the EDAC-SPH method that can be used for free-surface problems as
well as solid mechanics problems.

The method is implemented using the PySPH
framework~\cite{PR:pysph:scipy16,pysph2020}. The source code for all the
problems demonstrated in this manuscript is made available at
\url{https://gitlab.com/pypr/ctvf}. Every result produced in the manuscript is
fully automated using the \texttt{automan} package~\cite{pr:automan:2018}.

We next discuss the formulation for fluid mechanics as well as solid mechanics
along with the use of particle shifting. The consistent correction terms are
derived. We then consider a suite of benchmark problems for both fluids and
solids and compare our results with those of other methods where applicable.

\section{Governing equations}

For elastic dynamics we use the same equations as in
\cite{gray-ed-2001,zhang_hu_adams17} which we summarize below. The governing
equations of motion involve the conservation of mass, which in Lagrangian
form is,
\begin{equation}
  \label{eq:ce}
  \frac{d \rho}{d t} = - \rho \; \frac{\partial u_i}{\partial x_i},
\end{equation}
and conservation of linear momentum,
\begin{equation}
  \label{eq:me}
  \frac{d u_i}{d t} = \frac{1}{\rho} \; \frac{\partial \sigma_{ij}}{\partial x_j}
  + g_i,
\end{equation}
where $\rho$ is the density, $u_i$ is the $i$\textsuperscript{th} component of
the velocity field, $x_j$ is the $j$\textsuperscript{th} component of the
position vector, $g_i$ is the component of body force per unit mass and
$\sigma_{ij}$ is stress tensor.

The stress tensor is split into isotropic and deviatoric parts,
\begin{equation}
  \label{eq:stress_tensor_decomposition}
  \sigma_{ij} = - p \; \delta_{ij} + \sigma'_{ij},
\end{equation}
where $p$ is the pressure, $\delta_{ij}$ is the Kronecker delta function, and
$\sigma'_{ij}$ is the deviatoric stress.

The Jaumann's formulation for Hooke's stress provides us with the rate of
change of deviatoric stress,
\begin{equation}
  \label{eq:jaumann-stress-rate}
  \frac{d \sigma'_{ij}}{dt} = 2G (\dot{\epsilon}_{ij} - \frac{1}{3}
  \dot{\epsilon}_{kk} \delta_{ij}) + \sigma^{'}_{ik}  \Omega_{jk} +
  \Omega_{ik} \sigma^{'}_{kj},
\end{equation}
where $G$ is the shear modulus, $\dot{\epsilon}_{ij}$ is the strain rate tensor,
\begin{equation}
  \label{eq:strain-tensor}
  \dot{\epsilon}_{ij} = \frac{1}{2} \bigg(\frac{\partial u_i}{\partial x_j} +
  \frac{\partial u_j}{\partial x_i} \bigg),
\end{equation}
and $\Omega_{ij}$ is the rotation tensor,
\begin{equation}
  \label{eq:rotational-tensor}
  \Omega_{ij} = \frac{1}{2} \bigg(\frac{\partial u_i}{\partial x_j} -
  \frac{\partial u_j}{\partial x_i} \bigg).
\end{equation}

For a weakly-compressible or incompressible fluid, a viscous force is added:
\begin{equation}
  \label{eq:fluid-stress-decomposition}
  \sigma_{ij} = - p \delta_{ij} + 2 \eta \frac{\partial u_i}{\partial x_j}
\end{equation}
where $\eta$ is the kinematic viscosity of the fluid.

In both fluid and solid modelling the pressure is computed using an
isothermal equation of state, given as,
\begin{equation}
  \label{eq:pressure-equation}
  p = K \bigg(\frac{\rho}{\rho_{0}} - 1 \bigg),
\end{equation}
where $K = \rho_{0} c_0^2$ is the bulk modulus. Here, the constants $c_0$ and
$\rho_0$ are the reference speed of sound and density, respectively. For solids,
$c_0$ is computed as $\sqrt{\frac{E}{3 (1 - 2 \nu)\rho_{0}}}$, $\nu$ is the
Poisson ratio.

\section{Numerical method}

Following the TVF~\cite{Adami2013} and similar
formulations~\cite{oger_ale_sph_2016}, we move the particles with a
\emph{transport velocity}, $\tilde{\ten{u}}$. The material derivative in this
case is written as,
\begin{equation}
  \label{eq:modified-material-derivative}
  \frac{\tilde{d} }{d t} = \frac{\partial }{\partial t} +
  \tilde{u}_j \frac{\partial }{\partial x_j}.
\end{equation}

We therefore recast the governing equations to incorporate the transport
velocity starting with the conservation of mass, equation \eqref{eq:ce},
\begin{equation}
  \label{eq:ce-tvf-2}
  \frac{\tilde{d} \rho}{d t} = - \rho \frac{\partial u_j}{\partial x_j}+
  (\tilde{u}_j - u_j) \frac{\partial \rho} {\partial x_j}.
\end{equation}
Since,
\begin{equation}
  \label{eq:tmp:div-vv}
  (\tilde{u}_j - u_j)  \frac{\partial \rho}{\partial x_j} =
  \frac{\partial (\rho (\tilde{u}_j - u_j))}{\partial x_j} -
  \rho \frac{\partial (\tilde{u}_j - u_j)}{\partial x_j},
\end{equation}
we write equation~\eqref{eq:ce-tvf-2}, as
\begin{equation}
  \label{eq:ce-tvf}
  \frac{\tilde{d} \rho}{d t} =
  - \rho \frac{\partial \tilde{u}_j}{\partial x_j} +
  \frac{\partial (\rho (\tilde{u}_j - u_j))}{\partial x_j}.
\end{equation}

By combining the continuity equation~\eqref{eq:ce} and momentum
equation~\eqref{eq:me} one can obtain the conservative form of the momentum
equation as,
\begin{equation}
\begin{aligned}
  \label{eq:mom-eq-eulerian}
  \frac{\partial (\rho u_i)}{\partial t} +
  \frac{\partial} {\partial x_j} (\rho u_i u_j) &=
  \rho g_i + \frac{\partial \sigma_{ij}}{\partial x_j} \\
  &= \text{RHS}_i,
\end{aligned}
\end{equation}
where $g_i$ is the body force acceleration, and $\sigma_{ij}$ the stress tensor.
We write the left hand side in terms of a transport derivative as,
\begin{equation}
  \label{eq:tmp:mom1}
  \frac{\tilde{d} (\rho u_i)}{d t} +
  (u_j - \tilde{u}_j) \frac{\partial} {\partial x_j} (\rho u_i) +
  \rho u_i \frac{\partial u_j}{\partial x_j} = \text{RHS}_i.
\end{equation}
Similar to \cref{eq:tmp:div-vv}, we write,
\begin{equation}
  \label{eq:tmp:mom2}
  (\tilde{u}_j - u_j) \frac{\partial} {\partial x_j} (\rho u_i) =
  \frac{\partial}{\partial x_j} (\rho u_i (\tilde{u}_j - u_j)) - \rho u_i \frac{\partial}{\partial x_j} (\tilde{u}_j - u_j).
\end{equation}
Substituting, this in \cref{eq:tmp:mom1}, we have,
\begin{equation}
  \label{eq:tmp:mom3}
  \frac{\tilde{d} (\rho u_i)}{d t} =
  \frac{\partial}{\partial x_j} (\rho u_i (\tilde{u}_j - u_j)) - \rho u_i \frac{\partial}{\partial x_j} (\tilde{u}_j) + \text{RHS}_i.
\end{equation}
In \citet{Adami2013}, the second term is neglected but at this stage we do not
neglect this term. We simplify this further and write,
\begin{equation}
  \label{eq:tmp:mom4}
  \rho \frac{\tilde{d} u_i}{d t} + u_i \frac{\tilde{d} \rho}{d t} =
  \frac{\partial}{\partial x_j} (\rho u_i (\tilde{u}_j - u_j)) - \rho u_i \frac{\partial}{\partial x_j} (\tilde{u}_j) + \text{RHS}_i.
\end{equation}
Using, \cref{eq:ce-tvf}, we write
\begin{equation}
  \label{eq:tmp:mom5}
  \begin{aligned}[b]
    \rho \frac{\tilde{d} u_i}{d t} &= \frac{\partial}{\partial x_j} (\rho u_i
    (\tilde{u}_j - u_j)) - u_i \frac{\partial}{\partial x_j}(\rho (\tilde{u}_j
    - u_j)) + \text{RHS}_i\\
    &= \rho (\tilde{u}_j - u_j) \frac{\partial u_i}{\partial x_j} + \text{RHS}_i \\
    &= \rho \frac{\partial}{\partial x_j} (u_i (\tilde{u}_j - u_j)) -
    \rho u_i \frac{\partial}{\partial x_j} (\tilde{u}_j - u_j)
    + \text{RHS}_i.
  \end{aligned}
\end{equation}
We therefore write the momentum equation as,
\begin{equation}
  \label{eq:mom-tvf}
  \frac{\tilde{d} u_i}{d t} =
  \frac{\partial}{\partial x_j} (u_i (\tilde{u}_j - u_j))
  - u_i \frac{\partial}{\partial x_j} (\tilde{u}_j - u_j)
  + g_i
  +\frac{1}{\rho} \frac{\partial \sigma_{ij}}{\partial x_j}.
\end{equation}
We note that this equation encompasses both fluid dynamics as well as elastic
dynamics by simply changing the way $\sigma_{ij}$ is modeled. The first term
on the right-hand-side of \cref{eq:mom-tvf} is the additional artificial
stress term that is included in the TVF~\cite{Adami2013}. The second term
involves the divergence of the transport velocity field. In the case of the
TVF, the term includes a background pressure acceleration that is of the form,
\begin{equation}
  \label{eq:tvf-accel}
  \bigg(\frac{d \ten{u}_a}{dt}\bigg)_{c} = - p^0_a \sum_{b \in N(a)}
  \frac{m_b}{\rho_b^2} \nabla W(\ten{r}_{ab}, \tilde{h}_{ab}),
\end{equation}
where $p^0_a$ is the background pressure for the given particle $a$,
$\ten{r}_{ab} = \ten{r}_a - \ten{r}_b$, $\tilde{h}_{ab} = (h_a + h_b)/2$, and
index $b$ refers to the neighbors of particle $a$. The divergence of this term
results in the Laplacian of the kernel $W$. For most kernels used in SPH, this
term is certainly not zero and therefore this should not be ignored. We
investigate the importance of including these terms in \cref{sec:results}. We
note that in the case of elastic dynamics that these terms are negligible and
do not make a significant difference. This has also been pointed out by
\citet{zhang_hu_adams17}.

The Jaumann stress rate is also similarly modified to account for the
transport velocity as,
\begin{multline}
  \label{eq:modified-jaumann-stress-rate}
  \frac{\tilde{d} \sigma'_{ij}}{dt} = 2G (\dot{\epsilon}_{ij} - \frac{1}{3}
  \dot{\epsilon}_{kk} \delta_{ij}) + \sigma^{'}_{ik}  \Omega_{jk} +
  \Omega_{ik} \sigma^{'}_{kj} + \\
  \frac{\partial}{\partial x_k}\big(\sigma^{'}_{ij}  (\tilde{u}_k - u_k)\big)
  - \sigma^{'}_{ij} \frac{\partial}{\partial x_k} (\tilde{u}_k - u_k).
\end{multline}

\subsection{The EDAC-SPH method}
\label{sec:edac-tvf}

We apply the EDAC-SPH~\cite{edac-sph:cf:2019} in order to evolve the pressure
accurately and reduce the amount of noise in the pressure field. In the
original EDAC-SPH implementation, internal flows were evolved using the TVF
whereas for cases with a free-surface the traditional WCSPH was employed. In
this work we propose a unified approach by carefully incorporating
free-surfaces. The original EDAC-SPH scheme also did not accurately
incorporate the transport velocity which we include here. This allows us to
use the same scheme for both internal and external flows.

The $\delta$-SPH~\cite{antuono-deltasph:cpc:2010} implementation is in
principle similar to the EDAC-SPH method but requires the use of the kernel
gradient corrections which involve the solution of a small matrix ($3 \times
3$ in 3D), for each particle. The EDAC-SPH method does not require this and is
therefore simpler and in principle more efficient. In \cite{edac-sph:cf:2019},
the EDAC pressure evolution equation was,
\begin{equation}
  \label{eq:edac-p-evolve-orig}
     \frac{d p}{d t} = -\rho c_s^2 \text{div}(\ten{u}) + \nu_{edac}  \nabla^2 p,
\end{equation}
where $\nu_{edac}$ is a viscosity parameter for the smoothing of the pressure
and $c_s$ is the (artificial) speed of sound. We discuss this term later.
However, in the context of the consistent evolution using the transport
velocity, we note that the above should be evolved using,
\begin{equation}
  \label{eq:edac-p-evolve}
  \frac{\tilde{d} p}{d t} =
  (p -\rho c_s^2)
    \text{div}(\ten{u})
  - p \; \text{div}(\tilde{\ten{u}})
    + \text{div}(p (\tilde{\ten{u}} - \ten{u}))
    + \nu_{edac}  \nabla^2 p.
\end{equation}
The value of $\nu_{edac}$ is,
\begin{equation}
  \label{eq:nu-edac}
  \nu_{edac} = \frac{\alpha_{\textrm{edac}} h c_s}{8},
\end{equation}
where $h$ is the smoothing length of the kernel and a value of
$\alpha_{\textrm{edac}}=0.5$ is recommended as suggested in~\cite{PRKP:edac-sph-iccm2015}.

This along with the momentum equation and evolution of volume or density may
be employed. A state equation is often used even for elastic dynamics
problems, we propose to use the EDAC approach for elastic equations as well as
this reduces the amount of artificial viscosity that is needed.

\subsection{SPH discretization}

In the current work, both fluid and solid modelling uses the same continuity
and pressure evolution equation. The SPH discretization of the continuity
equation~\eqref{eq:ce-tvf} and the pressure evolution
equation~\eqref{eq:edac-p-evolve} respectively are,
\begin{equation}
  \label{eq:sph-discretization-continuity}
  \frac{\tilde{d}\rho_a}{dt} = \sum_{b} \; \frac{m_b}{\rho_{b}} \; (
  \rho_{a} \; \tilde{\ten{u}}_{ab} \; + \;
  (\rho \; (\tilde{\ten{u}} \; - \;
  \ten{u}))_{ab}) \; \cdot \nabla_{a} W_{ab},
\end{equation}

\begin{multline}
  \label{eq:sph-discretization-edac}
  \frac{\tilde{d}p_a}{dt} = \sum_{b} \; \frac{m_b}{\rho_{b}} \; \bigg(
  (p_{a} - \rho_{a} c_{s}^2) \; \ten{u}_{ab} \; + \;
  p_{a} \; \tilde{\ten{u}}_{ab} \; - \;
  (p \; (\tilde{\ten{u}} - \ten{u}))_{ab} \; + \; \\
  4 \; \nu_{edac}
  \frac{p_a - p_b}{(\rho_a + \rho_b) (r^2_{ab} + 0.01 h_{ab}^{2})} \ten{r}_{ab}
  \bigg) \; \cdot \nabla_{a} W_{ab}.
\end{multline}
Similarly, the discretized momentum equation for fluids is written as,
\begin{multline}
  \label{eq:sph-momentum-fluid}
  \frac{\tilde{d}\ten{u}_{a}}{dt} = - \sum_{b} m_b \bigg[
  \bigg(\frac{p_a}{\rho_a^2} + \frac{p_b}{\rho_b^2}\bigg) \ten{I} -
  \bigg(\frac{\ten{A}_a}{\rho_a^2} + \frac{\ten{A}_b}{\rho_b^2}
  \bigg) \bigg]
  \cdot \nabla_{a} W_{ab} \\
  + \ten{u}_{a} \sum_{b} \frac{m_b}{\rho_{b}} \; \tilde{\ten{u}}_{ab} \cdot
  \nabla_{a} W_{ab} + \sum_{b} m_b \frac{4 \eta \nabla W_{ab}\cdot
    \ten{r}_{ab}}{(\rho_a + \rho_b) (r_{ab}^2 + 0.01 h_{ab}^2)} \ten{u}_{ab} +
  \ten{g}_{a},
\end{multline}
where $\ten{A}_a = \rho_a \ten{u}_a \otimes (\ten{\tilde{u}}_a - \ten{u}_a)$,
$\ten{I}$ is the identity matrix, $\eta$ is the kinematic viscosity of the
fluid and \citet{morris1997modeling} formulation is used to discretize the
viscosity term.

We add to the momentum equation an additional artificial viscosity term
$\Pi_{ab}$~\cite{monaghan-review:2005} to maintain the stability of the
numerical scheme, given as,
\begin{align}
  \label{eq:mom-av}
  \Pi_{ab} =
  \begin{cases}
\frac{-\alpha h_{ab} \bar{c}_{ab} \phi_{ab}}{\bar{\rho}_{ab}}
  & \ten{u}_{ab}\cdot \ten{r}_{ab} < 0, \\
  0 & \ten{u}_{ab}\cdot \ten{r}_{ab} \ge 0,
\end{cases}
\end{align}
where,
\begin{equation}
  \label{eq:av-phiij}
  \phi_{ab} = \frac{\ten{u}_{ab} \cdot \ten{r}_{ab}}{r^2_{ab} + 0.01 h^2_{ab}},
\end{equation}
where $\ten{r}_{ab} = \ten{r}_a - \ten{r}_b$, $\ten{u}_{ab} = \ten{u}_a -
\ten{u}_b$, $h_{ab} = (h_a + h_b)/2$, $\bar{\rho}_{ab} = (\rho_a + \rho_b)/2$,
$\bar{c}_{ab} = (c_a + c_b) / 2$, and $\alpha$ is the artificial
viscosity parameter.

For solid mechanics the momentum equation is written as,
\begin{equation}
  \label{eq:sph-momentum-solid}
  \frac{\tilde{d}\ten{u}_{a}}{dt} = - \sum_{b} m_b \bigg[
  \bigg(\frac{p_a}{\rho_a^2} + \frac{p_b}{\rho_b^2}\bigg) \ten{I} -
  \bigg(\frac{\teng{\sigma}^{'}_{a}}{\rho_a^2} +
  \frac{\teng{\sigma}^{'}_{b}}{\rho_b^2} + \Pi_{ab} \ten{I} \bigg) \bigg]  \cdot \nabla_{a} W_{ab} +
  \ten{g}_{a},
\end{equation}
we have not considered the correction stress term $\ten{A}$ in momentum
equation of solid mechanics as it has a negligible effect.

In addition to these three equations, the Jaumann stress rate equation is also
solved. In the current work we use the momentum velocity $\ten{u}$ rather than
$\tilde{\ten{u}}$ as in the GTVF~\cite{zhang_hu_adams17} in the computation of
gradient of velocity. The SPH discretization of the gradient of velocity is
given as,
\begin{equation}
  \label{eq:sph-vel-grad}
  \nabla \ten{u}_a =
  - \sum_{b} \frac{m_b}{\rho_{b}} (\ten{u}_{a} - \ten{u}_{b}) \otimes (\nabla_{a} W_{ab}),
\end{equation}
where $\otimes$ is the outer product.

The SPH discretization of the modified Jaumann stress rate
\cref{eq:modified-jaumann-stress-rate} is given as,
\begin{multline}
  \label{eq:sph-modified-jaumann-stress}
  \frac{\tilde{d}\teng{\sigma}^{'}_{a}}{dt} = 2G (\dot{\teng{\epsilon}}_{a} -
  \frac{1}{3} \dot{\teng{\epsilon}}_{a} \ten{I}) + \teng{\sigma}^{'}_{a}
  \teng{\Omega}_{a}^{T} +
  \teng{\Omega}_{a} \teng{\sigma}^{'}_{a} + \\
  + \sum_{b} \; \frac{m_b}{\rho_{b}} \; (\teng{\sigma}^{'} \otimes (\tilde{\ten{u}} -
  \ten{u}))_{ab} \; \cdot \nabla_{a} W_{ab}
  + \teng{\sigma}^{'}_{a} \sum_{b} \; \frac{m_b}{\rho_{b}} \;
  (\tilde{\ten{u}} - \ten{u})_{ab} \; \cdot \nabla_{a} W_{ab}.
\end{multline}

\subsection{Particle transport}

The particles in the current scheme are moved with the transport velocity,
\begin{equation}
  \label{eq:transport_velocity_position_derivative}
  \frac{d\ten{r}_a}{dt} = \ten{\tilde{u}}_a.
\end{equation}
The transport velocity is updated using,
\begin{equation}
  \label{eq:transport_velocity}
  \ten{\tilde{u}}_a(t + \Delta t) =\ten{u}_a(t) + \Delta t \; \frac{\tilde{d} \ten{u}_a}{dt} +
  \bigg(\frac{d \ten{u}_{a}}{dt}\bigg)_{\text{c}} \Delta t,
\end{equation}
where $\big(\frac{d \ten{u}_a}{dt}\big)_{\text{c}}$ is the homogenization
acceleration which ensures that the particle positions are homogeneous. In the
current work we have explored two kinds of homogenization accelerations, one
is a displacement based technique due to \citet{sun2017deltaplus}, which here
after we refer as SPST and the other is the iterative particle shifting
technique due to \citet{huang_kernel_2019} referred as IPST. These are
discussed in the following.

\subsubsection{Sun 2019 PST}
\label{sec:sunpst}

In \citet{sun2017deltaplus}, the particle shifting technique was implemented as
a particle displacement ($\delta \ten{r}$). This was modified in
\citet{sun_consistent_2019} to be computed as a change to the velocity. In the
present work we modify this to be treated as an acceleration to the particle
in order to unify the treatment of different PST methods.

Firstly, the velocity deviation based equation is given as,
\begin{equation}
  \label{eq:sun2019_pst}
  \delta \ten{u}_a = - \text{Ma} \; (2h) c_0 \sum_b \bigg[
  1 + R \bigg( \frac{W_{ab}}{W(\Delta x)} \bigg)^n  \bigg] \nabla_a W_{ab} V_b,
\end{equation}
it is modified to force based as,
\begin{equation}
  \label{eq:sun2019_pst}
  \bigg(\frac{d \ten{u}_a}{dt}\bigg)_{\text{c}} = - \frac{\text{Ma} \;
    (2h) c_0}{\Delta t} \sum_b \bigg[1 + R \bigg( \frac{W_{ab}}{W(\Delta x)} \bigg)^n
  \bigg] \nabla_a W_{ab} V_b,
\end{equation}
where $R$ is an adjustment factor to handle the tensile instability, and
$\text{Ma}$ is the mach number of the flow. $V_b$ is the volume of the
$b$\textsuperscript{th} particle. The acceleration is changed to account for
particles that are on the free surface. We use $R = 0.2$ and $n = 4$ as
suggested by \citet{sun_consistent_2019}.

\subsubsection{IPST}
\label{sec:ipst}

The Iterative PST of Huang \cite{huang_kernel_2019} builds on the work of
\citet{xu2009accuracy}. The method iteratively moves the particles every
timestep in order to achieve a uniform particle distribution determined by a
convergence criterion. The properties of the particles are corrected using a
Taylor series expansion.

In the original IPST, the shifting vector is computed as,
\begin{equation}
  \label{eq:ipst_step1}
  \delta r_a^m = U_{\max} \; \Delta t \; \sum_b (V_b \; \ten{n}_{ab} \; W_{ab})^m,
\end{equation}
where m is the number of iterations, $m=(1, 2, 3, \dots)$, $\ten{n}_{ab}$ is
the unit vector between particle $a$ and $b$. The particles are then moved
using this displacement using,
\begin{equation}
  \label{eq:ipst_step2}
  \ten{r}_a^{m+1} = \ten{r}_a^{m} + \delta \ten{r}_a^{m}.
\end{equation}
This is repeated until the convergence criterion is achieved. The convergence
criterion is defined as
\begin{equation}
  \label{eq:ipst_convergence_criterion}
  |\max(\chi_a^m) - \overline{\chi_a^m}| \leq \epsilon,
\end{equation}
where
\begin{equation}
  \label{eq:ipst_chi_0}
  \chi_a^m = h^2 \sum W^m_{ab},
\end{equation}
and $\overline{\chi_a^m}$ is the value of $\chi_a^m$ computed with the initial
configuration of the particles. For problems where there is no free surface
this value is a constant computed using the initial configuration. For
free-surface problems it is computed as the maximum value of $\chi_a^m$ at the
initial configuration, which corresponds to the a free surface particle.

The initial and final positions of the particles are used to determine an
acceleration on the particle that would produce such a displacement.  This is
computed as
\begin{equation}
  \label{eq:ipst_force}
  \bigg(\frac{d \ten{u}_a}{dt}\bigg)_{\text{c}} =
  2 \; \frac{\ten{r}_a^{M} - \ten{r}_a^0}{\Delta t^2},
\end{equation}
where $\ten{r}_a^M$ is the final position of the particle, $a$.

\subsection{Free surface identification algorithm}
\label{subsec:free-surface}

Free surfaces must be handled with care especially in the context of the PST
algorithm. The original TVF~\cite{Adami2013} is not designed to handle
free-surface problems. \citet{diff_smoothing_sph:lind:jcp:2012} was the first
to handle free-surfaces carefully in the context of the PST.
\citet{diff_smoothing_sph:lind:jcp:2012,oger_ale_sph_2016}, and
\citet{sun_consistent_2019} identify the particles that are on the
free-surface or near it and adjust the particle shifting algorithm so the free
surface particles remain intact. \citet{zhang_hu_adams17} on the other hand
relies on the pressure being zero at the free surface and scales the
homogenization force with the pressure.

Both \cite{oger_ale_sph_2016} and \cite{sun_consistent_2019} identify the
free-surface particles by computing the eigenvalues of the correction matrix
employed for the SPH scheme. This is based on the work of
\cite{marrone:sph:level-set:2010}. This method is computationally expensive
and in this work we use a much simpler approach that was introduced
in~\cite{muta_efficient_2020} to find the free-surface particles as well as
their normals. We first compute the normals of all the particles in the medium
whose free surface need to be identified. The normals are computed as,
\begin{equation}
  \label{eq:normal-approx}
  \ten{n}^*_a = \sum_b -\frac{m_b}{\rho_b} \nabla_a W_{ab}
\end{equation}
If the magnitude of the resulting vector is less than $\frac{1}{4h_a}$, then
the $\ten{n}^*$ is set to zero otherwise we normalize the vector by its
magnitude. Then, we smooth these normals using an SPH approximation,
\begin{equation}
  \label{eq:normal}
  \hat{\ten{n}}_a = \sum_b \frac{m_b}{\rho_b}\ten{n}^*_b W_{ab}.
\end{equation}
Finally, we normalize $\hat{\ten{n}}_a$ so they are unit vectors. We note that
for particles that are isolated and have no neighbors the above algorithm will
not work. We identify all such particles by computing the summation density of
all particles and any particles with a summation density that is lower than
half the fluid density are marked as free-surface particles. For the quintic
spline kernel used in this work we find that the cutoff value of half the
fluid density is effective in differentiating particles that are away from the
bulk fluid.

\begin{figure}[!htpb]
  \centering
  \begin{subfigure}{0.48\textwidth}
    \centering
    \includegraphics[width=1\linewidth]{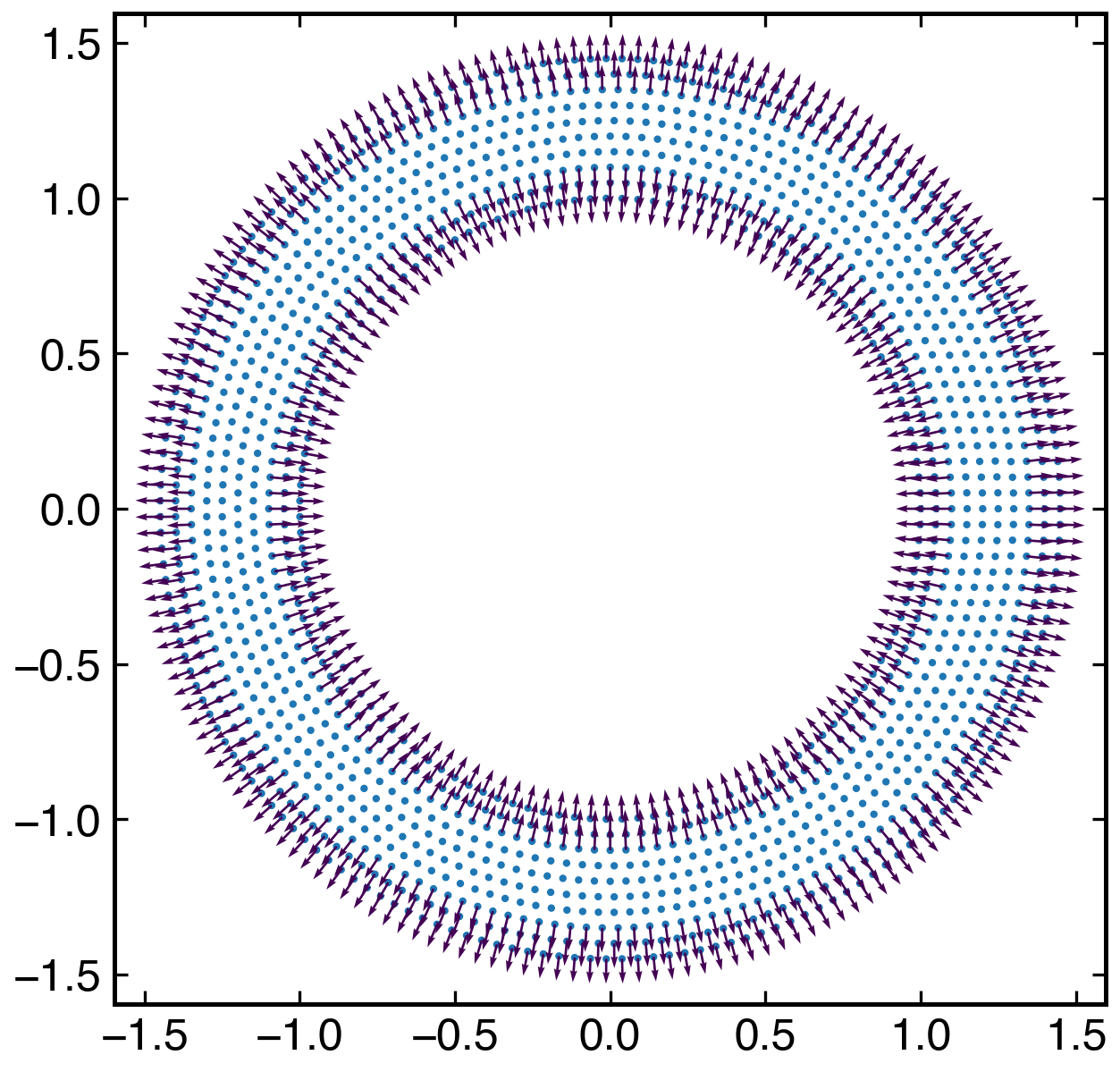}
    \subcaption{}%
    \label{fig:free_surface_circle_normals}
  \end{subfigure}
  \begin{subfigure}{0.48\textwidth}
    \centering
    \includegraphics[width=1\linewidth]{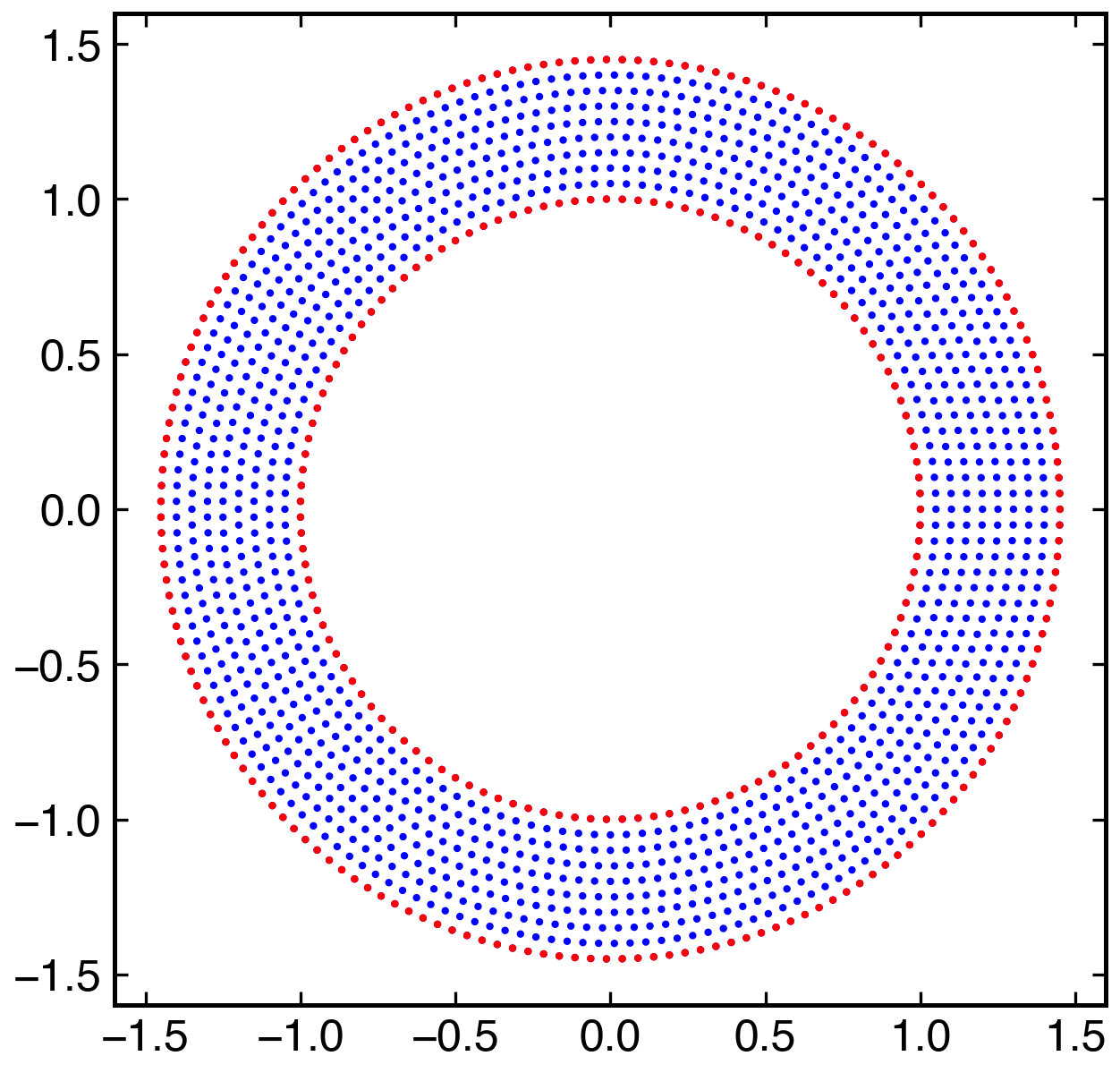}
    \subcaption{}%
    \label{fig:free_surface_circle_boundary_particles}
  \end{subfigure}
  \caption{Identification of free surface particles of a circular ring of
    fluid. Depicts (a) normals of the fluid particles, (b) boundary particles
    of the fluid particles}
\label{fig:boundary-particles-circle}
\end{figure}

\begin{figure}[!htpb]
  \centering
  \begin{subfigure}{0.48\textwidth}
    \centering
    \includegraphics[width=1\linewidth]{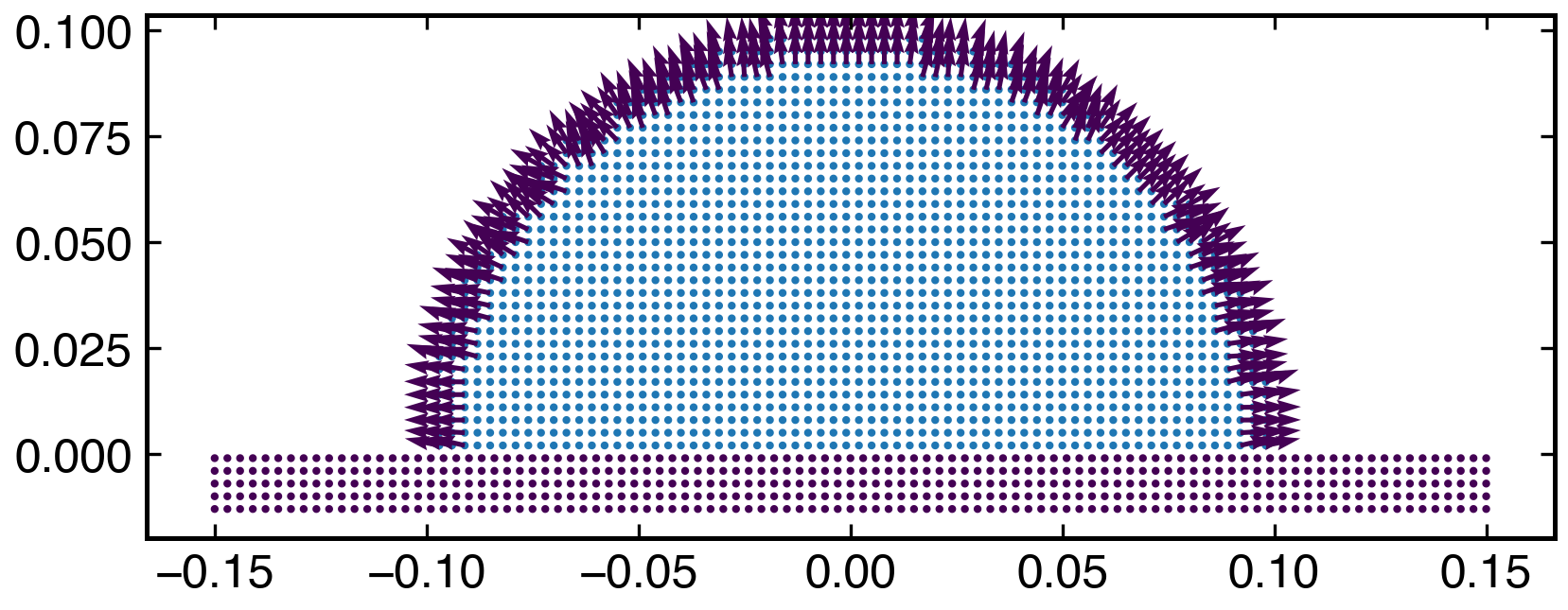}
    \subcaption{}%
    \label{fig:boundary-particles-circle-wall-normals}
  \end{subfigure}
  \begin{subfigure}{0.48\textwidth}
    \centering
    \includegraphics[width=1\linewidth]{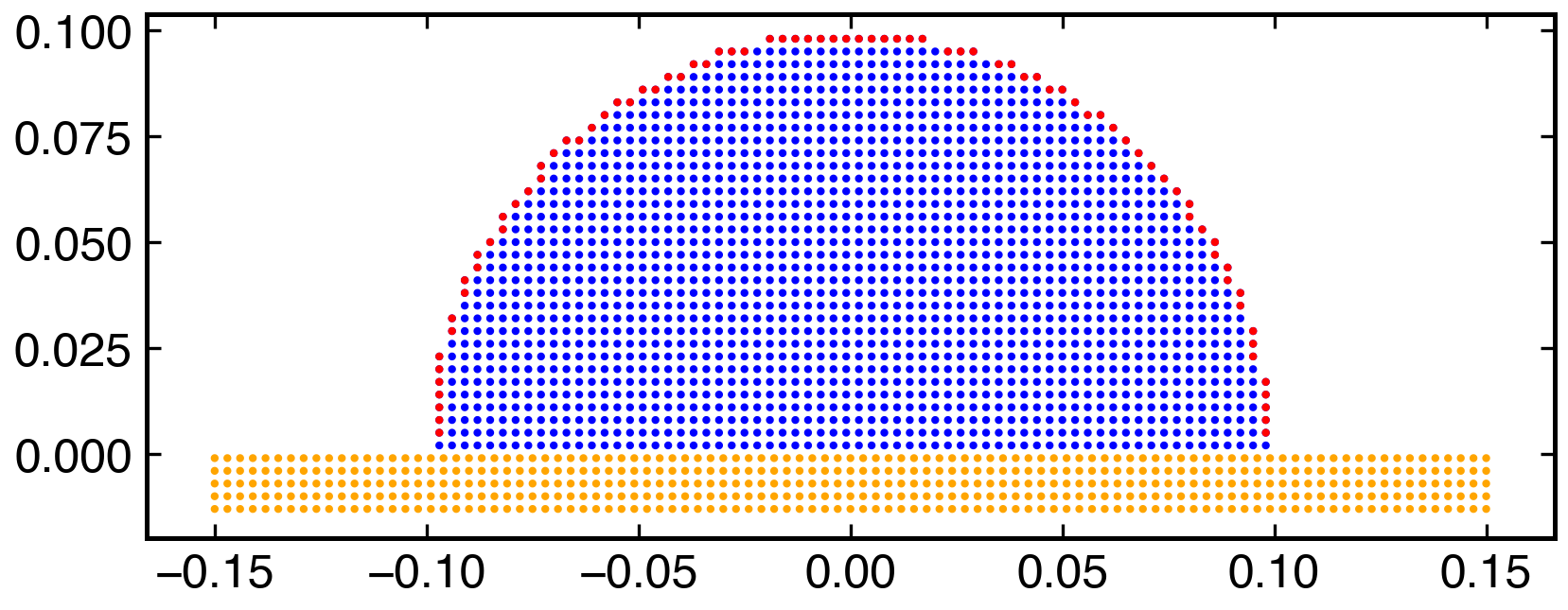}
    \subcaption{}%
    \label{fig:boundary-particles-circle-wall-bp}
  \end{subfigure}
  \caption{Identification of free surface particles of a fluid resting on a
    wall. Depicts (a) normals of the fluid particles, (b) boundary particles
    of the fluid particles}
\label{fig:boundary-particles-circle-wall}
\end{figure}

\begin{figure}[!htpb]
  \centering
  \includegraphics[width=1\linewidth]{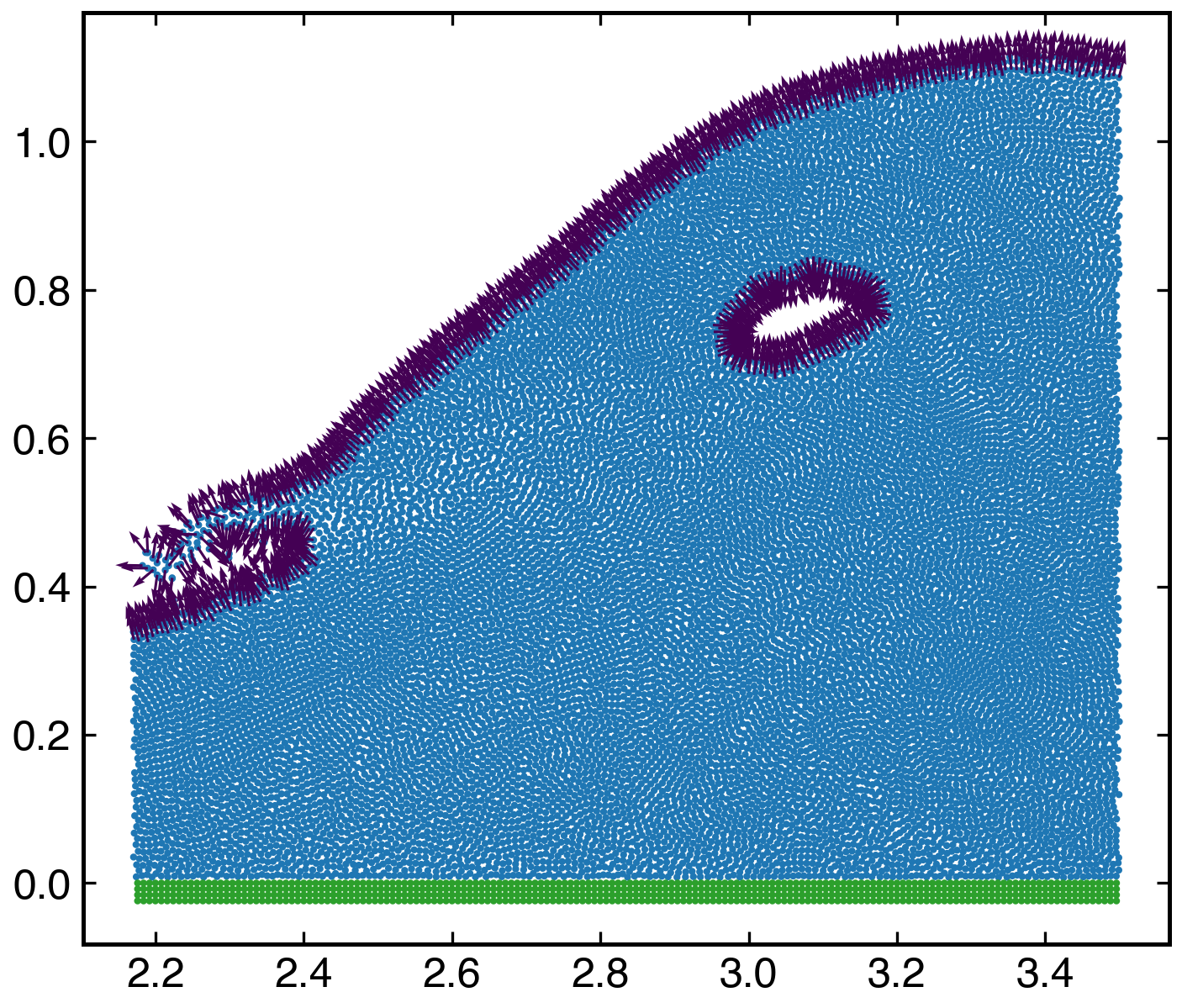}
  \caption{Identification of normals of fluid in a dam break simulation. Shows
    us the normals of all the fluid particles}
\label{fig:normals}
\end{figure}

\begin{figure}[!htpb]
  \centering
  \includegraphics[width=1\linewidth]{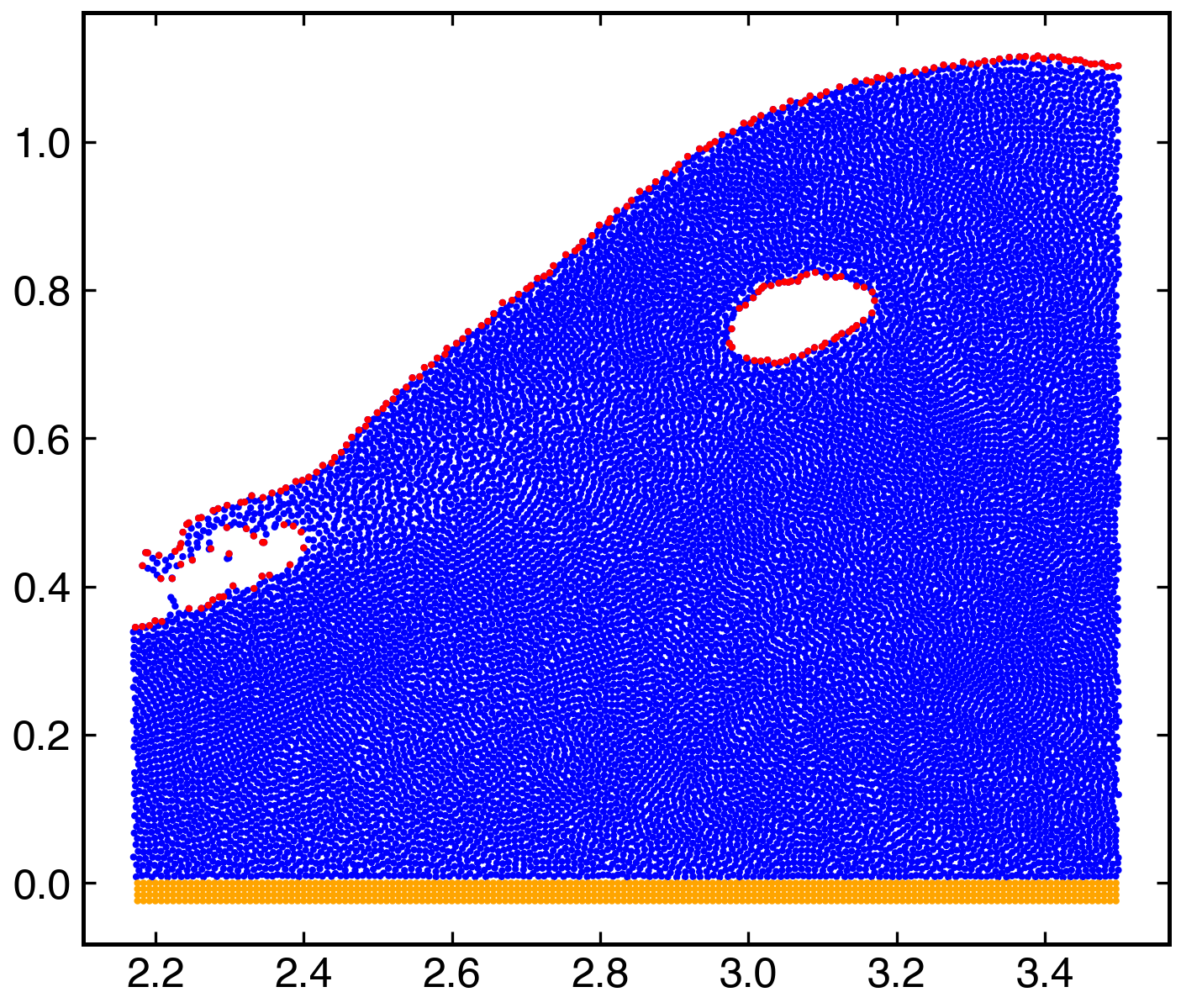}
  \caption{Identification of boundary particles of fluid at an instance in a
    dam break. Shows boundary particles of all the fluid particles}
\label{fig:boundary-particles}
\end{figure}

The current algorithm is tested with to two simple cases. We first consider a
circular ring of fluid. As can be seen from
\cref{fig:free_surface_circle_normals}, three layers of particles which are
near the free surface have a normal. From these normals we need to find the
surface particles. We loop over all the particles in the medium, and any
particles which have a non-zero normal are considered for further analysis.
For each of these potential surface particles we find the angle between the
particle and each of its neighbors. For a 2d case, if the angle between the
normal of the particle and that of the line joining the particle to its
neighbor is less than 60 degrees then this particle is \emph{not} considered
as a free-surface particle. If no such neighbor exists, then the particle is
considered to be a free-surface particle. As can be seen in
\cref{fig:free_surface_circle_boundary_particles}, the free surface particles
are correctly identified. As a second test case we consider a patch of fluid
resting on a wall. As a first step we compute the normals of the particles,
see \cref{fig:boundary-particles-circle-wall-normals} and then loop over all
the particles and by considering only the particles which have non-zero
normals, we determine the boundary particles, as in
\cref{fig:boundary-particles-circle-wall-bp}. Finally this is applied to the
case of a dam break. As can be seen from \cref{fig:normals} and
\cref{fig:boundary-particles}, the free surface particles are identified
correctly.

\subsection{PST close to the free surface}
\label{sec:pst-free-surf}

Near the free-surface the PST has to be performed with some care for both
fluids and solids. This is because of the lack of support for the particles
near the free-surface. After the free surface particles are identified by
using the algorithm described in \cref{subsec:free-surface}, we mark the
particles which are in close proximity to the free surface particles. This is
done through a variable associated with each particle called $h_{b}$, which is
initialized to the initial smoothing length of the particles.

We loop over all the particles that are not on the boundary, and their $h_b$
is adjusted to the distance to the closest boundary particle divided
appropriately by a kernel-dependent factor such that the kernel support is up
to the closest boundary particle. In the current work, we have used a quintic
spline kernel for which the factor is 3. The algorithm is shown in
\cref{alg:hb} and depicted in \cref{fig:pst_free_surf}. We note that this
$h_b$ is only used for the PST force/displacement computation. This process
allows us to ensure that the homogenization force does not push these
particles towards the free-surface.

\begin{algorithm}[!ht]
  \caption{Algorithm to set $h_b$}%
  \label{alg:hb}
  \begin{algorithmic}[1]
    \For{particle $i$ in all particles}
    \If{$i$ is a boundary particle}
    \State{set $h_{b, i} = 0$}
    \Else
    \State{set $h_{b,i} = h$}
    \EndIf
    \EndFor%
    \For{particle $i$ in all non-boundary particles}
    \If{particle $i$ has a boundary particle in its neighborhood}
     \State{$x_{\text{dist}, i} \leftarrow$ Distance to nearest boundary
       particle}
     \State{Set $h_{b,i} = \frac{x_{\text{dist, i}}}{3}$}
    \EndIf
    \EndFor%
  \end{algorithmic}
\end{algorithm}

\begin{figure}[!htpb]
  \centering
  \begin{subfigure}{0.3\textwidth}
    \centering
    \includegraphics[width=1\linewidth]{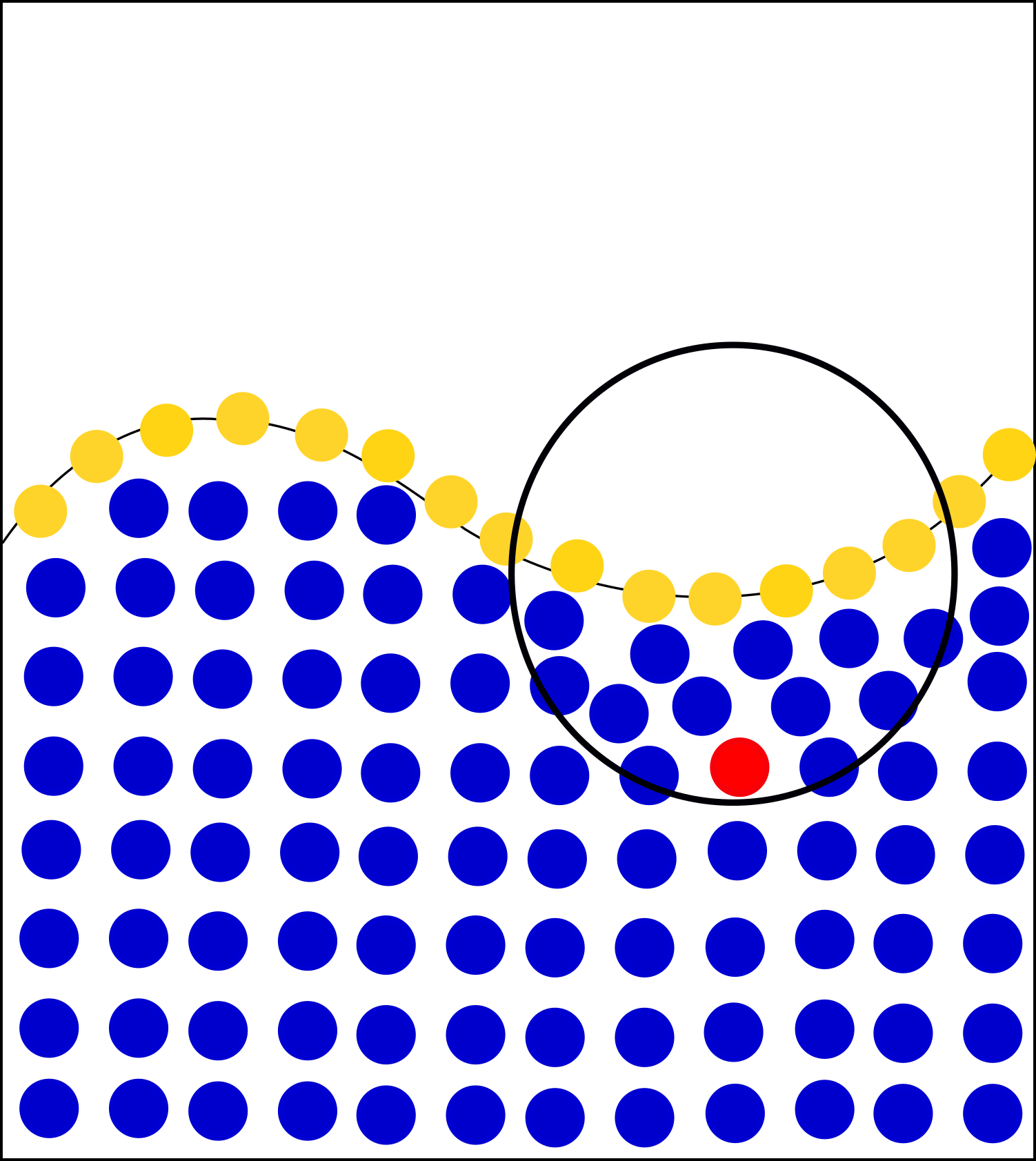}
    \subcaption{}%
    \label{fig:pst_free_surf}
  \end{subfigure}
  \begin{subfigure}{0.1\textwidth}
    \centering
    \includegraphics[width=1\linewidth]{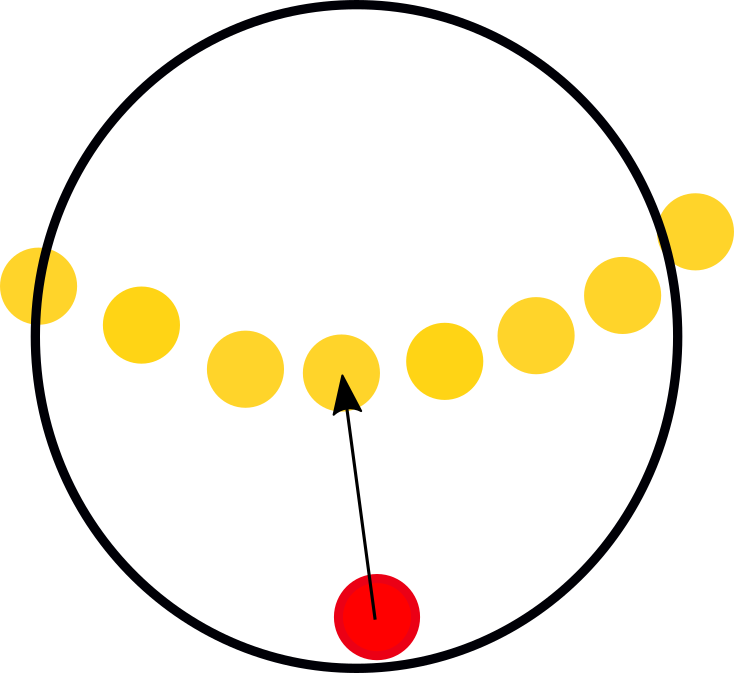}
    \subcaption{}%
    \label{fig:pst_free_surf_zoomed}
  \end{subfigure}
  \caption{
    Set $h_b$ of the particles. (a) Particles with a free surface whose
    free particles are identified (b) Minimum distance between the particle in
    the vicinity of the free surface to the free surface particle.}
\label{fig:pst_free_surf}
\end{figure}

In the PST of \citet{sun_consistent_2019}, the shifting acceleration is
adjusted using
\begin{equation}
 \label{eq:shifting_force_free_surface_adjust_sun2019}
 \bigg(\frac{d \ten{u}_a}{dt}\bigg)_{\text{c}} =\begin{cases}
   0& \text{if boundary},\\
   \big(\frac{d \ten{u}_a}{dt}\big)_{\text{c}}  - (\big(\frac{d \ten{u}_a}{dt}\big)_{\text{c}} \cdot \ten{n}_a) \ten{n}_a& \text{if $h_b < h$},\\
   \big(\frac{d \ten{u}_a}{dt}\big)_{\text{c}}& \text{if $h_b = h$}.
 \end{cases}
\end{equation}
Whereas while using IPST~\cite{huang_kernel_2019}, rather than adjusting the
final shifting acceleration, we adjust the increment in the position of
\cref{eq:ipst_step2},
\begin{equation}
 \label{eq:shifting_force_free_surface_adjust_ipst}
 \delta \ten{r}_a^{m} =\begin{cases}
   0& \text{if boundary},\\
   \delta \ten{r}_a^{m} - (\delta \ten{r}_a^{m}  \cdot \ten{n}_a) \ten{n}_a& \text{if $h_b < h$},\\
   \delta \ten{r}_a^{m}& \text{if $h_b = h$}.
 \end{cases}
\end{equation}

\subsection{Boundary conditions}

The ghost particle approach of \citet{Adami2012} is used to model the
boundaries. We use three layers of ghost particles to model the solid wall.
The properties of the solid wall are interpolated from the fluid particles.

When computing the divergence of the velocity field on fluid particles, we
enforce a no-penetration boundary condition and not a no-slip boundary
condition. The velocity of the fluid is projected onto the ghost particles
using,
\begin{equation}
  \label{eq:v-ghost}
  \ten{\hat{u}}_a = \frac{\sum_b\ten{u}_b W_{ab}}{\sum_b W_{ab}},
\end{equation}
\begin{equation}
  \label{eq:v-hat-ghost}
  \ten{\check{u}}_a = \frac{\sum_b\tilde{\ten{u}}_b W_{ab}}{\sum_b W_{ab}},
\end{equation}
where $\ten{u}_b$, $\ten{\tilde{u}}_b$ are the momentum and transport velocity
of the fluid respectively and $W_{ab}$ is the kernel value between the fluid
particle and the ghost particle.

The normal component of this projected velocity is then reflected and set as
the ghost particle velocity,
\begin{equation}
  \label{eq:free-slip-bc-u}
  \ten{u}_{\text{Ga}} = 2 \ten{\hat{n}}((\ten{u}_{\text{p}} - \ten{\hat{u}}_{\text{a}})\cdot \ten{\hat{n}}) + \ten{\hat{u}}_{\text{a}},
\end{equation}
where $\ten{u}_{\text{p}}$ is the local velocity of the boundary and
$\ten{\hat{n}}$ is the unit normal to the boundary particle $a$. Similarly the
transport velocity of the ghost particle is set as,
\begin{equation}
  \label{eq:free-slip-bc-u}
  \tilde{\ten{u}}_{\text{Gi}} = 2 \ten{\hat{n}}((\ten{u}_{\text{p}} - \ten{\check{u}}_{\text{i}})\cdot \ten{\hat{n}}) + \ten{\check{u}}_{\text{i}},
\end{equation}

When the viscous force is computed, the no slip boundary condition is used,
where the velocity on the boundary set as,
\begin{equation}
  \label{eq:no-slip-bc-u}
  \ten{u}_{\text{Ga}} = 2 \ten{u}_{\text{p}} - \ten{\hat{u}}_{\text{a}},
\end{equation}
a similar form is used for the transport velocity here too,
\begin{equation}
  \label{eq:no-slip-bc-uhat}
  \tilde{\ten{u}}_{\text{Ga}} = 2 \ten{u}_{\text{p}} - \ten{\check{u}}_{\text{a}}.
\end{equation}

The pressure of the boundary particle is extrapolated from its surrounding
fluid particles by the following equation,
\begin{equation}
  \label{eq:pressure-bc}
  p_w = \frac{\Sigma_f p_f W_{wf} + (\ten{g} - \ten{a}_{\ten{w}}) \cdot \Sigma_f
    \rho_f \ten{r}_{wf} W_{wf}}{\Sigma_f W_{wf}},
\end{equation}
where $\ten{a}_w$ is the acceleration of the wall. The subscript $f$ denotes
the fluid particles and $w$ denotes the wall particles.

For solid mechanics problems, in addition to the extrapolation of pressure, we
also extrapolate the deviatoric shear stress on to the boundary particles
using,
\begin{equation}
  \label{eq:shear-stress-bc}
  \sigma_{ij}^{'} = \frac{\Sigma_s \sigma_{ij}^{'} \; W_{ws}}{\Sigma_s W_{ws}},
\end{equation}
where $s$ denotes the solid particles.

\subsection{Time integration}

We use the kick-drift-kick scheme for the time integration. We first move the
velocities of the particles to half time step,
\begin{equation}
  \label{eq:velocity-update-stage-1}
  \ten{u}_a^{n+\frac{1}{2}} = \ten{u}_a^{n} + \frac{\Delta t}{2} \bigg(\frac{\tilde{d}\ten{u}_{a}}{dt}\bigg)^n,
\end{equation}

\begin{equation}
  \label{eq:velocity-hat-update-stage-1}
  \ten{\tilde{u}}_a^{n+\frac{1}{2}} = \ten{u}_a^{n+\frac{1}{2}} + \frac{\Delta t}{2} \bigg(\frac{d\ten{u}_{a}}{dt}\bigg)^{n}_{c}.
\end{equation}
Then the time derivatives of density and deviatoric stresses are calculated
using the \cref{eq:sph-discretization-continuity} and
\cref{eq:jaumann-stress-rate}. The new time step density, deviatoric stresses
and particle position are updated by,
\begin{equation}
  \label{eq:density-update-stage-2}
  \rho_{a}^{n+1} = \rho_{a}^{n} + \Delta t \; \bigg(\frac{\tilde{d}\rho_{a}}{dt}\bigg)^{n+\frac{1}{2}},
\end{equation}

\begin{equation}
  \label{eq:pressure-update-stage-2}
  p_{a}^{n+1} = p_{a}^{n} + \Delta t \; \bigg(\frac{\tilde{d}p_{a}}{dt}\bigg)^{n+\frac{1}{2}},
\end{equation}

\begin{equation}
  \label{eq:stress-update-stage-2}
  \teng{\sigma}_{a}^{' \; n+1} = \teng{\sigma}_{a}^{' \; n} +
  \Delta t \; \bigg(\frac{\tilde{d}\teng{\sigma}^{'}_{a}}{dt}\bigg)^{n+\frac{1}{2}},
\end{equation}

\begin{equation}
  \label{eq:position-update-stage-2}
  \ten{r}_{a}^{n+1} = \ten{r}_{a}^{n} + \Delta t \; \ten{\tilde{u}}_{a}^{n+1}.
\end{equation}
Finally, at new time-step particle position, the momentum velocity is updated

\begin{equation}
  \label{eq:velocity-update-stage-3}
  \ten{u}_a^{n+1} = \ten{u}_a^{n+\frac{1}{2}} + \frac{\Delta t}{2} \bigg(\frac{\tilde{d}\ten{u}_{a}}{dt}\bigg)^{n+1}.
\end{equation}

For the numerical stability, the time step depends on the CFL condition as,
\begin{equation}
  \label{eq:time-step-cfl}
  \Delta t = \mathrm{min} \bigg( 0.25 \; \frac{h}{c + |U|} ,  0.25 \; \frac{h^2}{\nu},  0.25 \; \frac{h^2}{g} \bigg),
\end{equation}
where $|U|$ is the maximum velocity magnitude, $c$ is the speed of sound
typically chosen as $10 |U|$ for fluids in this work.

For solid mechanics, the timestep is set based on the following,
\begin{equation}
  \label{eq:time-step-body-force}
  \Delta t \leq 0.25 \; \bigg(\frac{h}{c_0 + |U|} \bigg),
\end{equation}
where $c_0$ is the speed of sound of the solid body.

\section{Results}
\label{sec:results}

We validate the proposed scheme using a suite of benchmark problems for both
fluid and solid mechanics. We first consider fluids where we look at the
Taylor-Green vortex problem, the lid-driven cavity,
and the two-dimensional dam-break problem. We then consider
problems in elastic dynamics like the oscillating plate, a uniaxial
compression problem, the collision of rubber rings, and a high-velocity impact
problem.

We show how the proposed method is an improvement on previous work. Every
result shown is produced using an automation framework~\cite{pr:automan:2018}.
The source code is available at \url{https://gitlab.com/pypr/ctvf}.

\FloatBarrier%

\subsection{Taylor-Green vortex}
\label{sec:tgv}

In the first benchmark, we test the accuracy of the correction terms and
evaluate the different particle shifting schemes introduced in the proposed
scheme by simulating a Taylor-Green vortex. It consists of a periodic unit box
with no solid boundaries. Taylor-Green vortex problem has an exact solution
given as,
\begin{align}
  \label{eq:tgv_sol}
  u &= - U e^{bt} \cos(2 \pi x) \sin(2 \pi y) \\
  v &=   U e^{bt}\sin(2 \pi x) \cos(2 \pi y) \\
  p &=  -U^2 e^{2bt} (\cos(4 \pi x) + \cos(4 \pi y))/4,
\end{align}
where $U$ is chosen as $1$ m\,s\textsuperscript{-1}, $b=-8\pi^2/Re$, $Re=U L /\nu$,
and $L=1$ m. We initialize the fluid using this at $t=0$ and compare the
results with the exact solution. The Reynolds number, $Re$, is initially
chosen to be $100$. The quintic spline with $h/\Delta x = 1.0$ is used. We use
summation density to compute the density and evolve pressure with
\cref{eq:sph-discretization-edac}. No artificial viscosity is used for this
problem.
The decay rate of the velocity is studied using the evolution of maximum
velocity $|\ten{u}_{\max}|$ in time. We compute the $L_1$ error in the
velocity magnitude as,
\begin{equation}
  \label{eq:tg:l1}
  L_1 = \frac{\sum_i |\ten{u}_{i, computed}| - |\ten{u}_{i, exact}|}
  {\sum_i |\ten{u}_{i, exact}|},
\end{equation}
where $\ten{u}_{i, exact}$ is found at the position of the $i$'th particle.

\begin{figure}[!htpb]
  \centering
  \begin{subfigure}{0.48\textwidth}
    \centering
    \includegraphics[width=1\linewidth]{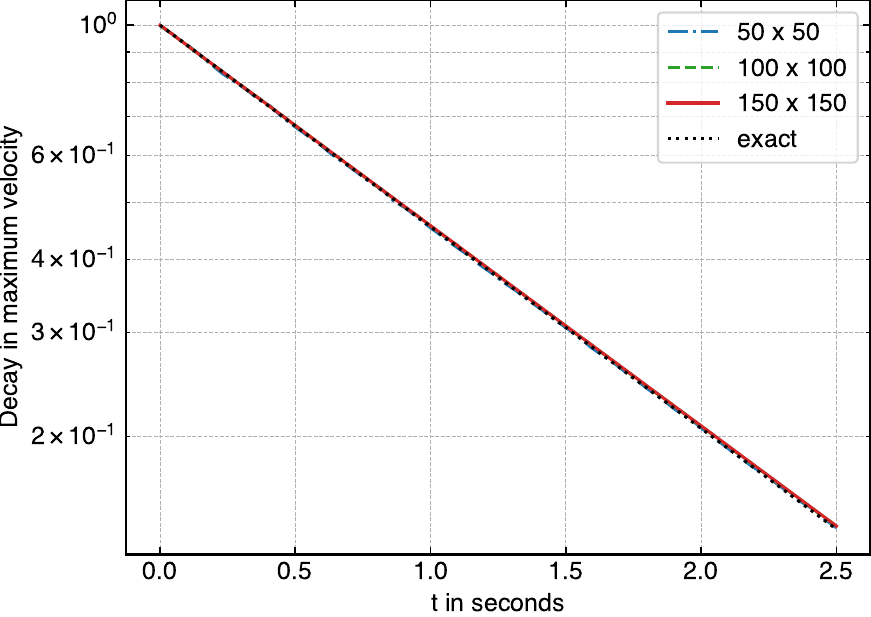}
    \subcaption{}%
    \label{fig:tg_sun2019:tg_decay}
  \end{subfigure}
  \begin{subfigure}{0.48\textwidth}
    \centering
    \includegraphics[width=1\linewidth]{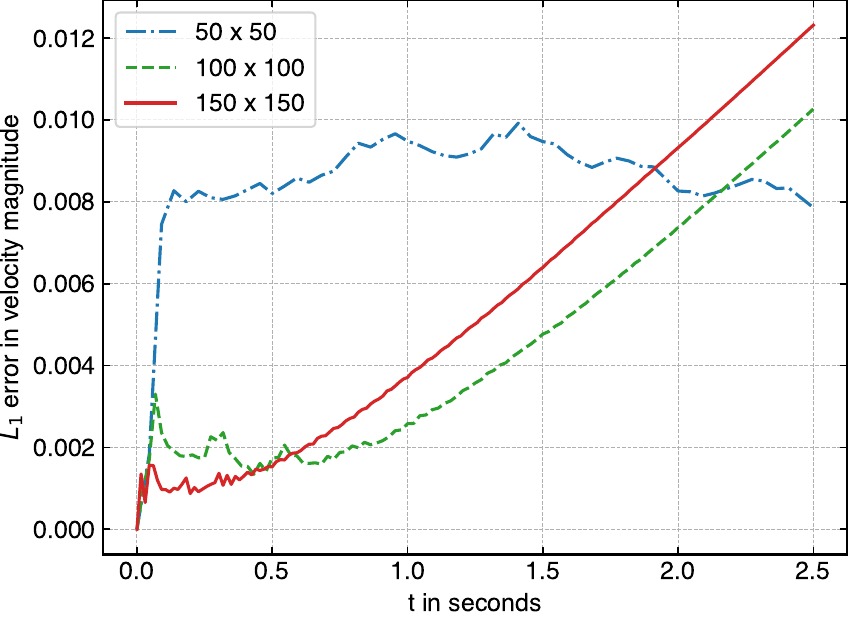}
    \subcaption{}%
    \label{fig:tg_sun2019:tg_l1}
  \end{subfigure}
  \caption{Taylor-Green vortices for an initial particle distribution of
    $50\times50$, $100\times100$ and $150\times150$ is simulated with a
    Reynolds number of 100 using SPST. Plots shown are (a) decay
    in maximum velocity (b) $L_1$ error in velocity magnitude.}
\label{fig:tg:sun2019}
\end{figure}

\begin{figure}[!htpb]
  \centering
  \begin{subfigure}{0.48\textwidth}
    \centering
    \includegraphics[width=1\linewidth]{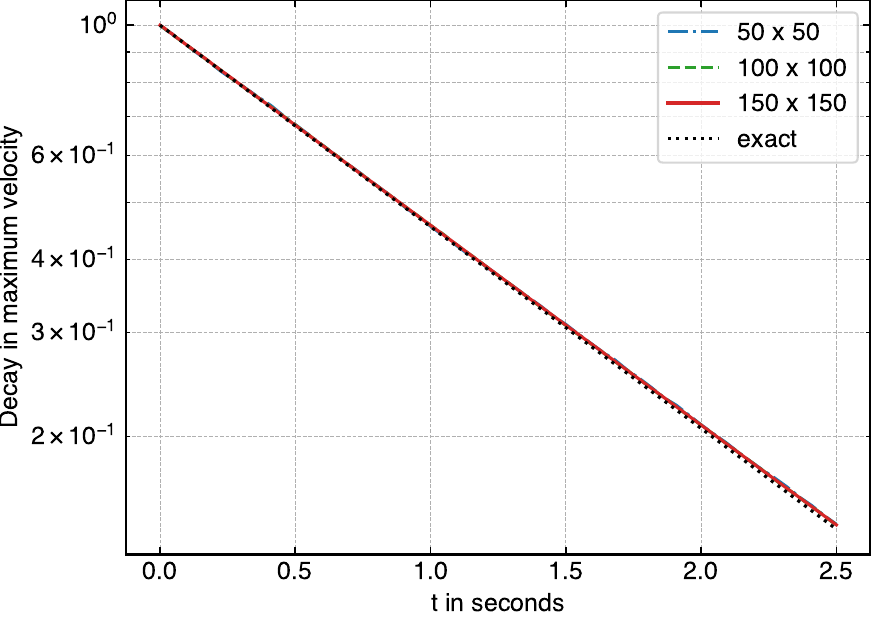}
    \subcaption{}%
    \label{fig:tg_ipst:tg_decay}
  \end{subfigure}
  \begin{subfigure}{0.48\textwidth}
    \centering
    \includegraphics[width=1\linewidth]{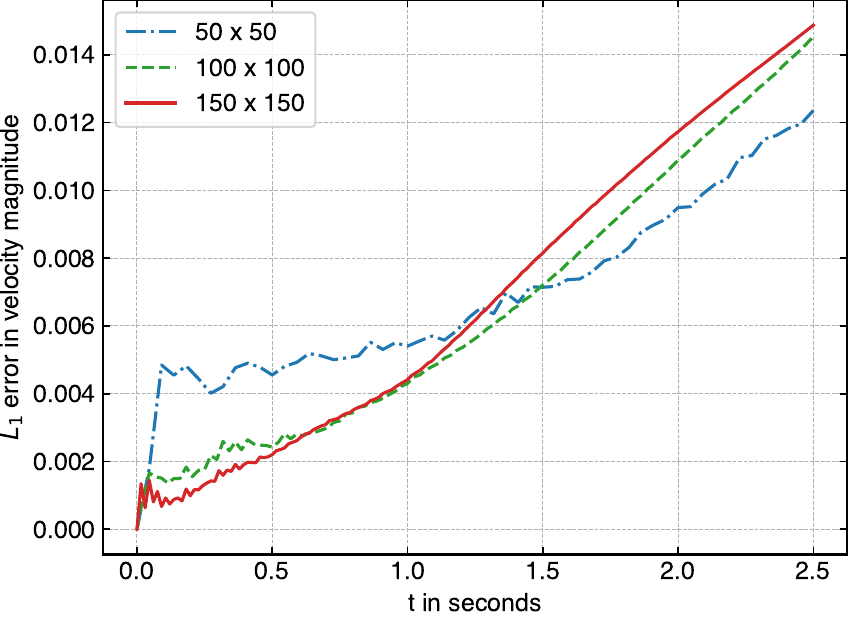}
    \subcaption{}%
    \label{fig:tg_ipst:tg_l1}
  \end{subfigure}
  \caption{Taylor-Green vortices for an initial particle distribution of
    $50\times50$, $100\times100$ and $150\times150$ is simulated with a
    Reynolds number of 100 using IPST. Plots shown are (a) decay in
    maximum velocity (b) $L_1$ error in velocity magnitude.}
\label{fig:tg:ipst}
\end{figure}

\begin{figure}[!htpb]
  \centering
  \begin{subfigure}{0.48\textwidth}
    \centering
    \includegraphics[width=1\linewidth]{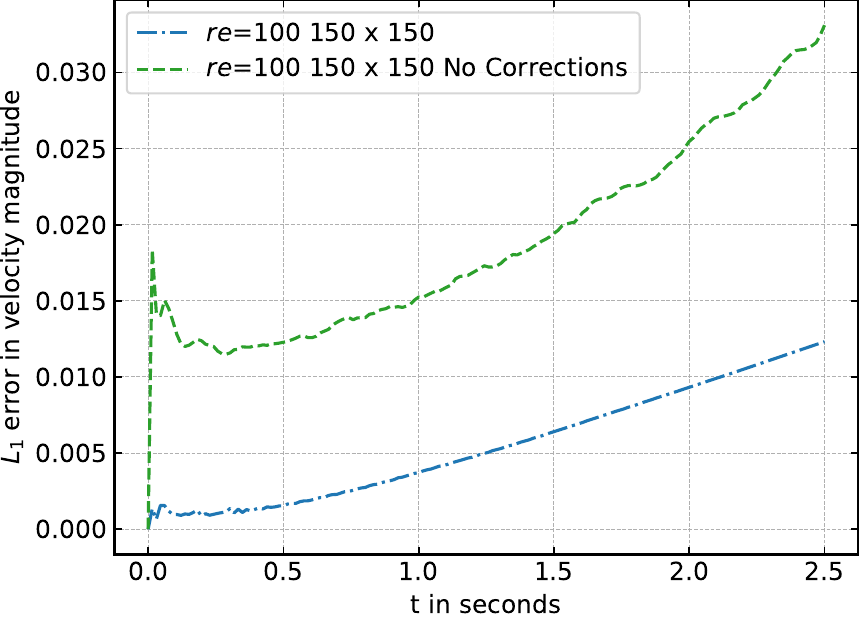}
    \subcaption{}%
    \label{fig:tg_sun2019_correcions:tg_l1_re_100}
  \end{subfigure}
  \begin{subfigure}{0.48\textwidth}
    \centering
    \includegraphics[width=1\linewidth]{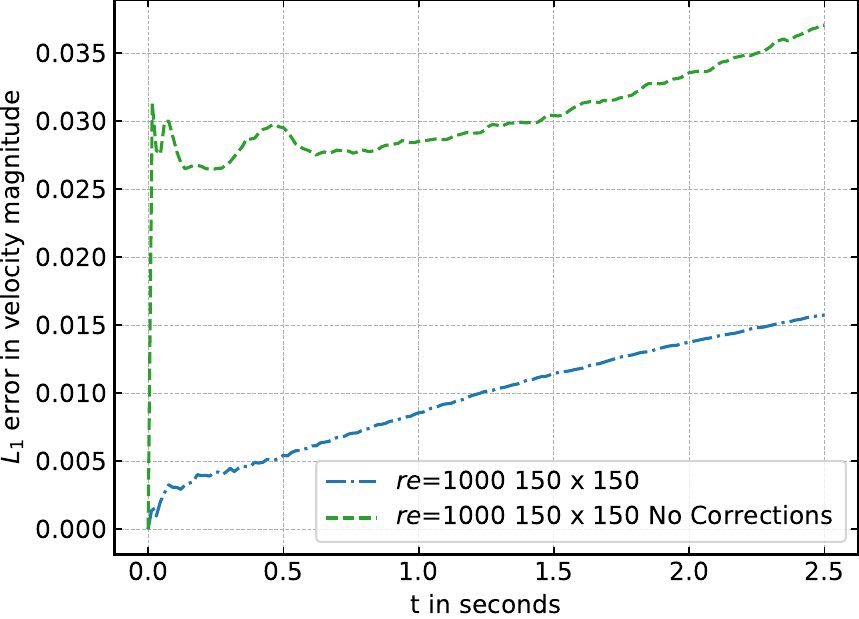}
    \subcaption{}%
    \label{fig:tg_sun2019_correcions:tg_l1_re_1000}
  \end{subfigure}
  \caption{$L_1$ error for nx with 150 $\times$ 150 with and without
    corrections with SPST with a Reynolds number of a) 100 and b) 1000}
\label{fig:tg_sun2019_corrections:tg_l1}
\end{figure}

\begin{figure}[!htpb]
  \centering
  \begin{subfigure}{0.48\textwidth}
    \centering
    \includegraphics[width=1\linewidth]{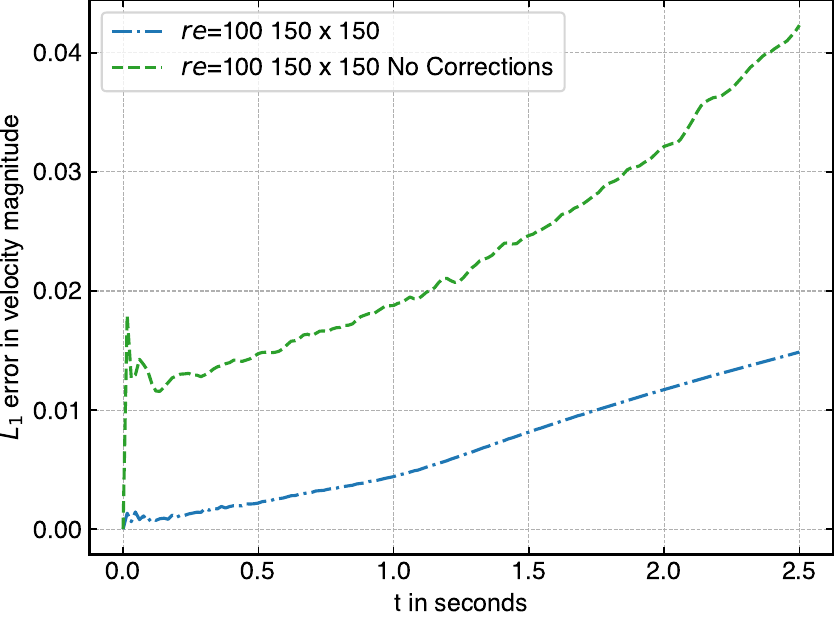}
    \subcaption{}%
    \label{fig:tg_ipst_correcions:tg_l1_re_100}
  \end{subfigure}
  \begin{subfigure}{0.48\textwidth}
    \centering
    \includegraphics[width=1\linewidth]{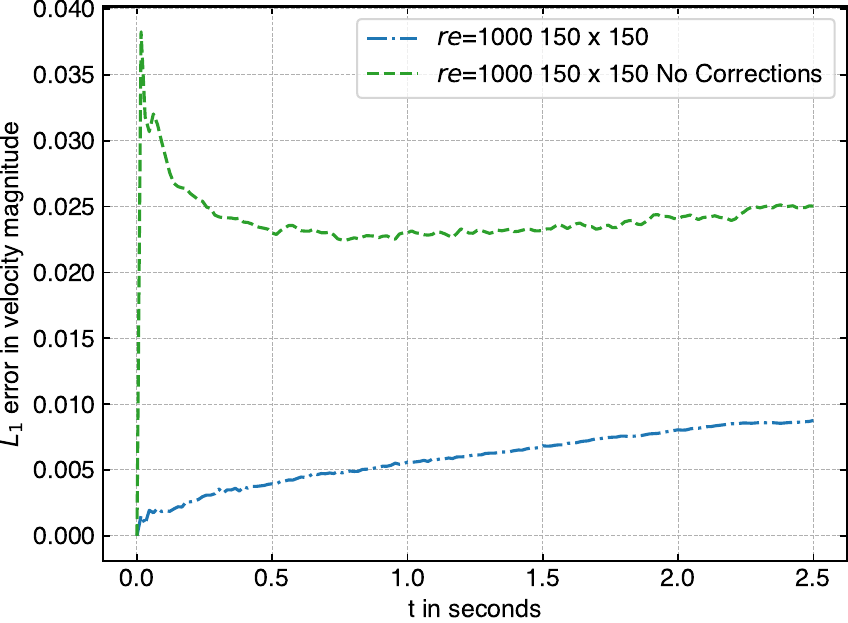}
    \subcaption{}%
    \label{fig:tg_ipst_correcions:tg_l1_re_1000}
  \end{subfigure}
  \caption{$L_1$ error for nx with 150 $\times$ 150 with and without
    corrections with IPST with a Reynolds number of a) 100 and b)
    1000}
\label{fig:tg_ipst_corrections:tg_l1}
\end{figure}

In \cref{fig:tg_sun2019:tg_decay} we compare the decay of $|\ten{u}_{\max}|$
with that of the exact solution for the case where we use SPST for particle
shifting. As can be seen, the results are in excellent agreement with the
expected decay. The same is seen in \cref{fig:tg_ipst:tg_decay} for the case
using IPST. This shows the accuracy and robustness of the scheme with respect
to changing the PST method. \Cref{fig:tg_sun2019:tg_l1} and
\cref{fig:tg_ipst:tg_l1} show the $L_1$ error of velocity magnitude for
various resolutions simulated using the two PST techniques.
\Cref{fig:tg_sun2019_corrections:tg_l1} depicts the $L_1$ error of velocity
magnitude for a Reynolds number of 100 and 1000 using SPST with and without
correction terms. \Cref{fig:tg_ipst_corrections:tg_l1} shows the same but
using the IPST. The improvement due to the correction terms is clearly seen
as a significant reduction in the error.

\begin{figure}[!htpb]
  \centering
  \begin{subfigure}{0.48\textwidth}
    \centering
    \includegraphics[width=1\linewidth]{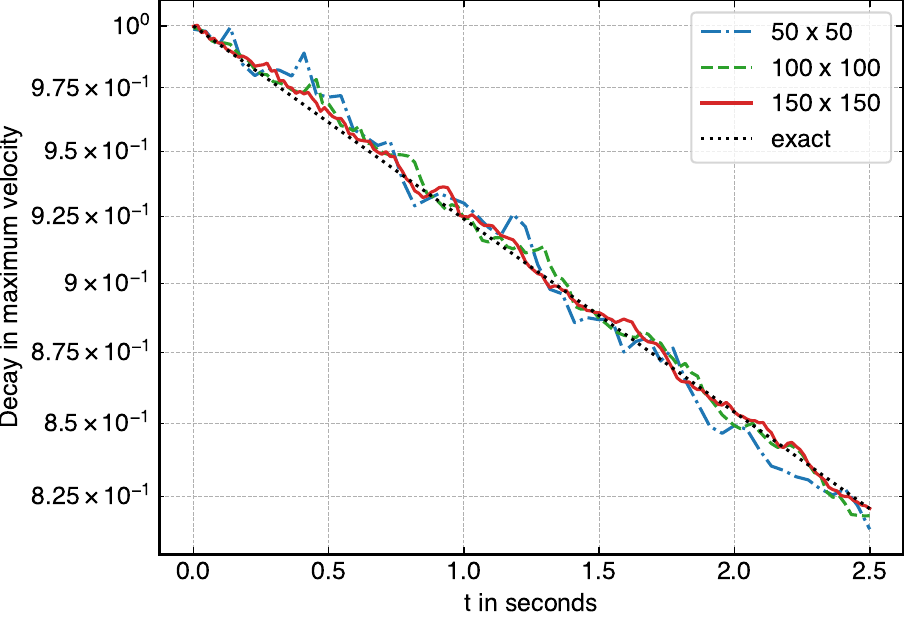}
    \subcaption{}%
    \label{fig:tg_re_1000_ipst:tg_decay}
  \end{subfigure}
  \begin{subfigure}{0.48\textwidth}
    \centering
    \includegraphics[width=1\linewidth]{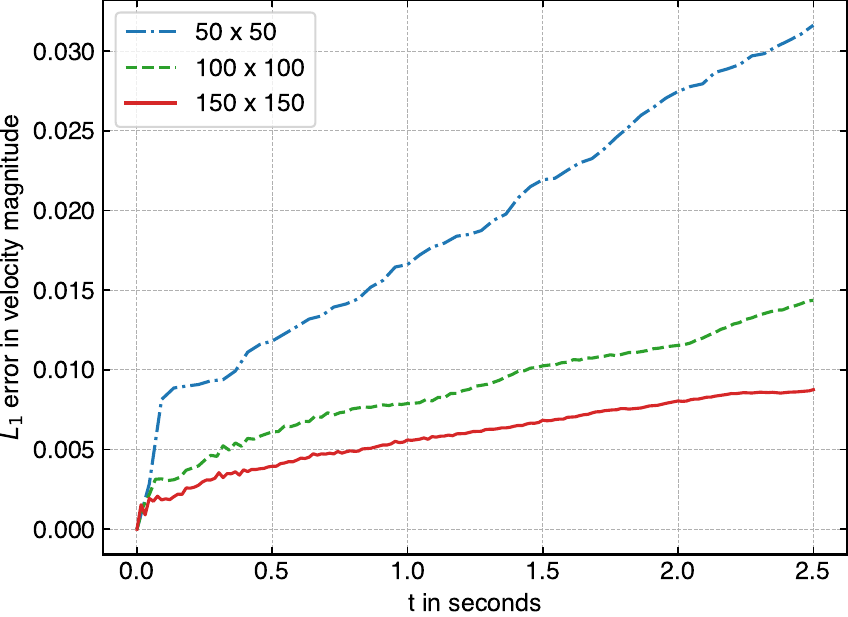}
    \subcaption{}%
    \label{fig:tg_re_1000_ipst:tg_l1}
  \end{subfigure}
  \caption{Taylor-Green vortices for an initial particle distribution of
    $50\times50$, $100\times100$ and $150\times150$ is simulated with a
    Reynolds number of 1000 using IPST. Plots shown are (a) decay in
    maximum velocity (b) $L_1$ error in velocity magnitude.}
\label{fig:tg_re_1000:ipst}
\end{figure}

\begin{figure}[!htpb]
  \centering
  \begin{subfigure}{0.48\textwidth}
    \centering
    \includegraphics[width=1\linewidth]{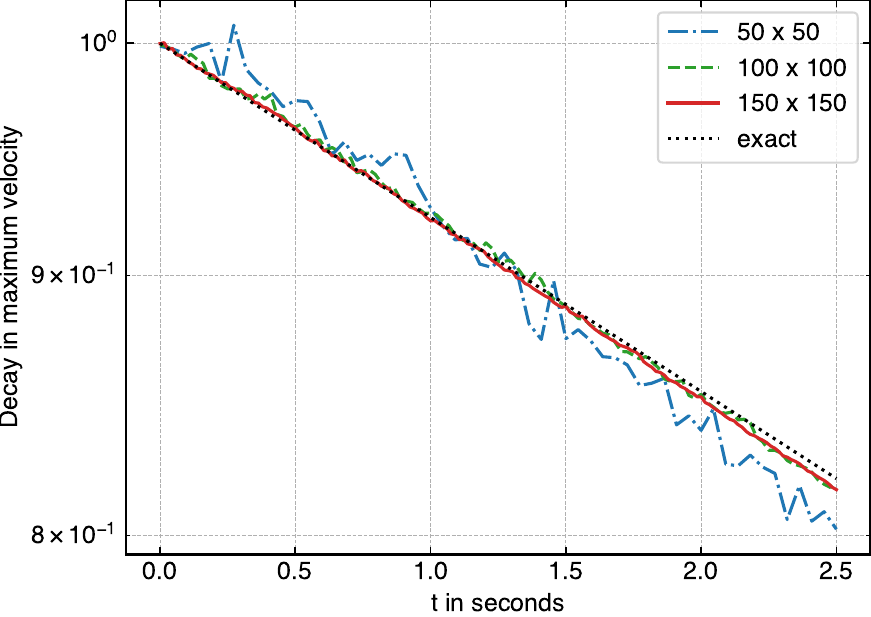}
    \subcaption{}%
    \label{fig:tg_re_1000:tg_decay}
  \end{subfigure}
  \begin{subfigure}{0.48\textwidth}
    \centering
    \includegraphics[width=1\linewidth]{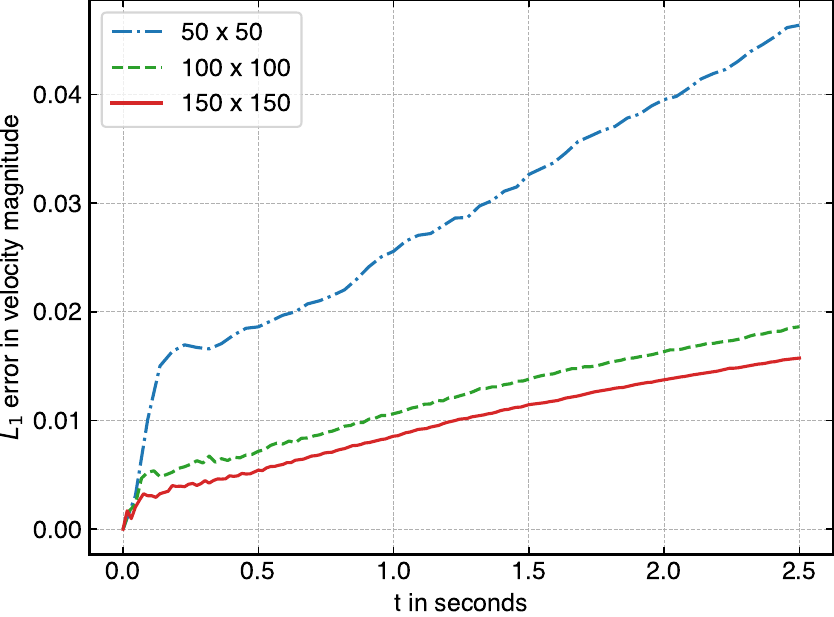}
    \subcaption{}%
    \label{fig:tg_re_1000:tg_l1}
  \end{subfigure}
  \caption{Taylor-Green vortices for an initial particle distribution of
    $50\times50$, $100\times100$ and $150\times150$ is simulated with a
    Reynolds number of 1000 using SPST. Plots shown are (a) decay in
    maximum velocity (b) $L_1$ error in velocity magnitude.}
\label{fig:tg_re_1000:sun2019}
\end{figure}

\begin{figure}[!htpb]
  \centering
  \includegraphics[width=0.5\linewidth]{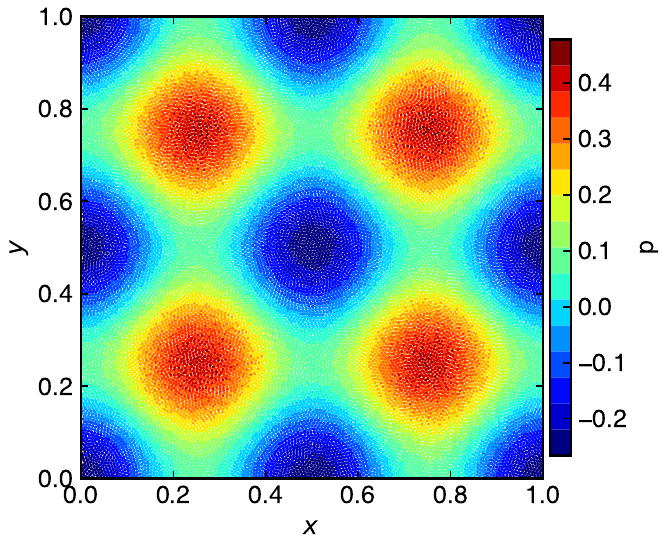}
  \caption{Particle plot of Taylor green vortices for a Reynolds number of
    1000 with a resolution of $150\times150$. The colors represent the
    pressure.}%
  \label{fig:tg_re_1000:pplot}
\end{figure}

One can see that the IPST has lower errors at initial times. However,
we do note that there appears to be a lack of convergence in the result as the
resolution is increased. As the number of particles is increased the $L_1$
error does not correspondingly reduce. This is due to the low Reynolds number
and the discretization of the viscous term that is being used. We show the
results of the velocity decay and the $L_1$ error when a Reynolds number of
1000 is used in \cref{fig:tg_re_1000:ipst} for IPST and in
\cref{fig:tg_re_1000:sun2019} with SPST. In this case the convergence
is clearly seen as the resolution is increased. \Cref{fig:tg_re_1000:pplot}
shows distribution of particles with the color representing pressure. The
Reynolds number of 1000 with a resolution of 150$\times$150. We can see
that the pressure distribution is smooth.

%
%
%
%
%

An inspection of figures \ref{fig:tg_sun2019:tg_l1}, \ref{fig:tg_ipst:tg_l1}
and \ref{fig:tg_re_1000_ipst:tg_l1}, \ref{fig:tg_re_1000:tg_l1} suggests that the
IPST  appears to be better than that of SPST.

\FloatBarrier%
\subsection{Lid driven cavity}
\label{sec:ldc}

We evaluate the ability of the proposed scheme to handle solid wall boundary
conditions by simulating a lid-driven cavity. The lid-driven cavity is a
classic problem that can be challenging to simulate in the context of the SPH.
It has been simulated by \cite{Adami2013}, \cite{huang_kernel_2019},
\cite{edac-sph:cf:2019} to note a few. A rectangular cavity with length 1 m
which is filled with fluid is constrained by four walls. Top wall has a
velocity of $U = 1 $ m\,s\textsuperscript{-1}. A unit density is assumed for the
fluid. The speed of sound of the fluid particle is set to $c = 10 U_{max}$. We
use the summation density to compute the density. The viscosity of the fluid
is set through the Reynolds number of the flow, $\nu = \frac{Re}{U}$. No
artificial viscosity is used in the current problem.

\begin{figure}[!htpb]
  \centering
  \includegraphics[width=1\linewidth]{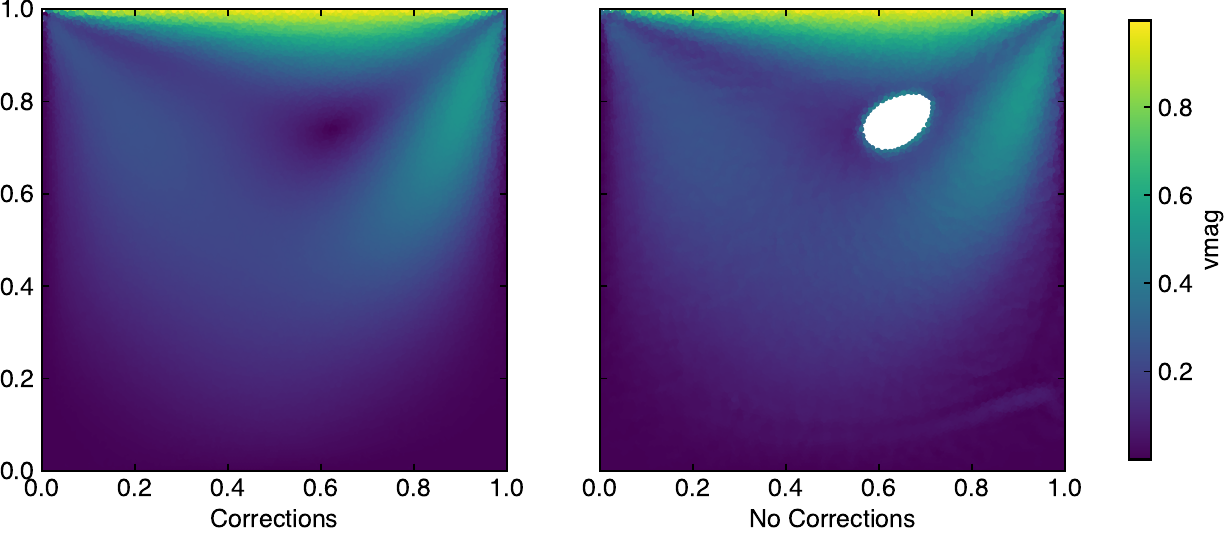}
  \caption{ Particle plot of cavity with a $Re=100$ with particle arrangement
    of $150 \times 150$, left side with corrections and right side without
    correction terms.}%
  \label{fig:ldc:particle_plots_re100_compare}
\end{figure}
We first simulate the cavity problem with a Reynolds number of 100 with and
without corrections. In \cref{fig:ldc:particle_plots_re100_compare} we can see
that an unphysical void is produced when no corrections are employed. This is
eliminated with the current scheme.

\begin{figure}[!htpb]
  \centering
  \includegraphics[width=0.5\linewidth]{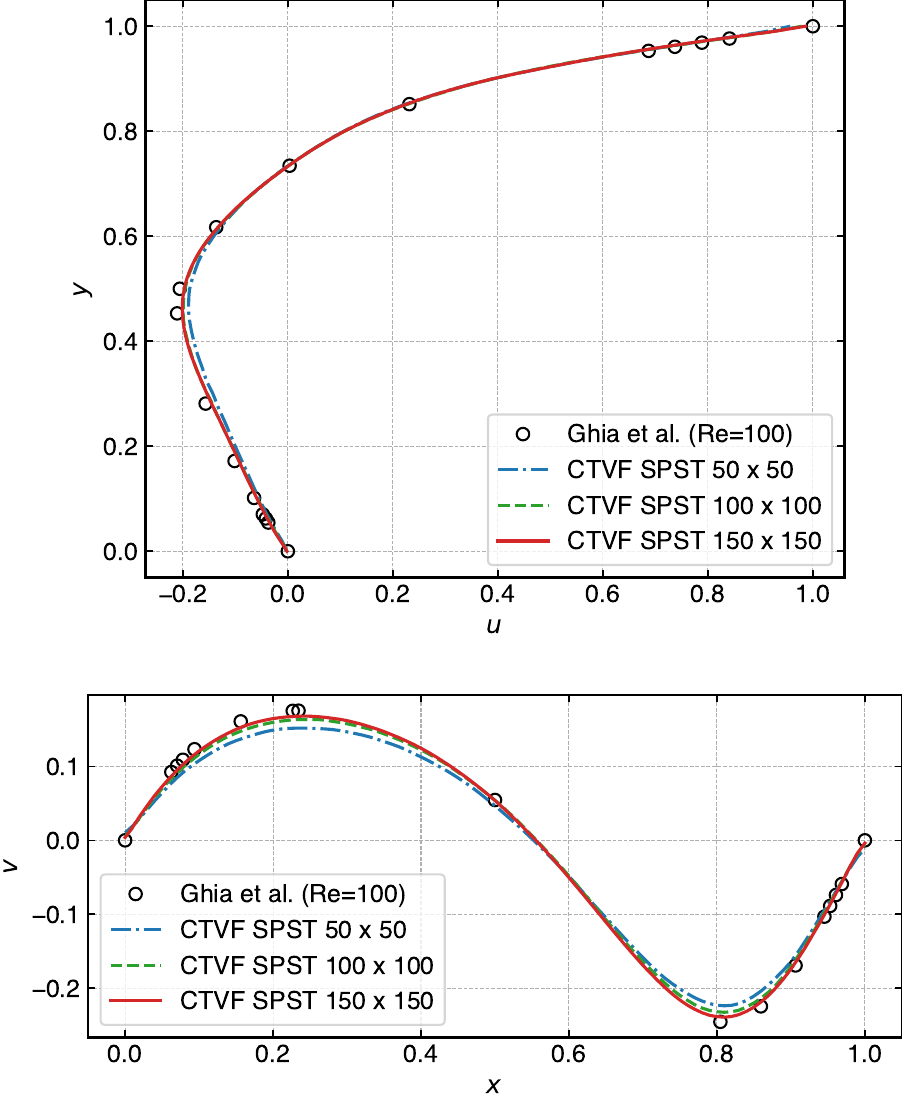}
  \caption{Velocity profiles $u$ vs.\ $y$ and $v$ vs.\ $x$ for the
    lid-driven-cavity problem at $Re=100$ with three initial particle
    arrangement of $50 \times 50$, $100 \times 100$, and $150 \times
    150$.}%
  \label{fig:ldc:uv_re100}
\end{figure}
\begin{figure}[!htpb]
  \centering
  \includegraphics[width=0.5\linewidth]{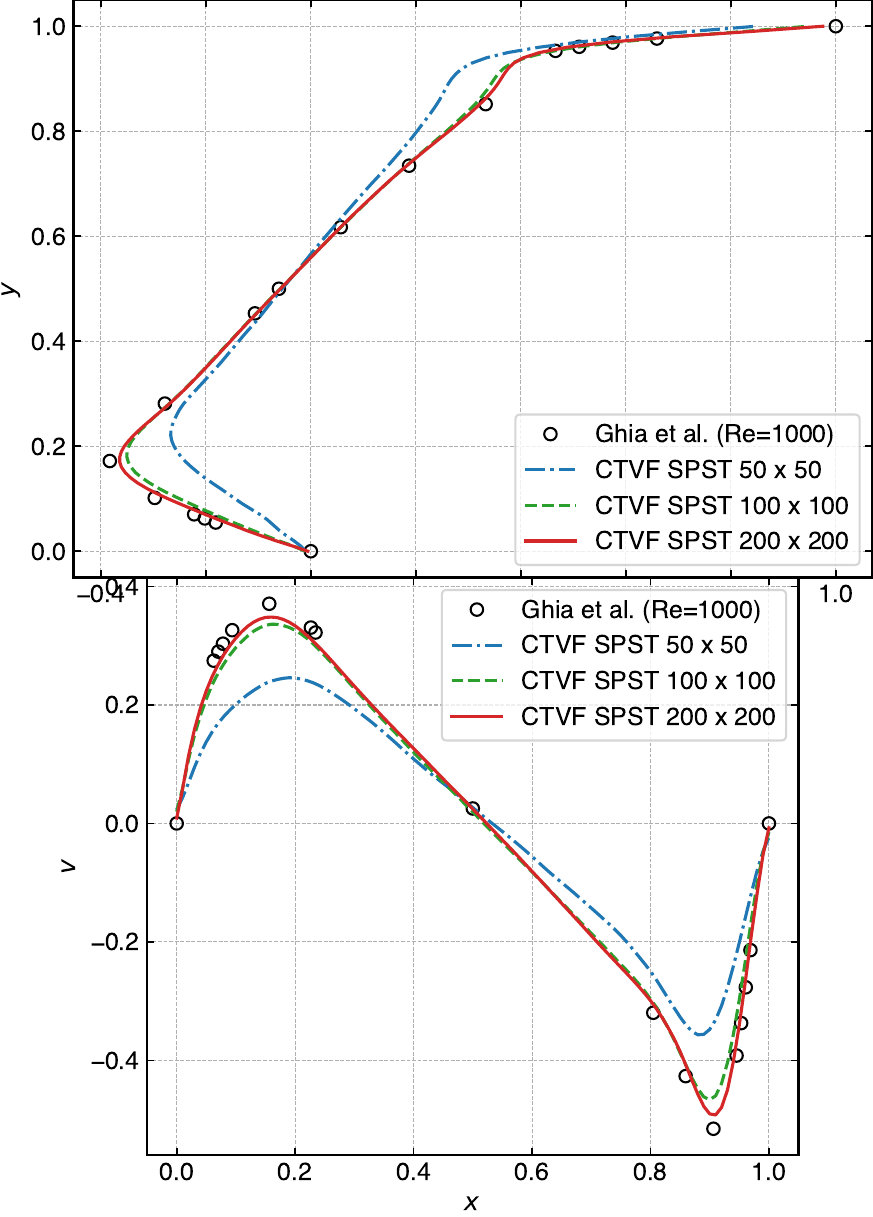}
  \caption{Velocity profiles for the lid-driven-cavity using the steady state
    simulation procedure for $Re = 1000$ with initial partial arrangement of
    $50 \times 50$, $100 \times 100$, and $200 \times 200$ compared with
    the results of~\cite{ldc:ghia-1982}.}%
\label{fig:ldc:uv_re1000}
\end{figure}

We now study convergence of the method as we vary the
resolution. \Cref{fig:ldc:uv_re100} and \cref{fig:ldc:uv_re1000} show the
center-line velocities $u$ versus $y$ and $v$ versus $x$ for the Reynolds
numbers 100 and 1000 respectively. For the $Re=100$ case we use three
different resolutions of $50\times 50, 100 \times 100$ and $150 \times
150$. For the $Re=1000$ case, we use an initial $50 \times 50$,
$100 \times 100$, and $200 \times 200$ grid of particles. These are compared
against the results of \cite{ldc:ghia-1982}. As we can see that the current
scheme is able to predict the velocity profiles well.

\FloatBarrier%

\subsection{2D Dam-break}

We apply the proposed scheme to free surface flows by simulating a
dam-break. This problem has been extensively studied before for example in
\cite{muta_efficient_2020}, \cite{zhang_hu_adams17}, and
\cite{edac-sph:cf:2019}.

A block of fluid having width 1m and a height of $2$ m is allowed to settle
under the influence of gravity inside a tank of length $4$ m. The fluid block
is initially placed to the left of the tank. The acceleration due to gravity
is $g=9.81$ m\,s\textsuperscript{-2}. To simulate the free surface flows we
use the continuity equation to evolve the density using
\eqref{eq:sph-discretization-continuity} and the
\eqref{eq:sph-discretization-edac} to evolve the pressure. We use free slip
boundary conditions to compute the divergence of the velocities and a no-slip
boundary condition while computing the viscous forces. The value of
$\alpha=0.05$ is used for the artificial viscosity \cref{eq:mom-av} term.

\begin{figure}[!htpb]
  \centering
  \includegraphics[width=0.8\textwidth]{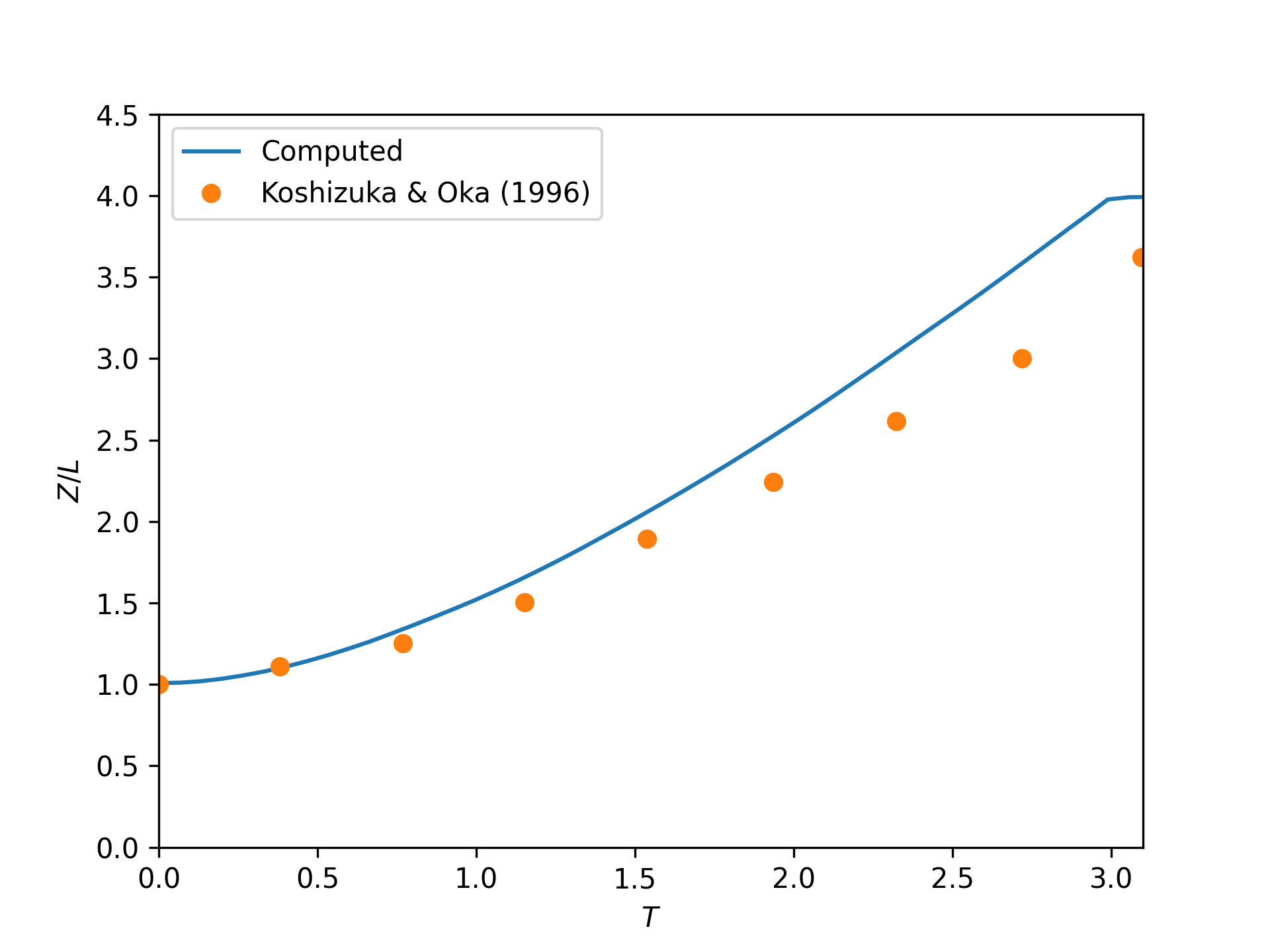}
  \caption{Position of the toe of the water versus time of CTVF as compared
    with the simulation of ~\cite{koshizuka1996moving}. Z is the
    distance of toe of the dam from the left wall and L is the initial width
    of the dam}
\label{fig:dam-break}
\end{figure}

\Cref{fig:dam-break} compares the position of the toe of the fluid block with
time against \cite{koshizuka1996moving}, where the authors use the moving
particle semi-implicit scheme to simulate the same.

\begin{figure}[!htpb]
  \centering
  \includegraphics[width=\textwidth]{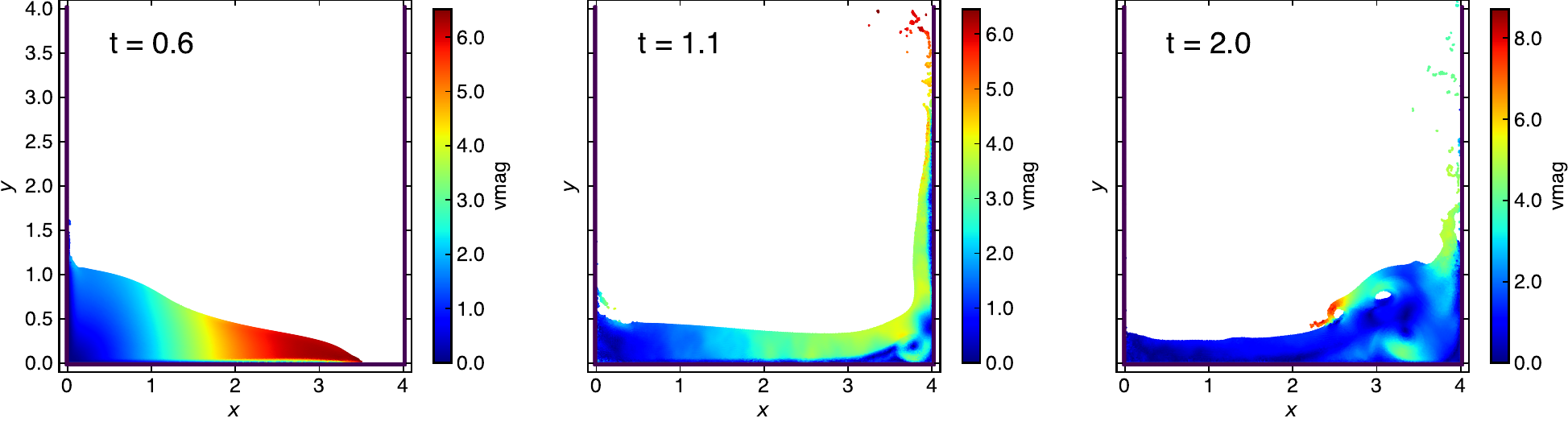}
  \caption{Particle plots of fluid in dam break at time $t=0.6, 1.1, 2.0$
    second, showing velocity magnitude as contour.}
\label{fig:dam-break-plots-vmag}
\end{figure}

The evolution of the fluid at three different time instants t$=0.6, 1.1, 2.0$
seconds, is shown in \cref{fig:dam-break-plots-vmag}. As can be seen from
\cref{fig:dam-break-plots-vmag}, at time $2.0$ seconds we have captured the void
created due to the splashing of the fluid. The colors in
\cref{fig:dam-break-plots-vmag} shows the velocity magnitude.

\FloatBarrier%
\subsection{Oscillating plate}
\label{sec:oscillating-plate}
In this section, we test the improvement due to the correction terms while
simulating elastic solids. We show the elimination of tensile instability
while extending the transport velocity formulation~\cite{Adami2013} scheme to
more particle shifting techniques. We consider a thin oscillating
plate that is clamped on one side. \citet{landau1960} provide an analytical
solution for this problem. This is also simulated numerically in
\cite{gray-ed-2001} and \cite{zhang_hu_adams17}.

An oscillating plate with a length of $0.2$ m and a height of $0.02$ m is
initially given with a velocity profile of,
\begin{equation*}
  v_y(x) = V_f \, c_0 \frac{F(x)}{F(L)},
\end{equation*}
where $V_f$ varies for different cases. $L$ is the length of the plate. $F(x)$
is given by,
\begin{multline}
  F(x) = (\cos(kL) + \cosh(kL)) \, (\cosh(kx) - \cos(kx)) + \\
  (\sin(kL) - \sinh(kL)) \, (\sinh(kx) - \sin(kx)).
\end{multline}
In the present example $kL$ is 1.875. The material properties of the plate are
as follows, Young's modulus $E=2.0\times 10^6$ Pa, a Poisson's ratio of
$\nu=0.3975$. $c_0$ is speed of sound, and a density of $\rho=1000$
kg\,m\textsuperscript{-3}, as done in~\cite{gray-ed-2001}.  In all the cases
simulated here, we use an $\alpha$ of $1$ for artificial viscosity.

\begin{figure}[!htpb]
  \centering
  \begin{subfigure}{0.48\textwidth}
    \centering
    \includegraphics[width=0.8\textwidth]{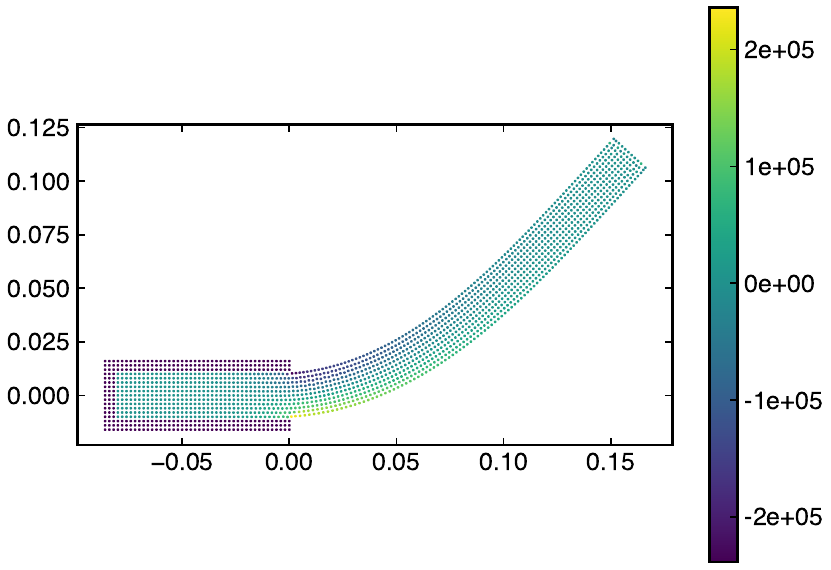}
    \subcaption{}%
    \label{fig:oscillating-plate:gtvf-sun2019-a}
  \end{subfigure}
  \begin{subfigure}{0.48\textwidth}
    \centering
    \includegraphics[width=0.8\textwidth]{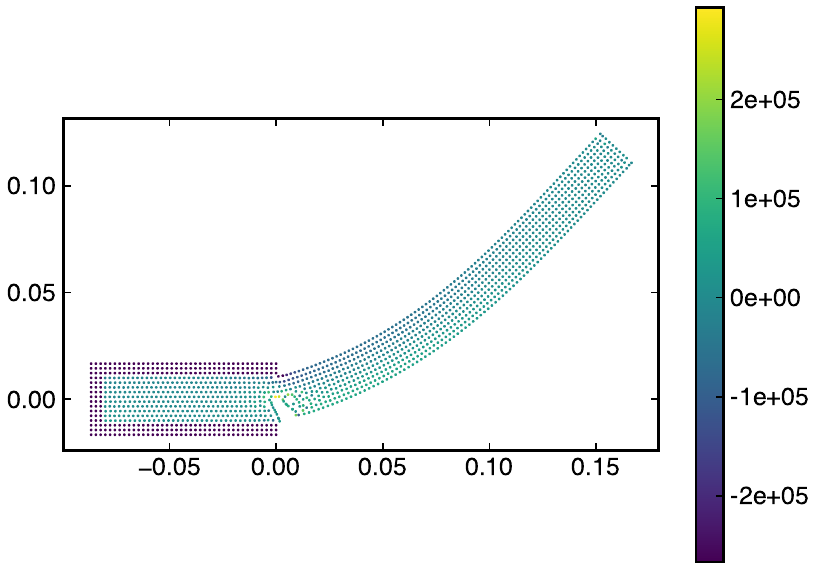}
    \subcaption{}%
    \label{fig:oscillating-plate:gtvf-sun2019-b}
  \end{subfigure}
\caption{Oscillating plate with a length of $0.2$m and height of $0.02$m when
  simulated with GTVF Scheme. Figure in left is original GTVF scheme and right
  is while using SPST with GTVF scheme.}
\end{figure}

The GTVF~\cite{zhang_hu_adams17} eliminates the tensile instability while
using the special PST proposed in the original paper. We show that the
GTVF~\cite{zhang_hu_adams17} scheme is unable to eliminate numerical fracture
when a different PST algorithm is employed. Instead of using
the standard GTVF homogenization acceleration we use Sun's particle shifting
technique (SPST). This results in a numerical fracture, as seen in
\cref{fig:oscillating-plate:gtvf-sun2019-b}. We reproduced the same case with
original GTVF scheme, where no numerical fracture has found as seen in
\cref{fig:oscillating-plate:gtvf-sun2019-a}. This numerical fracture is
eliminated by the current scheme. This is due to the incorporation of the
additional terms in the current scheme as well as the use of momentum velocity
in the computation of the velocity gradient. Note that a particle spacing of
$\Delta x=0.002$ m and $V_f=0.05$ m\,s\textsuperscript{-1} has been used.

\begin{figure}[!htpb]
  \centering
  \includegraphics[width=0.8\textwidth]{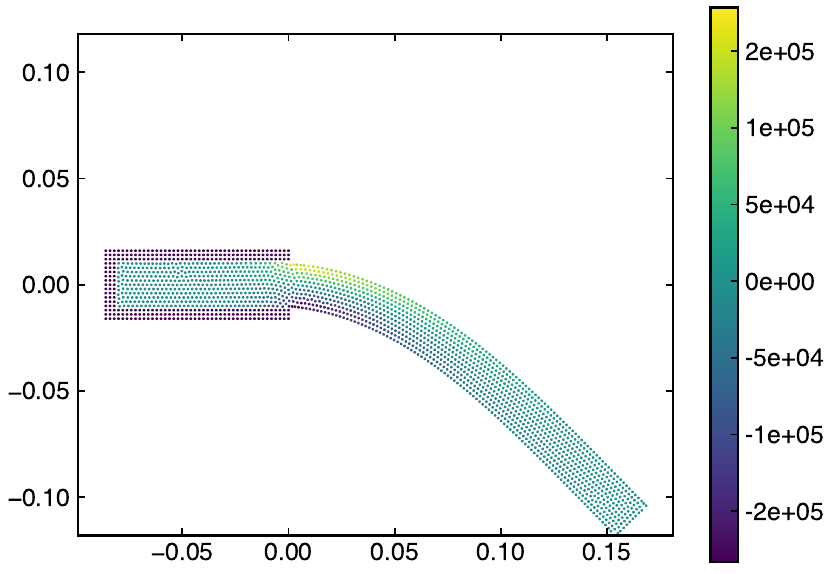}
  \caption{Oscillating plate at time $t=0.22$s with a length of $0.2$m and
    height of $0.02$m simulated with SPST with CTVF scheme.}
\label{fig:oscillating-plate:etvf-sun2019-l-0-2-h-0-22}
\end{figure}

\begin{figure}[!htpb]
  \centering
  \includegraphics[width=0.8\textwidth]{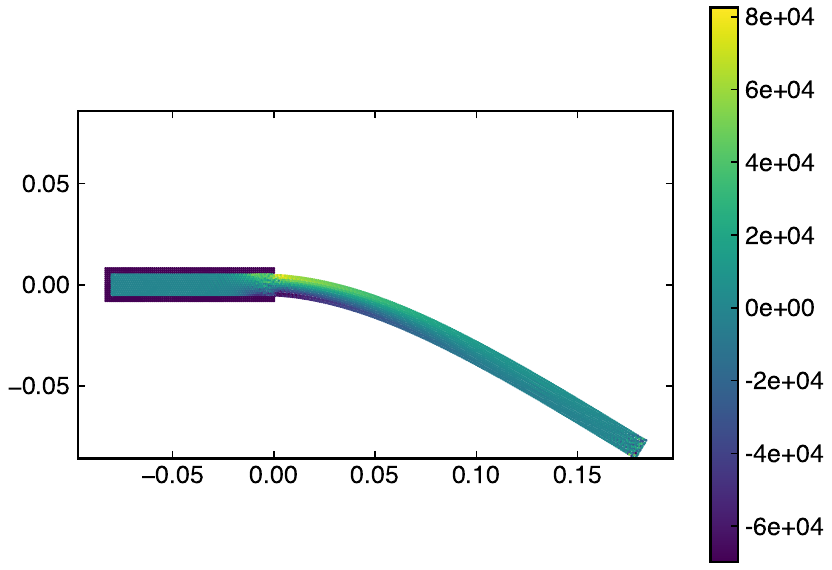}
  \caption{Oscillating plate at time $t=0.51$s with a length of $0.2$m and
    height of $0.01$m simulated with SPST with CTVF scheme.}
\label{fig:oscillating-plate:etvf-sun2019-l-0-2-h-0-01}
\end{figure}

This is further demonstrated by a case where an oscillating plate of length of
$0.2$ m and a height of $0.02$ m is simulated for a time of $0.22$ seconds.
Similarly, another case where plate of height $0.01$ m and a width of $0.2$ m is
run for a time of $0.51$ s.
\Cref{fig:oscillating-plate:etvf-sun2019-l-0-2-h-0-22} and
\cref{fig:oscillating-plate:etvf-sun2019-l-0-2-h-0-01} shows particles of the
plate at time $t=0.22$ s and $0.51$ s of these two cases respectively. As we can
see from the figure that the plate is free of numerical fracture, thus the
tensile instability is eliminated.

\begin{figure}[!htpb]
  \centering
  \includegraphics[width=0.7\textwidth]{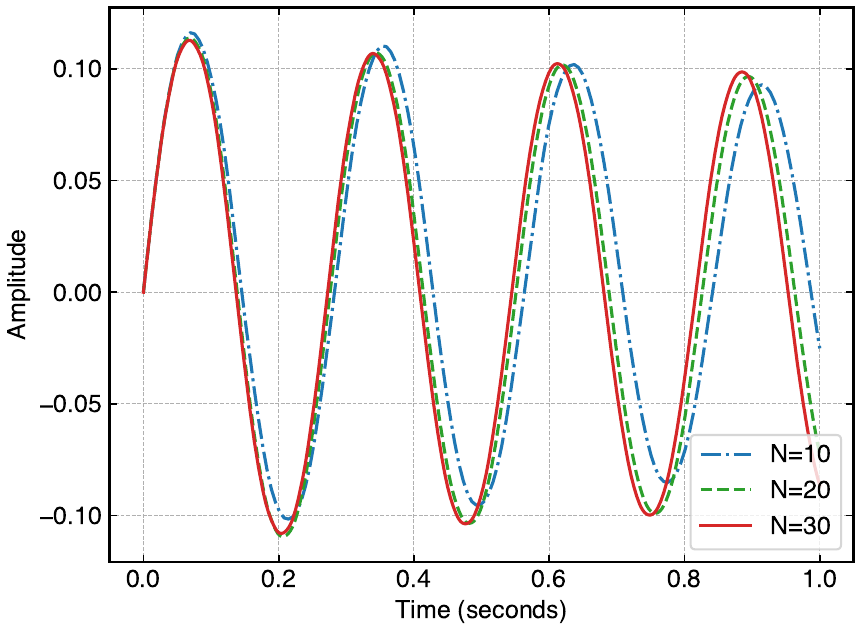}
  \caption{The vertical position of the particle at the end of the plate as a
    function of time. Here we consider a three particle variations, 10, 20 and
    30 particles across the plate width.}
\label{fig:oscillating:ipst_convergence_plot}
\end{figure}

The accuracy of the current scheme is evaluated by comparing with the
analytical results and with a convergence study. In
\cref{table:compare-analytical-with-simulated-h-l-0-1} we compare the time
period for the oscillation by the analytical and the numerical results with
varying $V_f$, where we consider an oscillating plate whose $H/L$ ratio is
$0.1$. The difference between the analytical result and the numerical result
is due to the fact that the analytical results are based on thin plate theory
where as the plate considered here has a finite thickness. Further, we can see
that the current numerical results are in agreement with the previously
reported numerical results~\cite{gray-ed-2001, zhang_hu_adams17}. In
\cref{fig:oscillating:ipst_convergence_plot}, we have performed a convergence
study of an oscillating plate, with a $\nu=0.3975$, and $V_f=0.05$
m\,s\textsuperscript{-1}, and IPST is used for particle homogenization. The trend of
the current scheme matches well with the other updated Lagrangian SPH
schemes~\cite{gray-ed-2001, zhang_hu_adams17}. Hence the current scheme is
able to work with different PST methods and remove the tensile instability.

\begin{table}[!htpb]
\centering
\begin{tabular}{c c c c c}
  \hline
  $V_f$ & 0.001 & 0.01 & 0.03 & 0.05 \\
  \hline
  $\text{T}_{\mathrm{CTVF}}$ & 0.284 & 0.283 & 0.283 & 0.284 \\
  $\text{T}_{\mathrm{GTVF}}$ & 0.284 & 0.283 & 0.284 & 0.285 \\
  $\text{T}_{\mathrm{analytical}}$ & 0.254 & 0.252 & 0.254 & 0.254
\end{tabular}
\caption{Comparison between the CTVF and the analytical solution for the time
  period of the oscillating plate with a length of $0.2$m and height of
  $0.02$m with various $V_f$}
\label{table:compare-analytical-with-simulated-h-l-0-1}
\end{table}
\FloatBarrier%
\subsection{Uniaxial compression}
\label{sec:uniaxial-compression}

This benchmark is used to test the proposed scheme. A uniaxial bar is
compressed by a moving piston on top of it. This problem has been simulated by
\citet{das2015evaluation}. We compare the von Mises stress at the center point
of the bar with the result of the FEM analysis and SPH provided in
\cite{das2015evaluation}.

\begin{figure}[!htpb]
  \centering
  \includegraphics[width=0.4\textwidth]{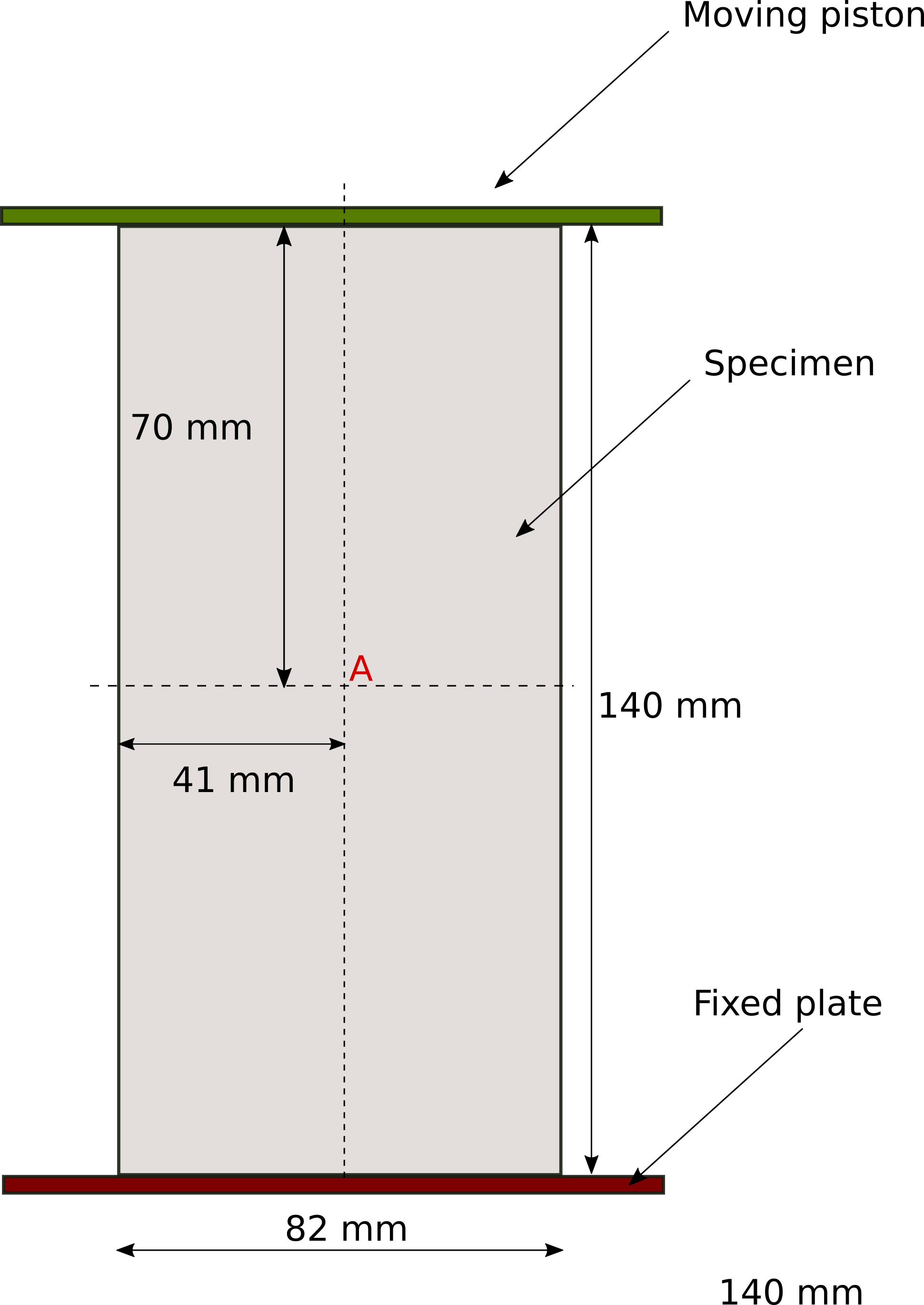}
  \caption{Test configuration of sand stone under uniaxial compression.}
\label{fig:uniaxial_test_configuration}
\end{figure}

The numerical model consists of three parts. It has an axially loaded
rectangular specimen of width $82$ mm and height of $140$ mm. The specimen has
the properties of a sand stone (Crossley Sandstone) with a Young's modulus of
$7.5$ GPa and Poisson ratio of $0.398$ and with a density of $2300$
kg\,m\textsuperscript{-3}. The speed of sound resulting from such properties is
$2303$ m\,s\textsuperscript{-1}. We run three particle resolutions,
$\Delta x = 0.5$ mm, $1$ mm and $2$ mm. The particles are placed on a regular
square grid pattern. The velocity of the top plate is
$1.5$ mm\,s\textsuperscript{-1}, which is used to apply the load on the specimen in
such a fashion, such that the loaded end is deformed at the required constant
rate. This is described in \cref{fig:uniaxial_test_configuration}. An $\alpha$
of $1$ is used in the current simulation for the artificial viscosity.

We use the von Mises stress as the criterion for analysing the stress field.
It combines the normal and shear components of the deviatoric stress tensor,
and is a commonly used criterion to assess failure strength of materials. The
von Mises stress $\sigma_{vm}$ can be expressed in 2D in terms of principle
stress $\sigma_1$ and $\sigma_2$ as

\begin{equation}
  \label{eq:von_mises_with_principal_stress}
  \sigma_{vm} = \sqrt{\left(\sigma_1^2 + \sigma_2^2 - \sigma_1 \ \sigma_2\right)}
\end{equation}
Where the principal stress are found by
\begin{eqnarray}
  \label{eq:principal_stress}
  \sigma_{1} = \frac{\sigma_{xx} + \sigma_{yy}}{2} + \sqrt{\left(\bigg(
  \frac{\sigma_{xx} + \sigma_{yy}}{2}\bigg)^2 + \sigma_{xy}^2\right)}\\
  \sigma_{2} = \frac{\sigma_{xx} + \sigma_{yy}}{2} -  \sqrt{\left(\bigg(
  \frac{\sigma_{xx} + \sigma_{yy}}{2}\bigg)^2 + \sigma_{xy}^2\right)}
\end{eqnarray}
\begin{figure}[!htpb]
  \centering
  \includegraphics[width=0.8\textwidth]{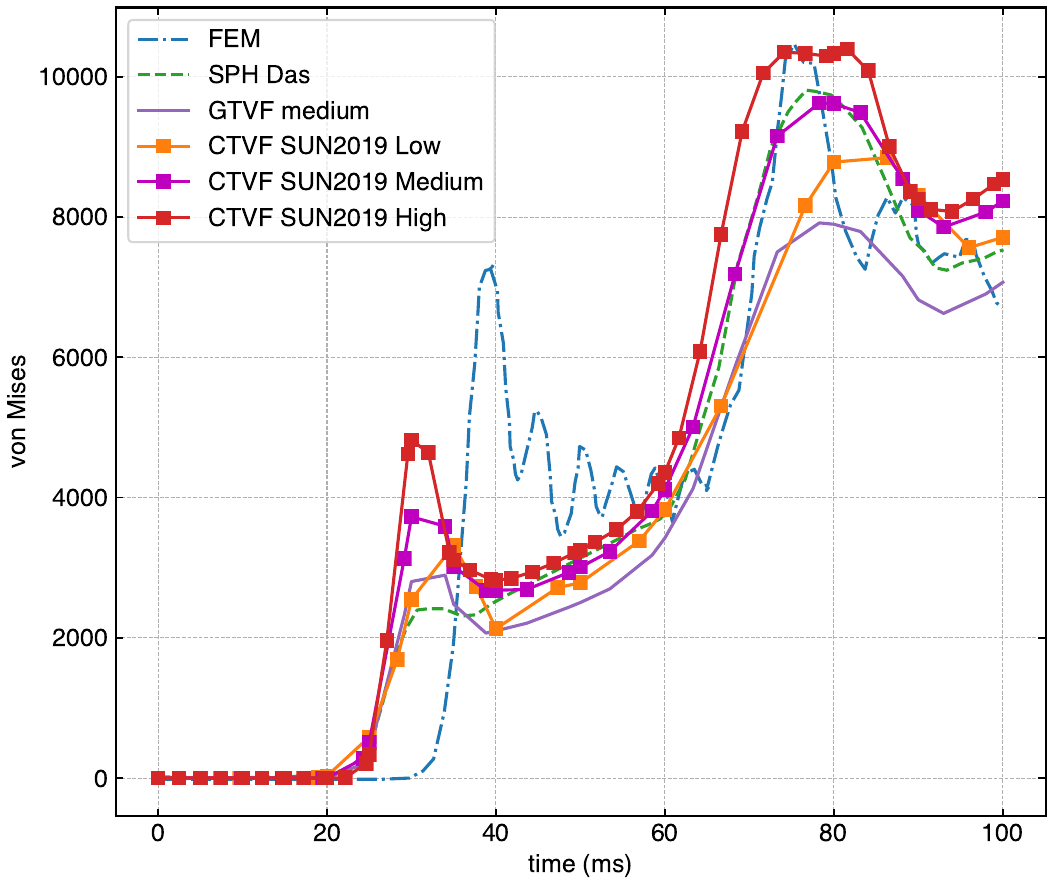}
  \caption{von Mises stress at point A in uniaxial compression with three
    different resolutions compared against those from
    \cite{das2015evaluation}.}
\label{fig:uniaxial}
\end{figure}

\Cref{fig:uniaxial} shows the von Mises stress versus time of the current
scheme, when simulated with three different resolutions compared against with
the finite element result and SPH result provided in \cite{das2015evaluation}.
It also shows the result with the GTVF scheme using the medium resolution. As
can be seen in \cref{fig:uniaxial} the GTVF result does not match very well
with FEM and SPH result provided by \citet{das2015evaluation}, and the current
scheme performs significantly better.

\FloatBarrier%
\subsection{Colliding Rings}
\label{colliding-rings}

Having shown the flexibility of proposed scheme to work with different PST
methods in \cref{sec:oscillating-plate}, in the current example, we compare
the robustness of the PST methods by investigating the collision of rubber
rings with different Poisson ratios. This was first studied in SPH by
\citet{swegle1995smoothed}.

The inner ring radius of the ring is $r_{min} = 0.03$ m and the outer ring
radius $r_{max} = 0.04$ m. Both the rings have the same material properties:
Young's modulus $E = 0.01$ GPa and density $\rho = 1.2 \times 10^{3}$
 kg\,m\textsuperscript{-3}. The initial speed of the rings are equal to
$v_0 = 0.12 c_0$ m\,s\textsuperscript{-1} with an initial inter particle spacing of
$\Delta x = 0.001$ m. Where $c_0$ is the speed of sound of the material. We use
an $\alpha=1$ for the artificial viscosity in the current simulation.

\begin{figure}[!htpb]
  \centering
%
  \begin{subfigure}{0.48\textwidth}
    \centering
    \includegraphics[width=1.0\textwidth]{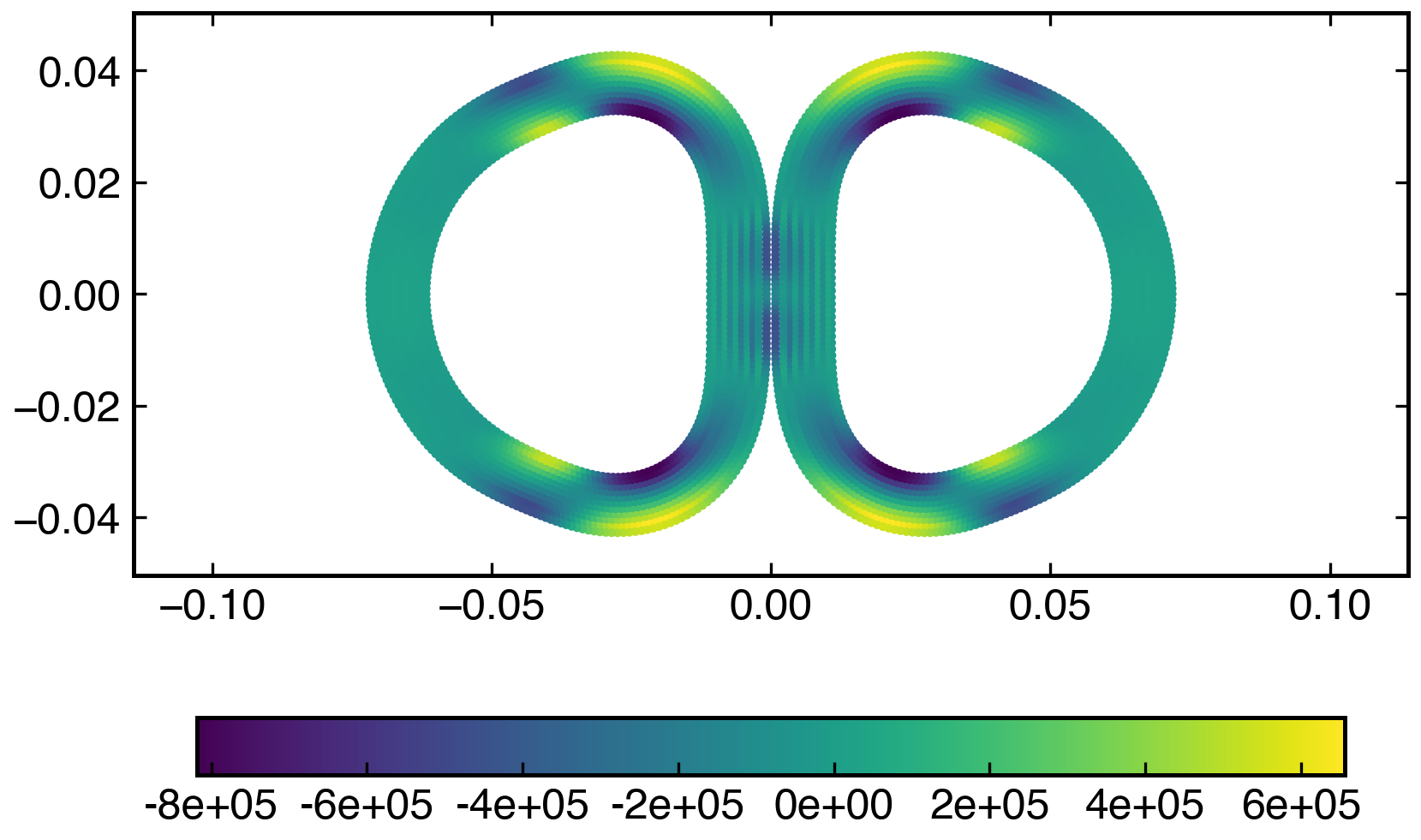}
    \subcaption{t = 2.5e-03 sec}\label{fig:rings:sun2019-nu-0.3975-1}
  \end{subfigure}
  \begin{subfigure}{0.48\textwidth}
    \centering
    \includegraphics[width=1.0\textwidth]{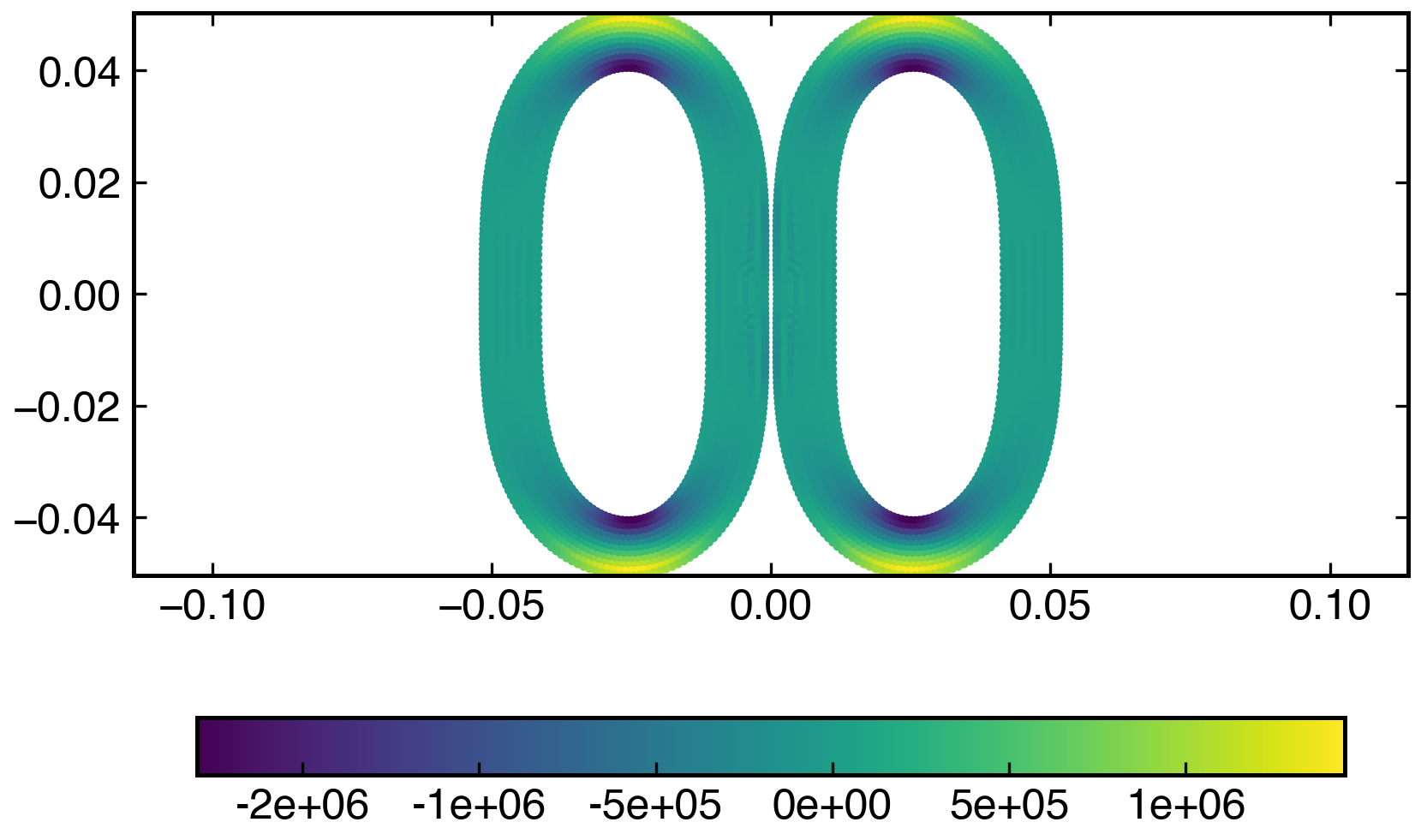}
    \subcaption{t = 4e-03 sec}\label{fig:rings:sun2019-nu-0.3975-2}
  \end{subfigure}

  \begin{subfigure}{0.48\textwidth}
    \centering
    \includegraphics[width=1.0\textwidth]{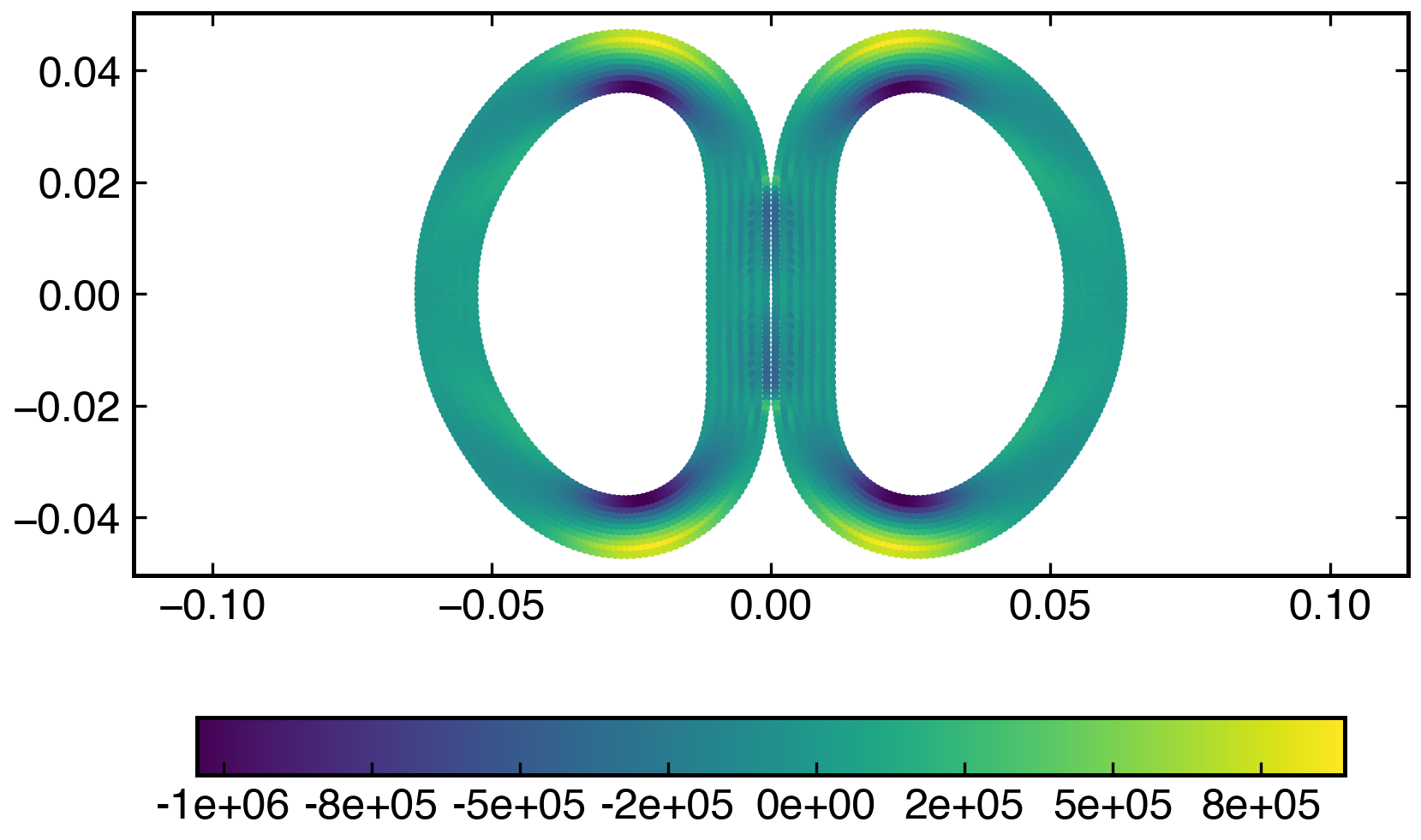}
    \subcaption{t = 7.3e-03 sec}\label{fig:rings:sun2019-nu-0.3975-3}
  \end{subfigure}
  \begin{subfigure}{0.48\textwidth}
    \centering
    \includegraphics[width=1.0\textwidth]{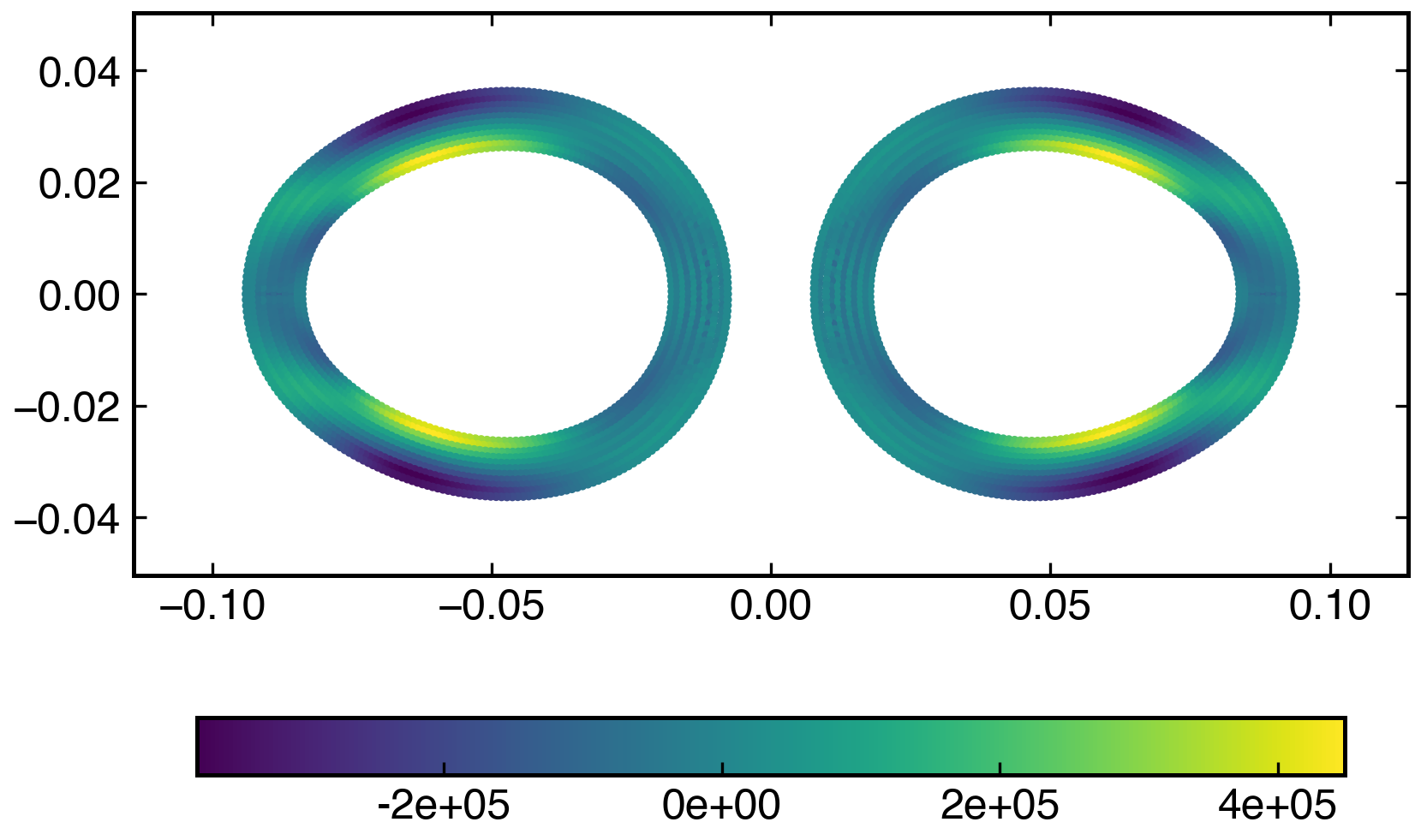}
    \subcaption{t = 1.45e-02 sec}\label{fig:rings:sun2019-nu-0.3975-4}
  \end{subfigure}
  \caption{Rings with a Poisson ratio of 0.3975 colliding head on, simulated with CTVF using SPST.}
\label{fig:rings:sun2019-nu-0.3975}
\end{figure}

Two different Poisson ratios are simulated. \Cref{fig:rings:sun2019-nu-0.3975}
shows the particle positions of rings with a Poisson ratio of $0.3975$ when
simulated with SPST. The recovery of the colliding rings without any tensile
instability can be seen.

\begin{figure}[!htpb]
  \centering
  %
  \begin{subfigure}{0.48\textwidth}
    \centering
    \includegraphics[width=1.0\textwidth]{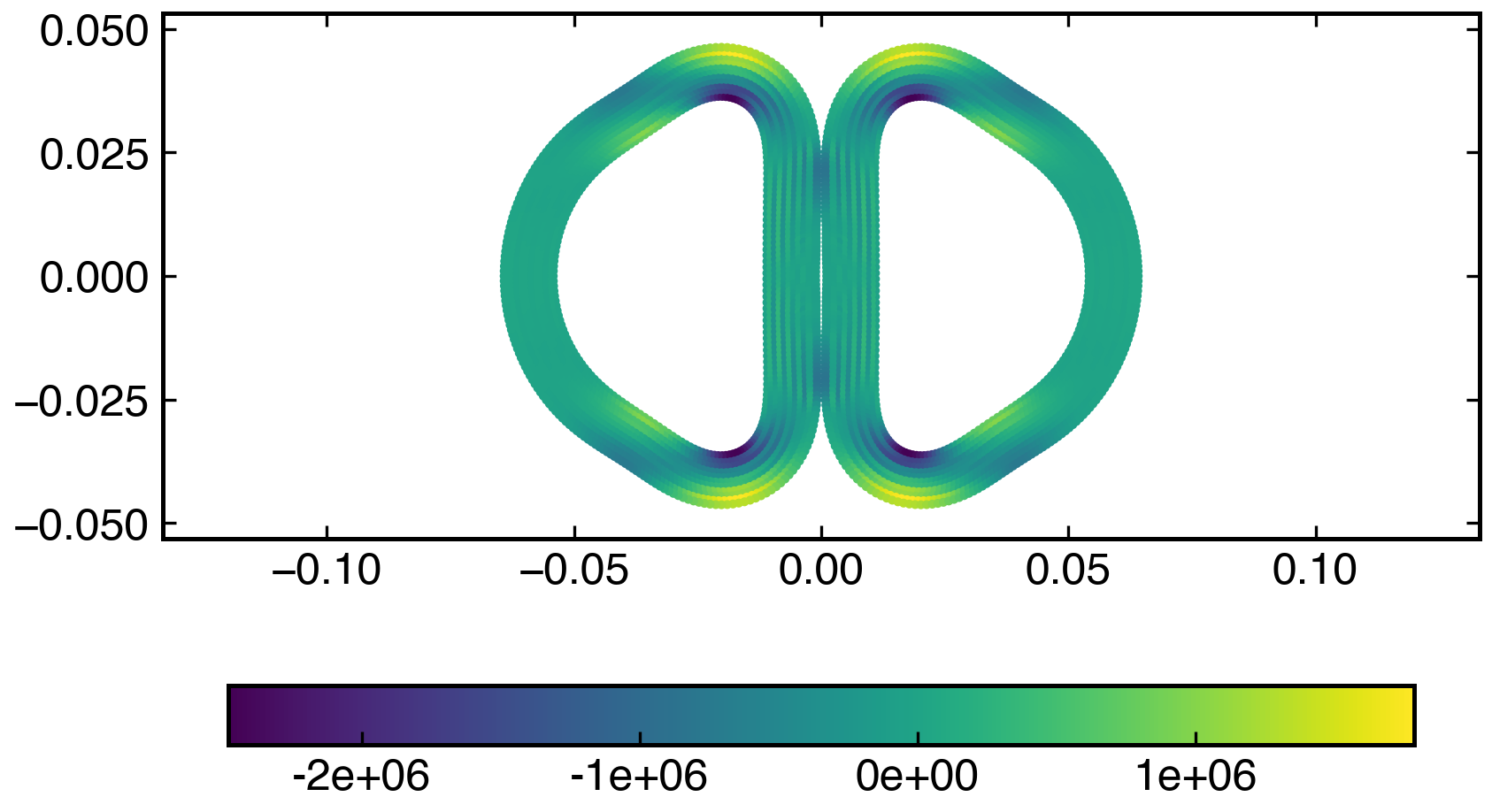}
    \subcaption{t = 2.5e-03 sec}\label{fig:rings:sun2019-nu-0.47-1}
  \end{subfigure}
  \begin{subfigure}{0.48\textwidth}
    \centering
    \includegraphics[width=1.0\textwidth]{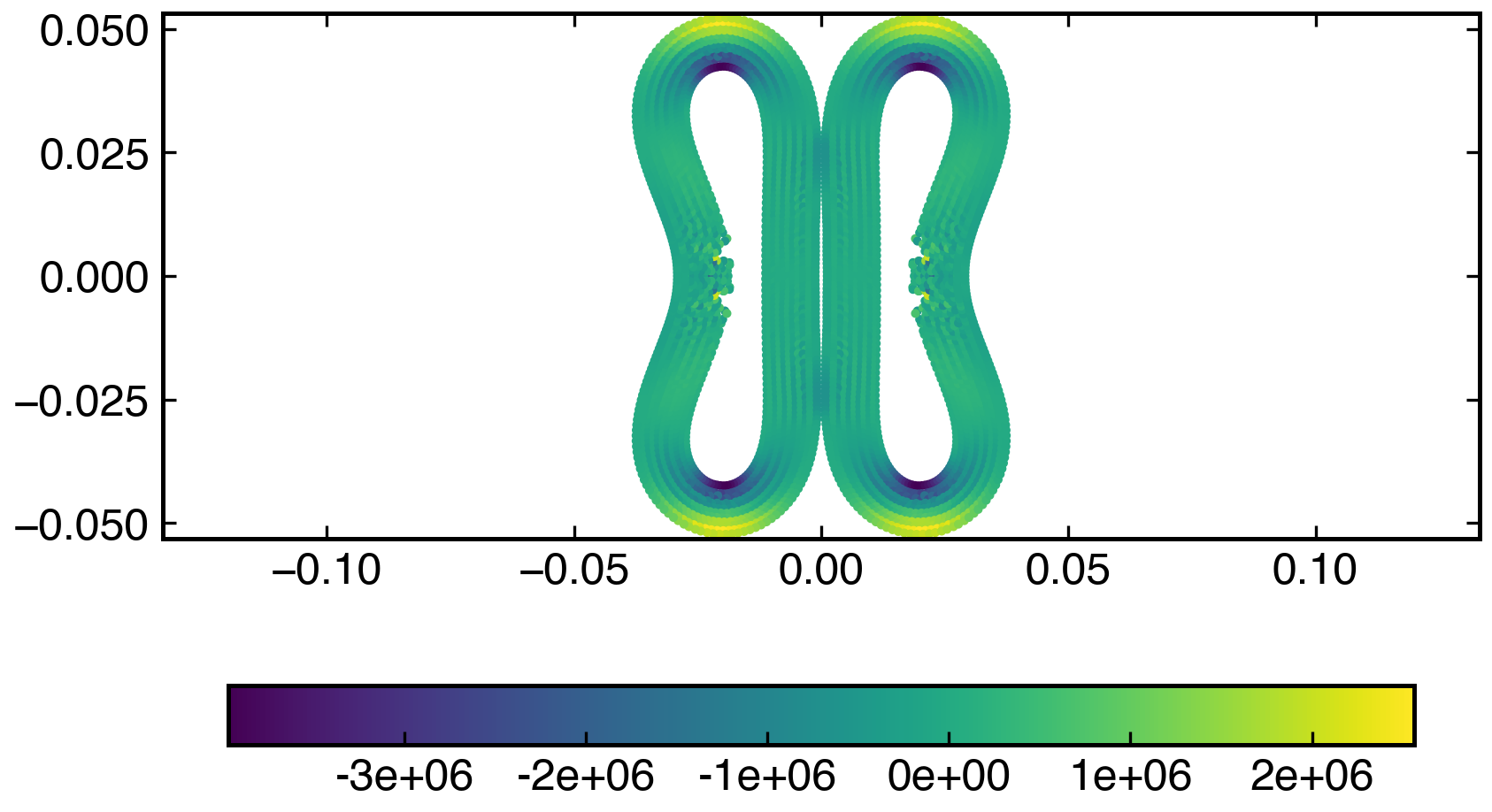}
    \subcaption{t = 4e-03 sec}\label{fig:rings:sun2019-nu-0.47-2}
  \end{subfigure}

  \begin{subfigure}{0.48\textwidth}
    \centering
    \includegraphics[width=1.0\textwidth]{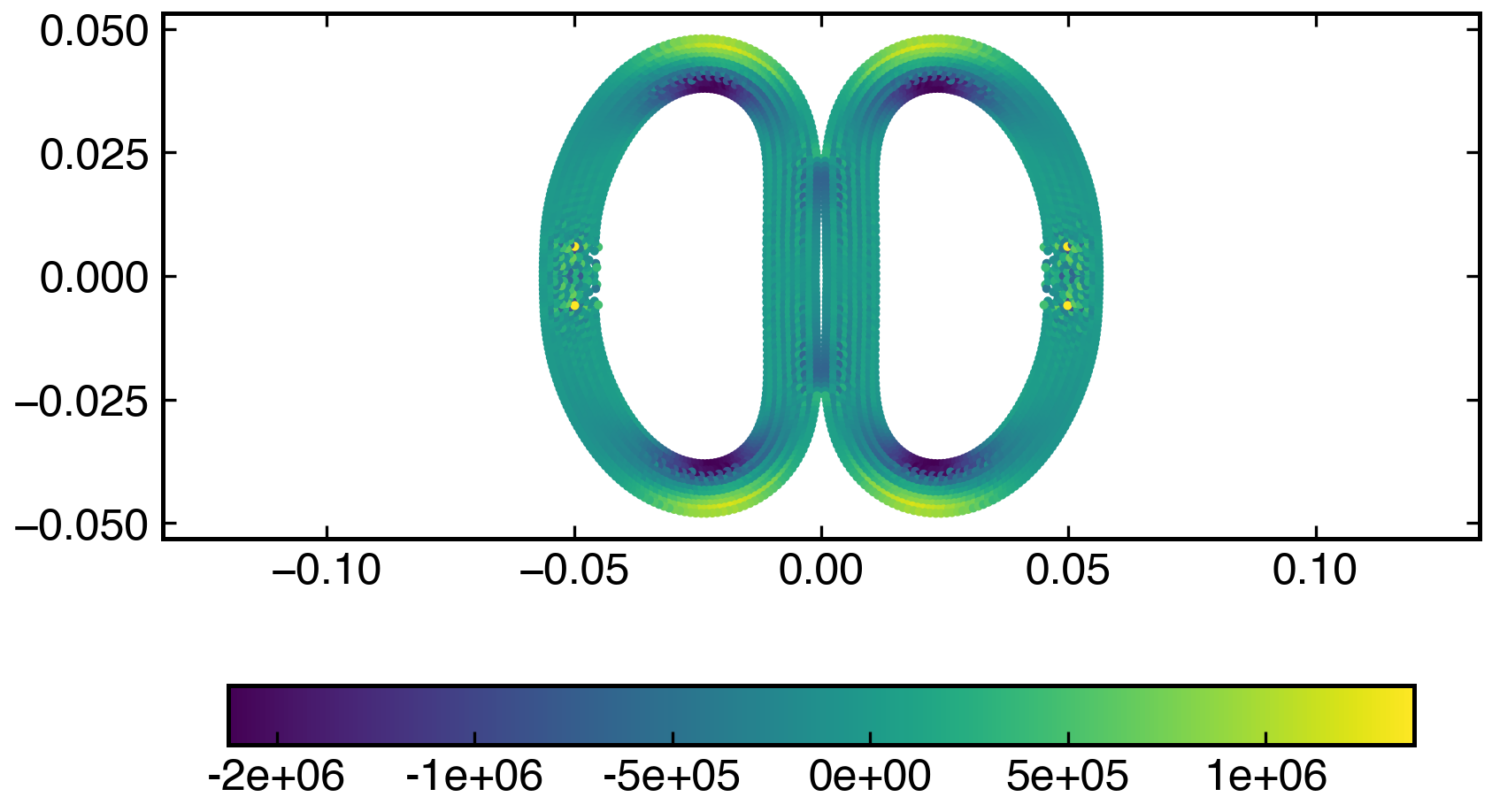}
    \subcaption{t = 7.3e-03 sec}\label{fig:rings:sun2019-nu-0.47-3}
  \end{subfigure}
  \begin{subfigure}{0.48\textwidth}
    \centering
    \includegraphics[width=1.0\textwidth]{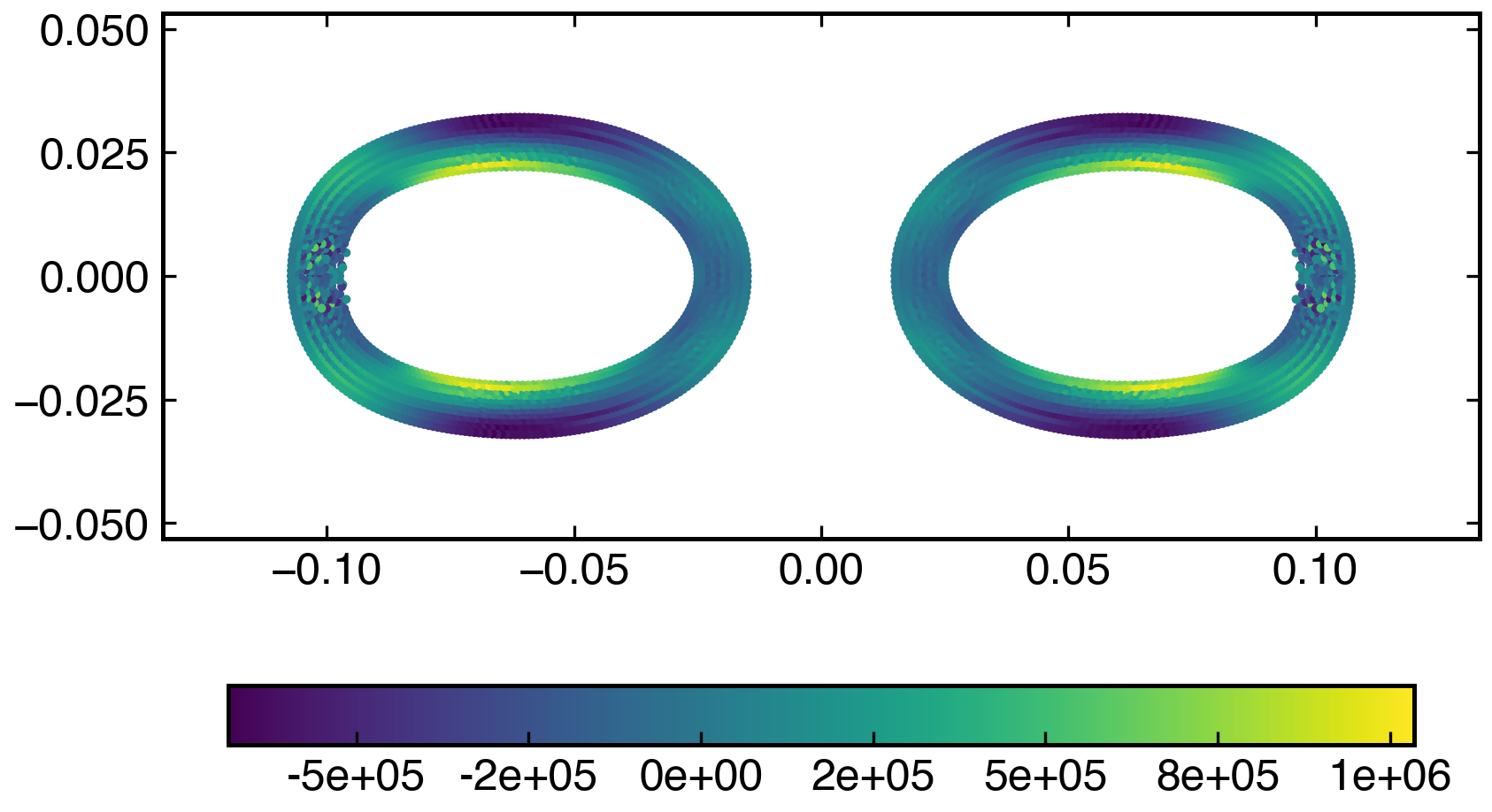}
    \subcaption{t = 1.45e-02 sec}\label{fig:rings:sun2019-nu-0.47-4}
  \end{subfigure}
  %
  \caption{Rings with a Poisson ratio of 0.47 colliding head on, simulated with CTVF using SPST.}
\label{fig:rings:sun2019-nu-0-47}
\end{figure}

\begin{figure}[!htpb]
  \centering
  %
  \begin{subfigure}{0.48\textwidth}
    \centering
    \includegraphics[width=1.0\textwidth]{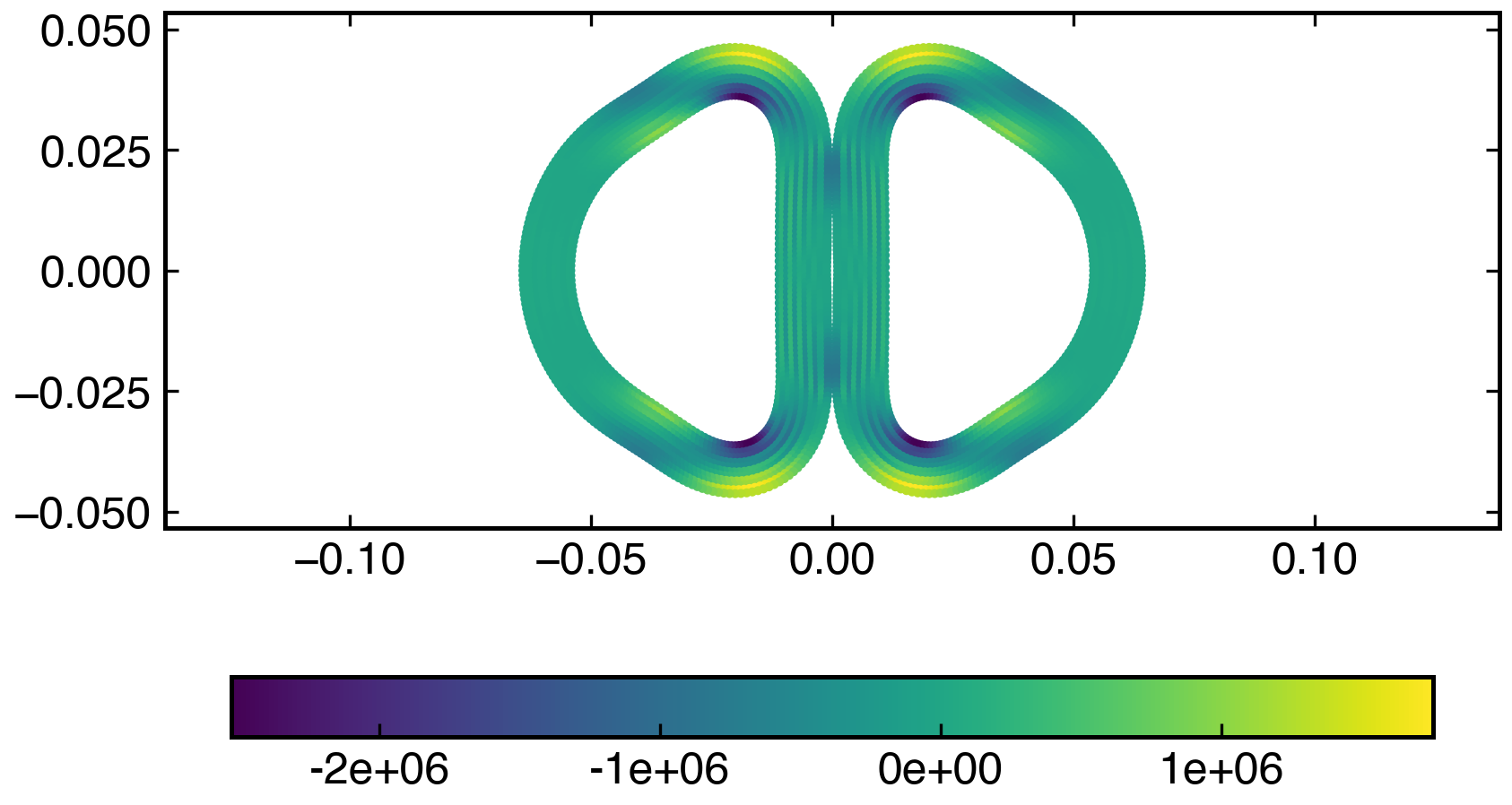}
    \subcaption{t = 2.5e-03 sec}\label{fig:rings:ipst-nu-0.47-1}
  \end{subfigure}
  \begin{subfigure}{0.48\textwidth}
    \centering
    \includegraphics[width=1.0\textwidth]{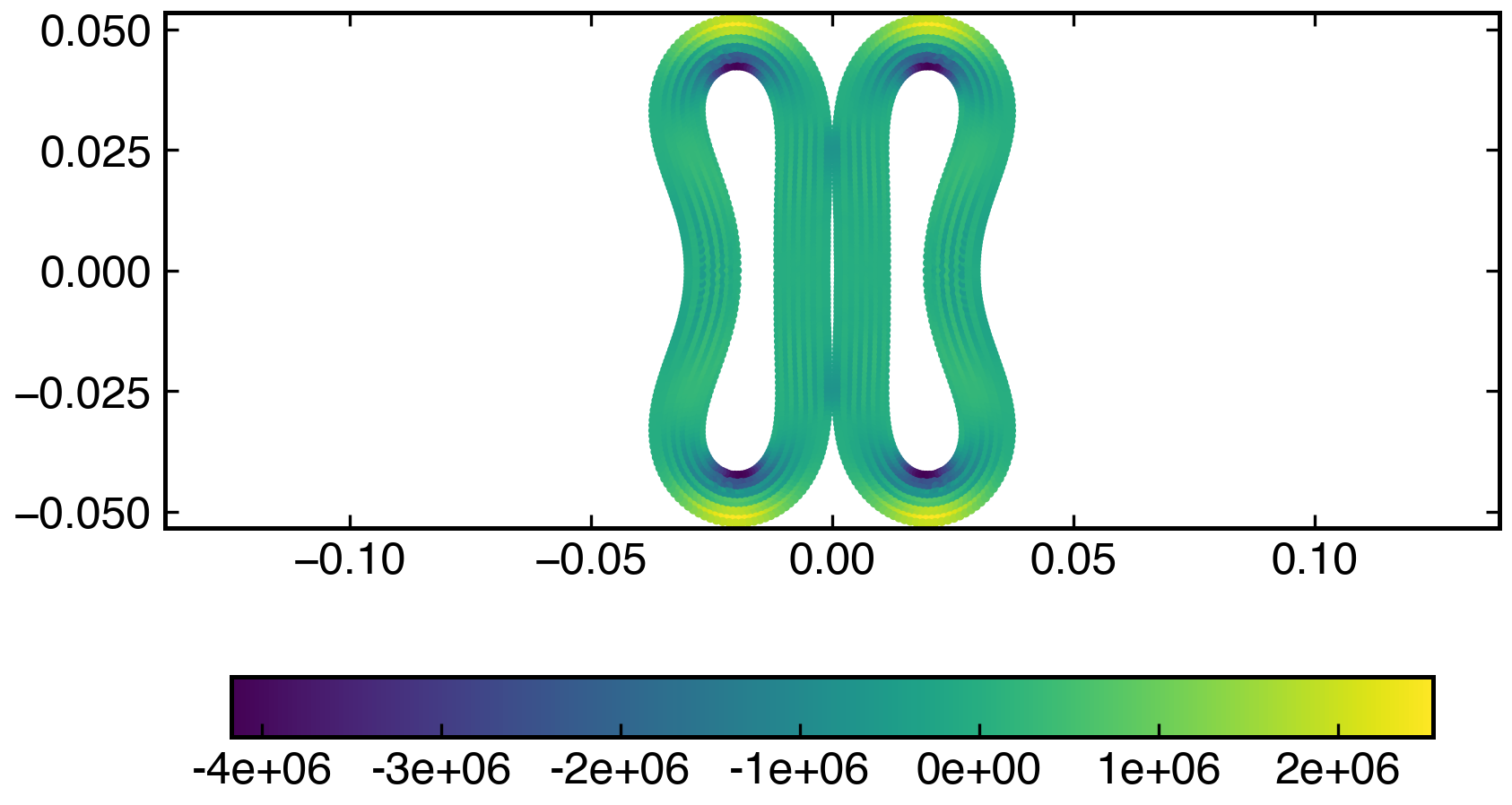}
    \subcaption{t = 4e-03 sec}\label{fig:rings:ipst-nu-0.47-2}
  \end{subfigure}

  \begin{subfigure}{0.48\textwidth}
    \centering
    \includegraphics[width=1.0\textwidth]{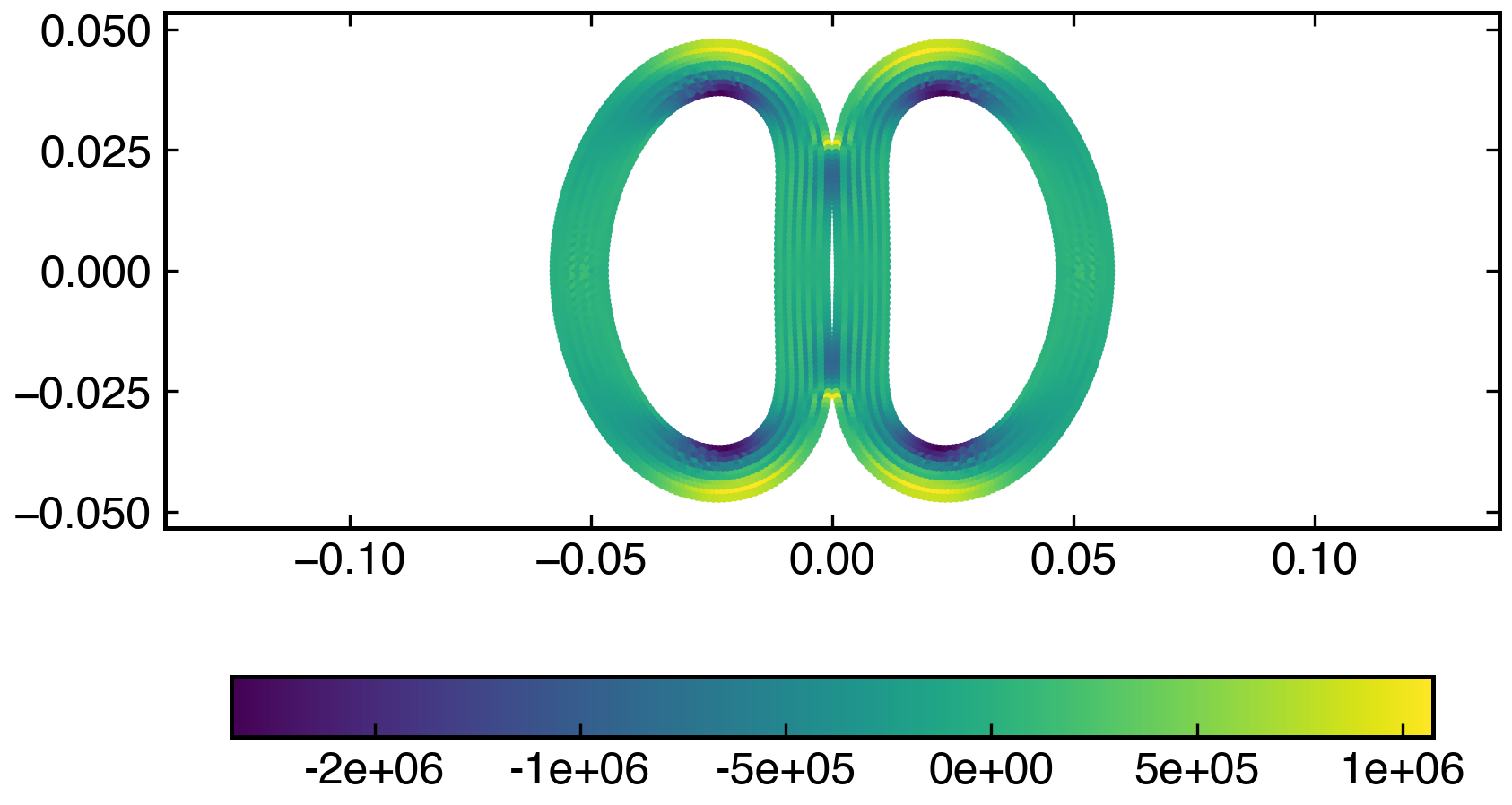}
    \subcaption{t = 7.3e-03 sec}\label{fig:rings:ipst-nu-0.47-3}
  \end{subfigure}
  \begin{subfigure}{0.48\textwidth}
    \centering
    \includegraphics[width=1.0\textwidth]{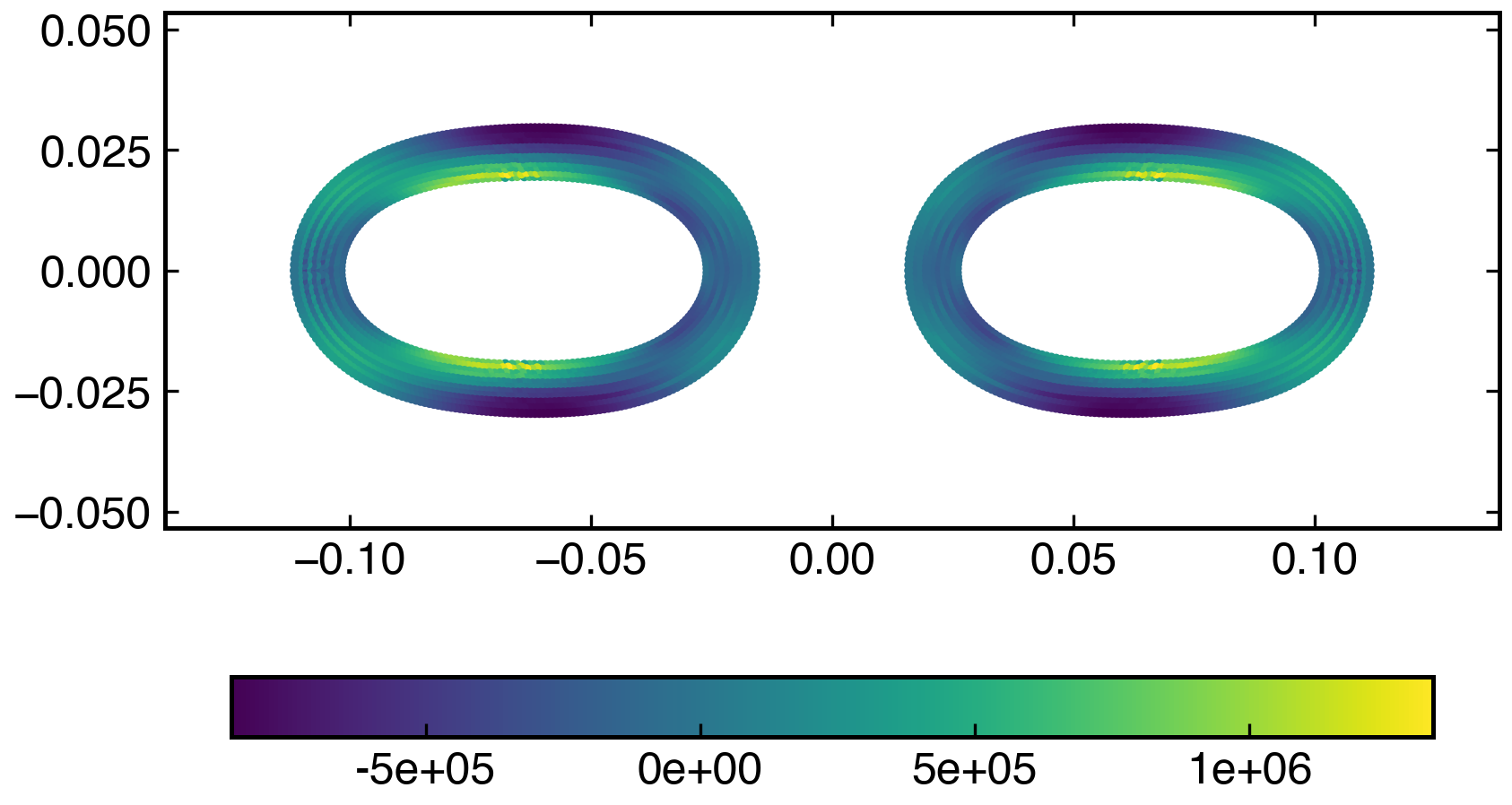}
    \subcaption{t = 1.45e-02 sec}\label{fig:rings:ipst-nu-0.47-4}
  \end{subfigure}
  %
  \caption{Rings with a Poisson ratio of 0.47 colliding head on, simulated with CTVF using IPST.}
\label{fig:rings:ipst-nu-0-47}
\end{figure}

\begin{figure}[!htpb]
  \centering
  \includegraphics[width=0.4\textwidth]{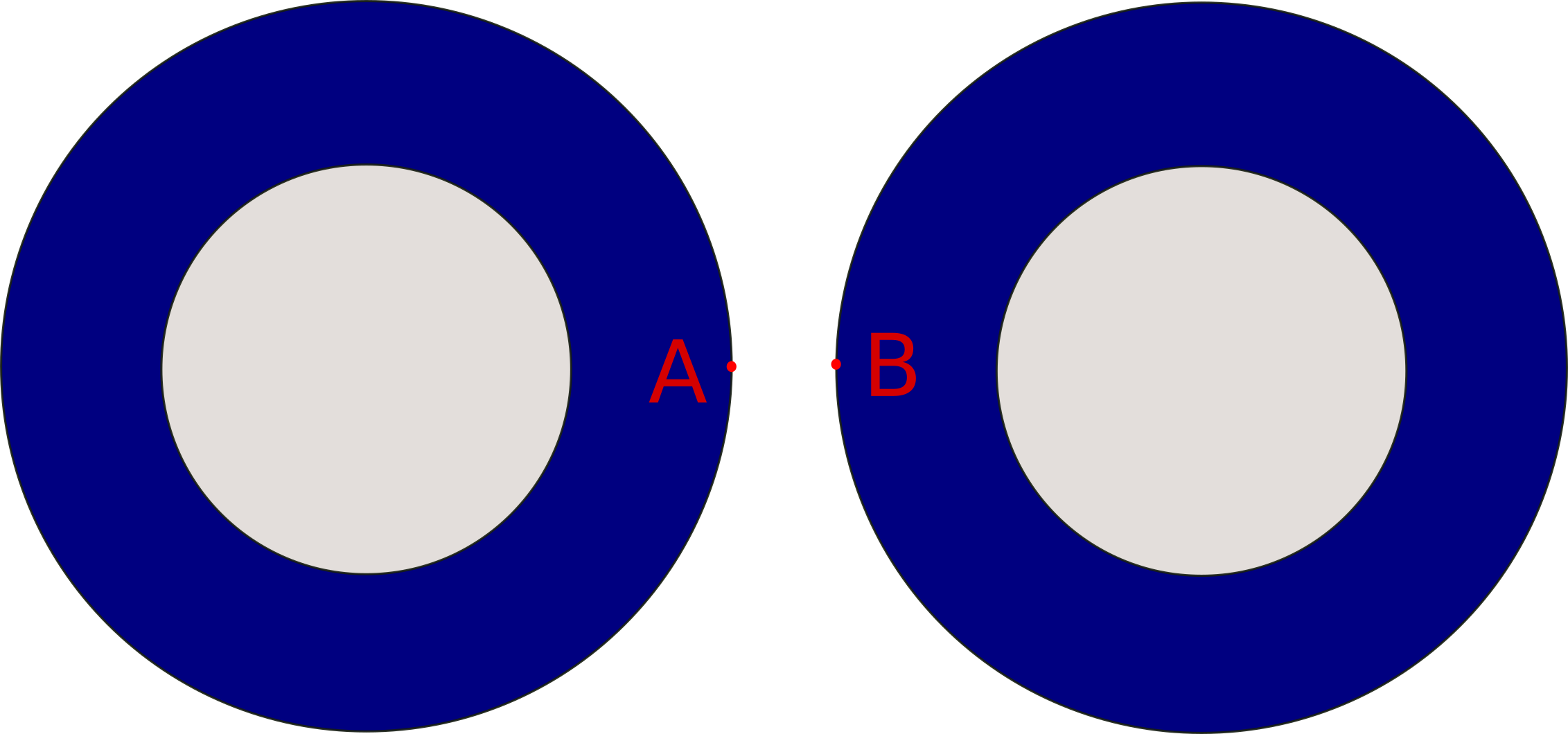}
  \caption{Schematic diagram of two rings colliding. Points A and B are marked.}
\label{fig:rings_initial}
\end{figure}

\begin{figure}[!htpb]
  \centering
  \begin{subfigure}{0.48\textwidth}
    \centering
    \includegraphics[width=1.0\textwidth]{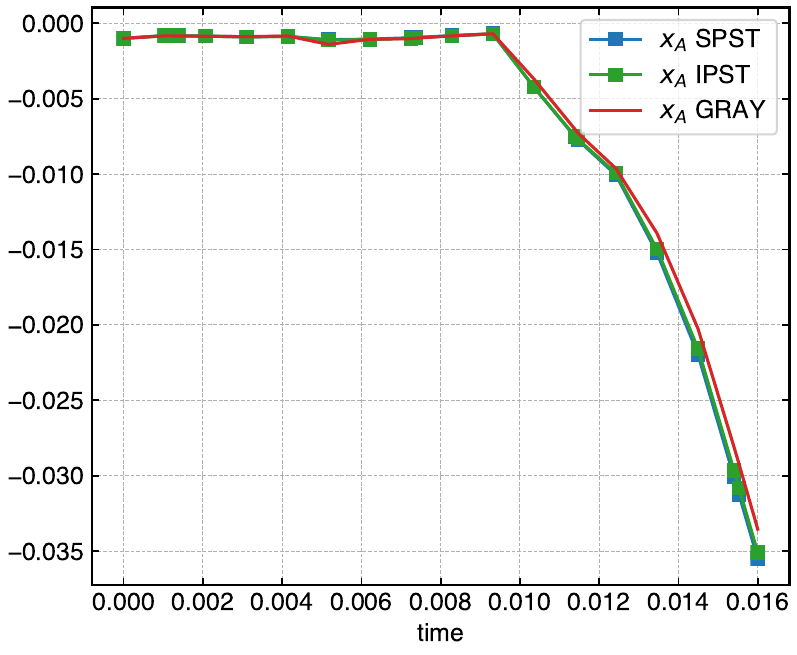}
    \subcaption{$x$ coordinate of the point A}\label{fig:rings-compare-x}
  \end{subfigure}
  \begin{subfigure}{0.48\textwidth}
    \centering
    \includegraphics[width=1.0\textwidth]{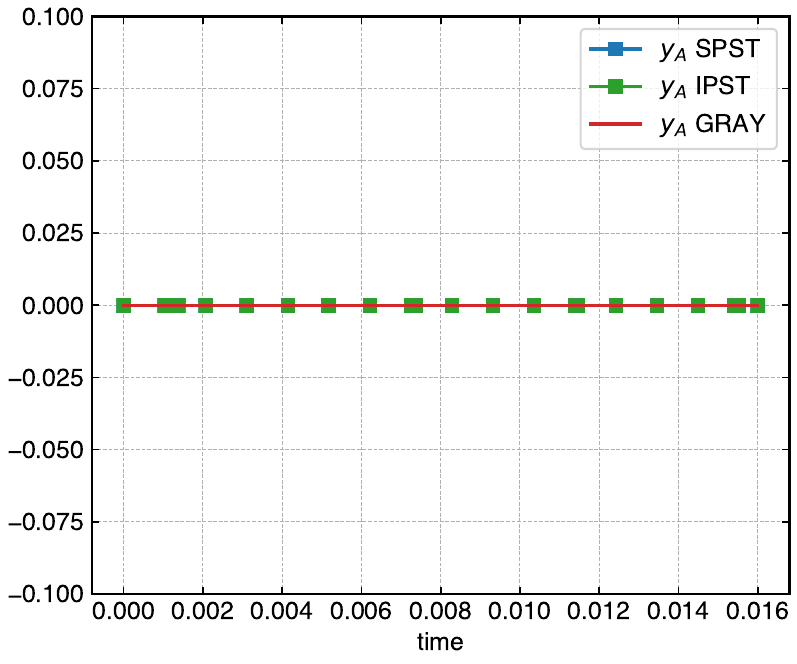}
    \subcaption{$y$ coordinate of the point A}\label{fig:rings-compare-y}
  \end{subfigure}
  \caption{The evolution of the $x$ and $y$ coordinates of points A and B for
    the CTVF using SPST, IPST, and compared with that of
    Gray~\cite{gray-ed-2001}.}
\label{fig:rings_compare}
\end{figure}

We also consider higher Poisson ratios, such as 0.47.
\Cref{fig:rings:sun2019-nu-0-47} shows the particle positions of rings when
simulated with SPST and \cref{fig:rings:ipst-nu-0-47} with IPST. Even though
both the particle shifting techniques are able to eliminate the numerical
fracture, IPST gives better results as in the distribution of particles
through out the simulation, see \cref{fig:rings:sun2019-nu-0.47-2} and
\cref{fig:rings:ipst-nu-0.47-2}. For the case where SPST is used, the final
particle distribution is not very uniform. This is not the case when IPST is
used. We can therefore say that IPST performs better than SPST.
In order to compare the different schemes quantitatively for this problem, we
plot the $x$ and $y$ positions of the point A of the left ring, as can be seen
in \cref{fig:rings_initial}. \Cref{fig:rings_compare} shows the results and as
can be seen excellent agreement of the different methods for this
problem.

\FloatBarrier%
\subsection{High velocity impact}

High-velocity impact problems are important in various contexts like space
debris applications. This case tests if the scheme is capable of hand large
deformation problems.

The projectile and the target are made of aluminium material. The projectile
is 10mm in diameter and the rectangular target has a size of $2 \times 50$ mm.
The projectile and the target have the following material properties: density
$\rho = 2785 $ kg\,m\textsuperscript{-3}, sound speed $c_0 = 5328$
 m\,s\textsuperscript{-1}, shear modulus $G=2.76 \times 10^{7}$ kPa, yield modulus
$Y_0 = 3.0 \times 10^{5}$ kPa, as studied in~\cite{zhang_hu_adams17}. The impact velocity is set to
$V_0 = 3100.0$m\,s\textsuperscript{-1}. The initial particle spacing is
$\Delta x = 0.5$ mm.
\begin{figure}[!htpb]
  \centering
  \begin{subfigure}{0.3\textwidth}
    \centering
    \includegraphics[width=1.0\textwidth]{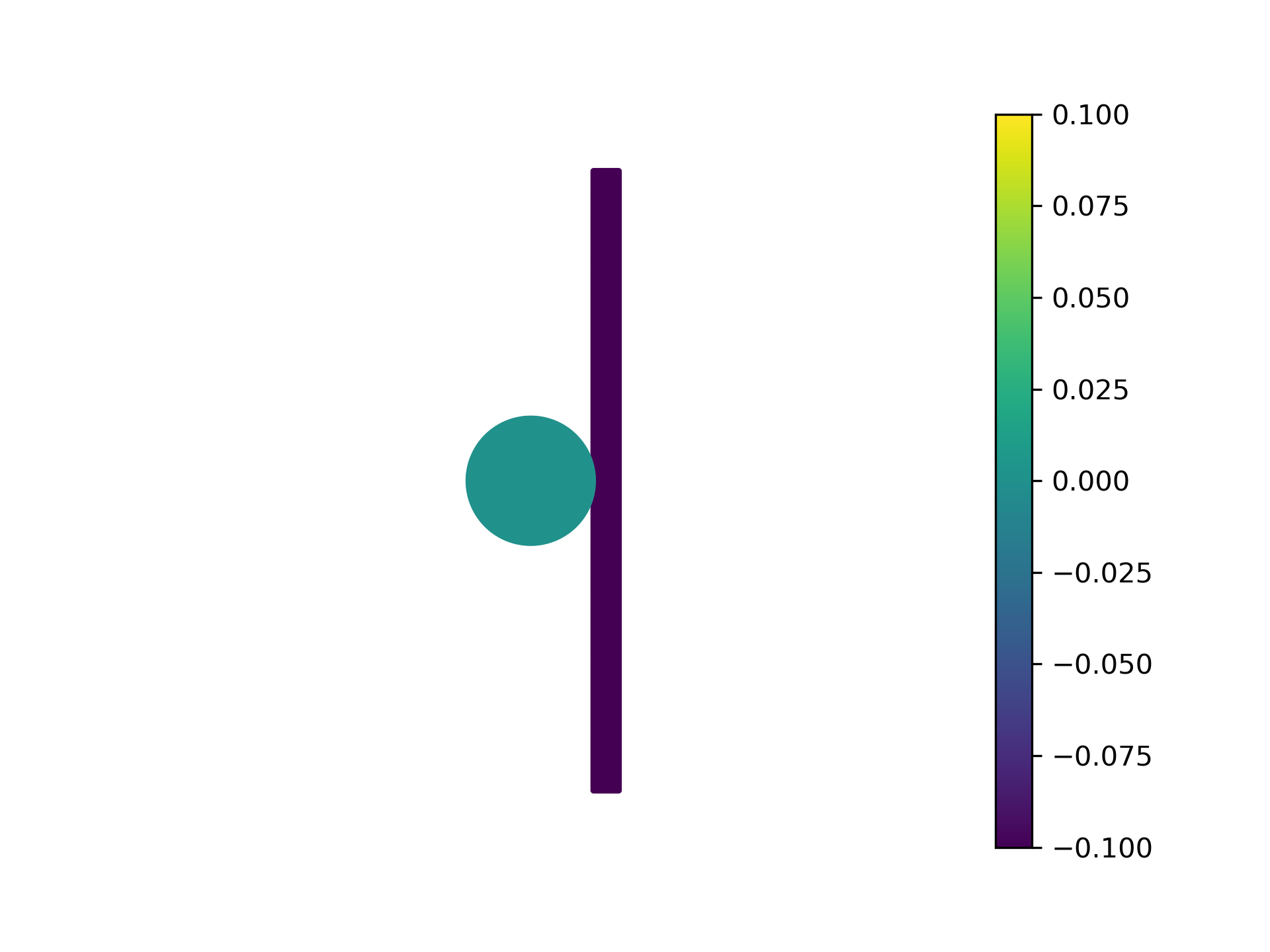}
    \subcaption{t = 0 sec}\label{}
  \end{subfigure}
  \begin{subfigure}{0.3\textwidth}
    \centering
    \includegraphics[width=1.0\textwidth]{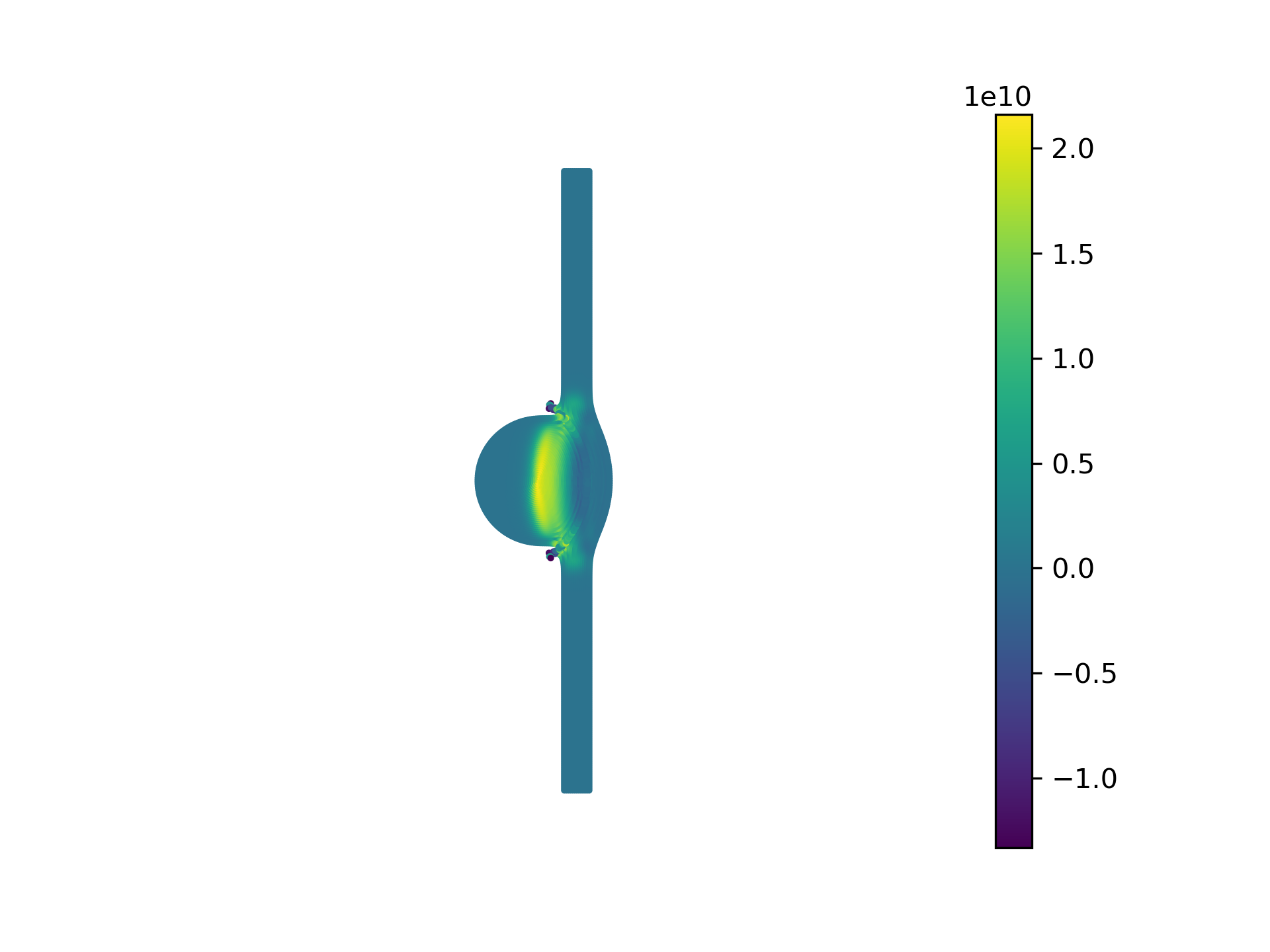}
    \subcaption{t = 2.5e-03 sec}\label{}
  \end{subfigure}
  \begin{subfigure}{0.3\textwidth}
    \centering
    \includegraphics[width=1.0\textwidth]{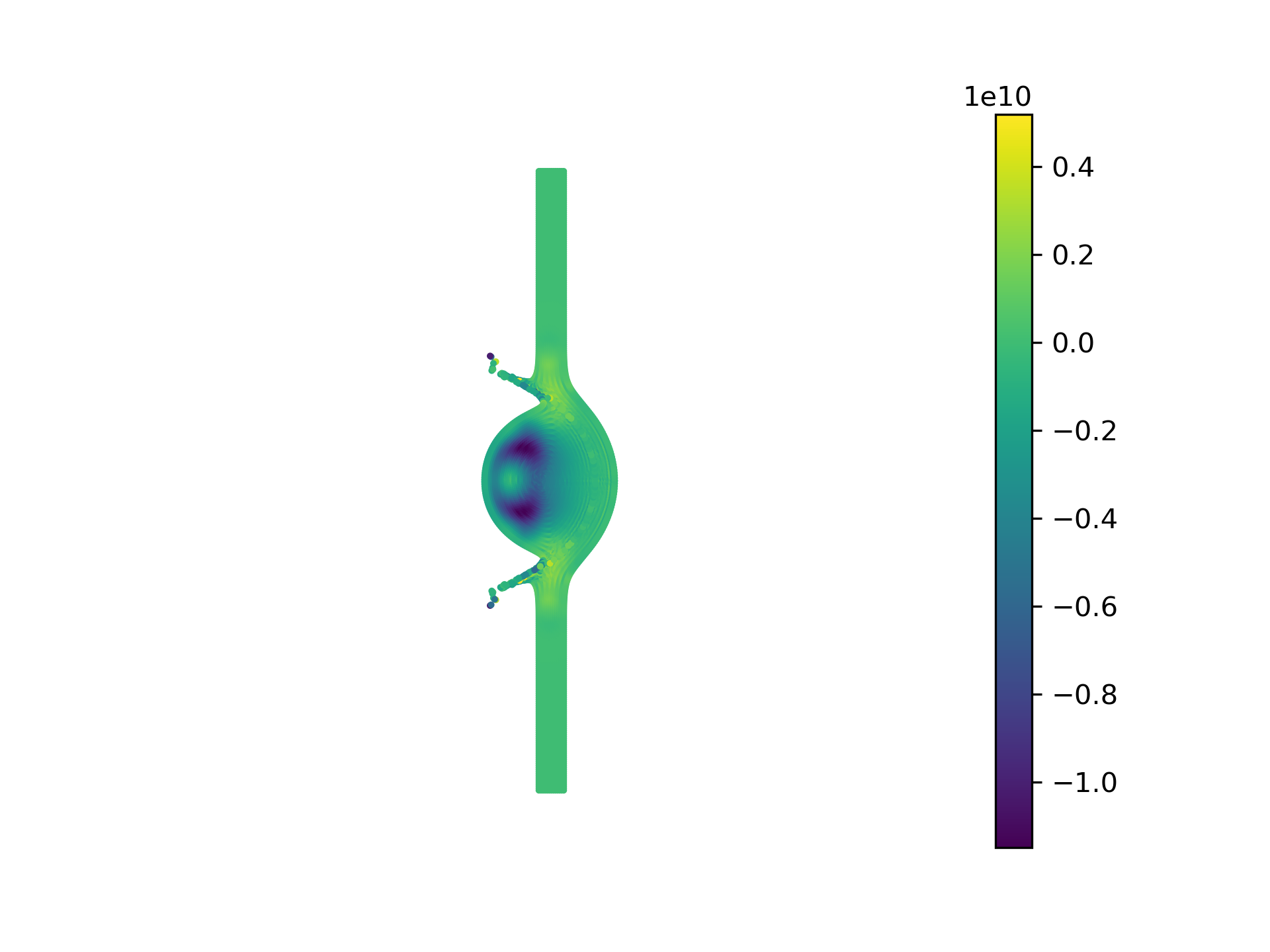}
    \subcaption{t = 4e-03 sec}\label{}
  \end{subfigure}

  \begin{subfigure}{0.3\textwidth}
    \centering
    \includegraphics[width=1.0\textwidth]{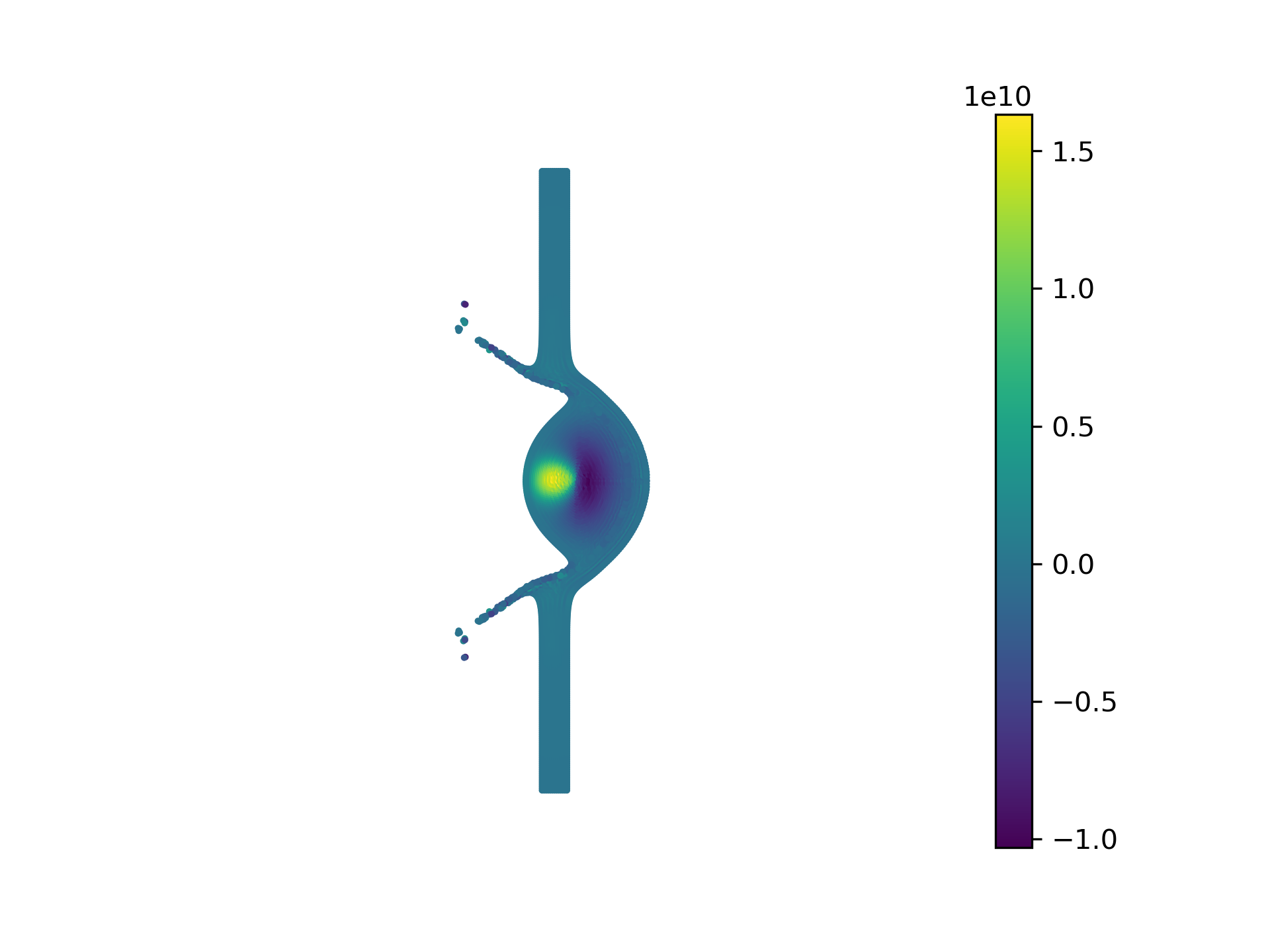}
    \subcaption{t = 7.3e-03 sec}\label{}
  \end{subfigure}
  \begin{subfigure}{0.3\textwidth}
    \centering
    \includegraphics[width=1.0\textwidth]{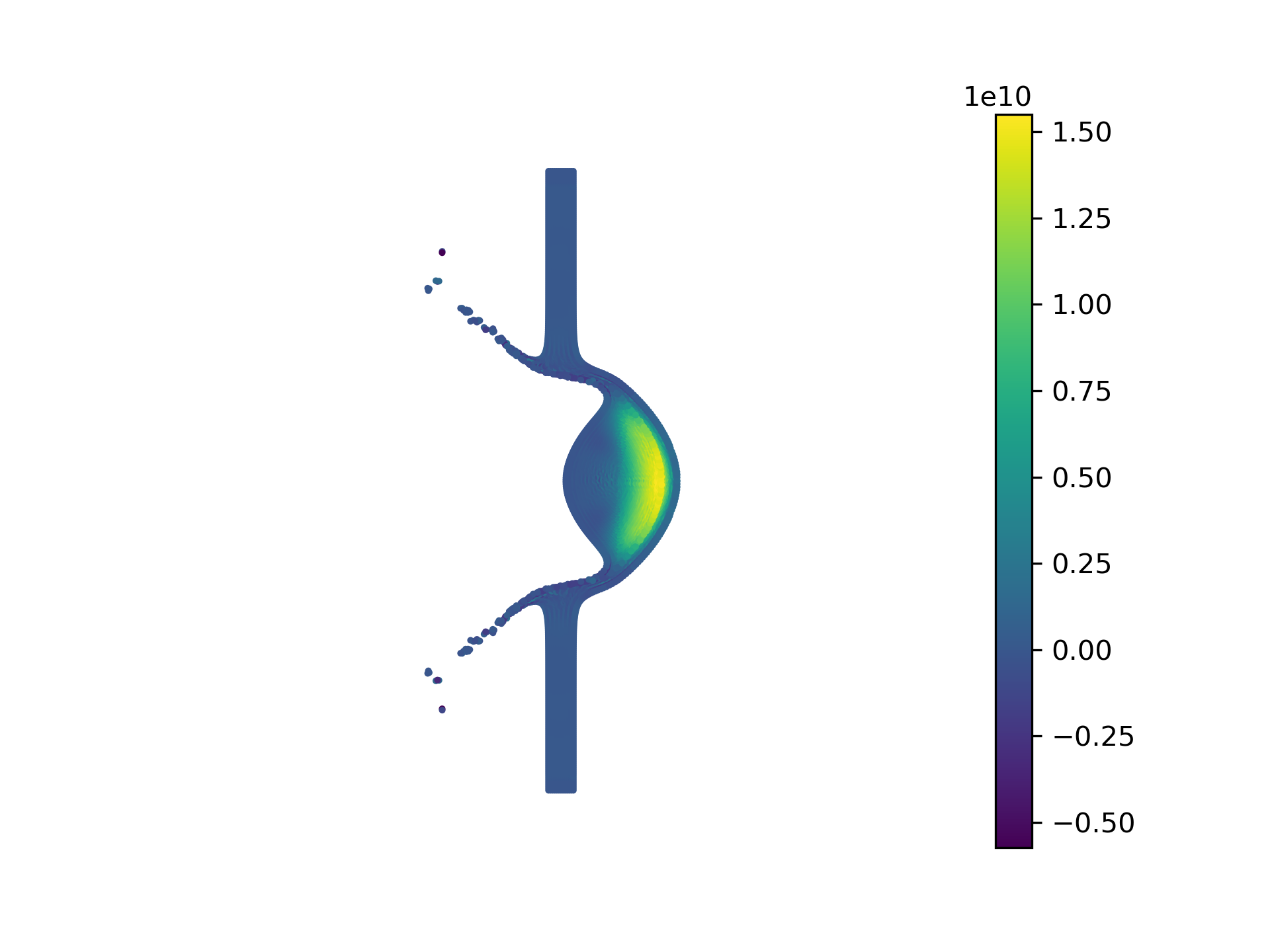}
    \subcaption{t = 1.45e-02 sec}\label{}
  \end{subfigure}
  \begin{subfigure}{0.3\textwidth}
    \centering
    \includegraphics[width=1.0\textwidth]{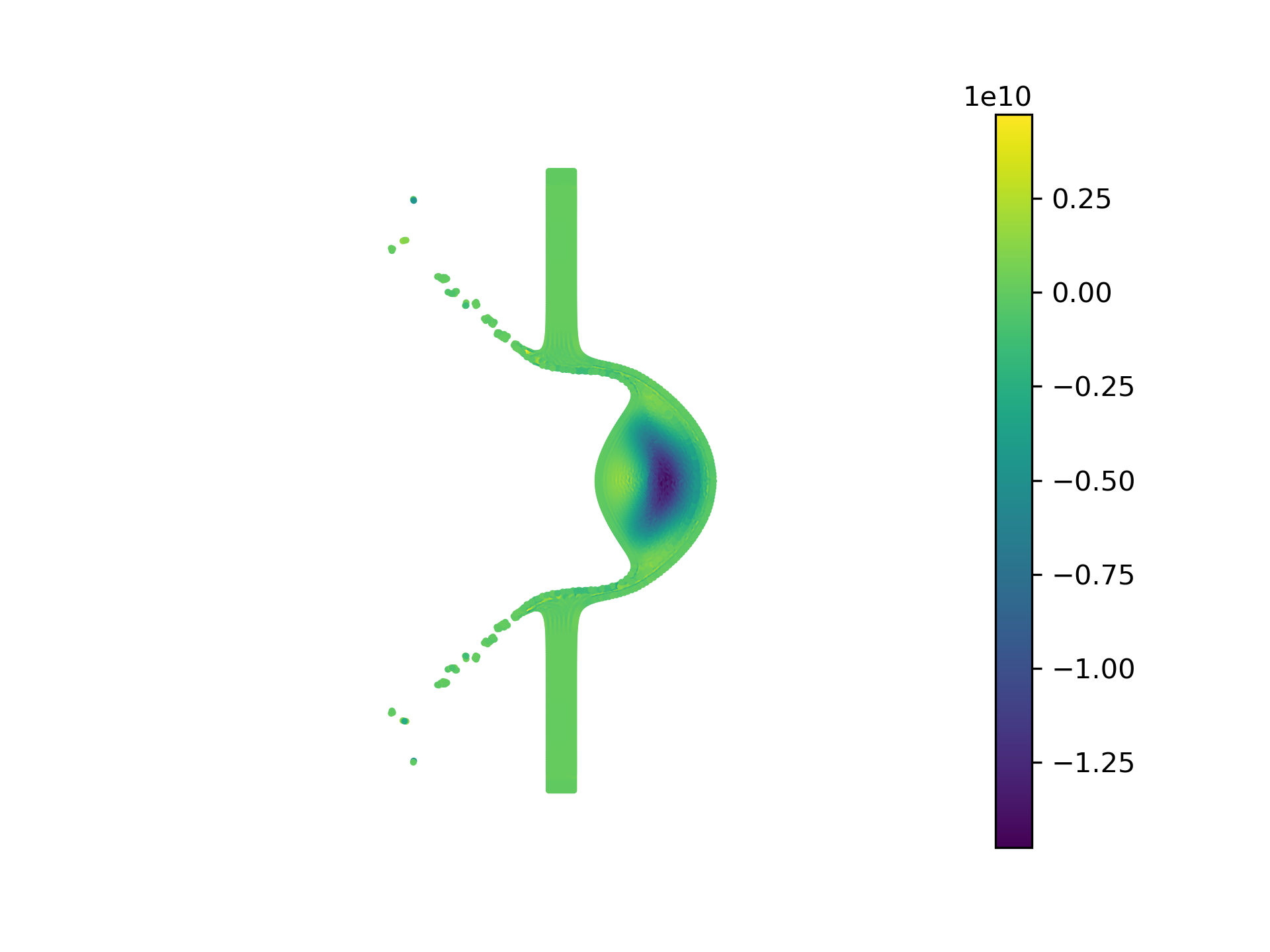}
    \subcaption{t = 1.5e-02 sec}\label{}
  \end{subfigure}

  \begin{subfigure}{0.3\textwidth}
    \centering
    \includegraphics[width=1.0\textwidth]{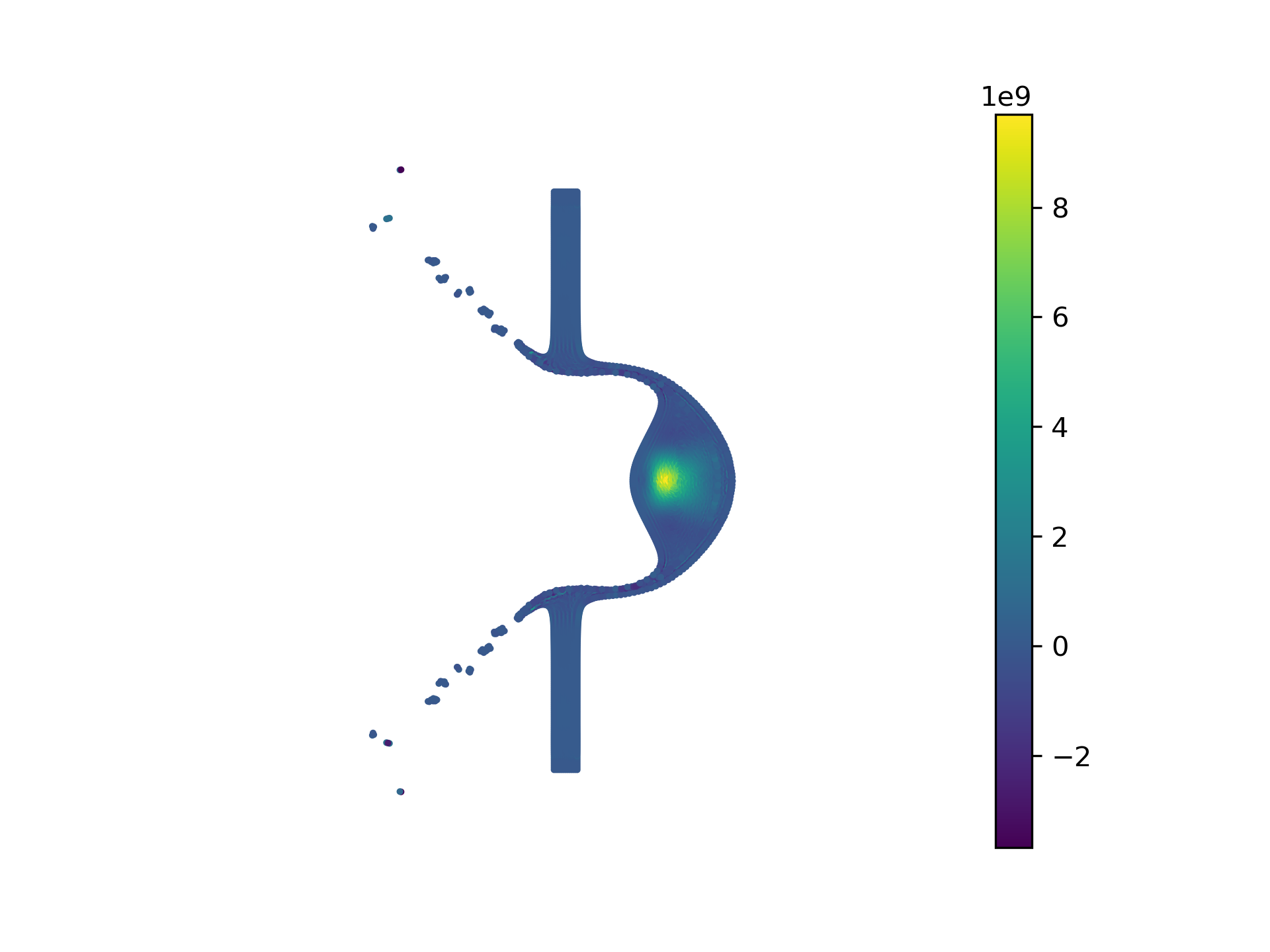}
    \subcaption{t = 1.5e-02 sec}\label{}
  \end{subfigure}
  \begin{subfigure}{0.3\textwidth}
    \centering
    \includegraphics[width=1.0\textwidth]{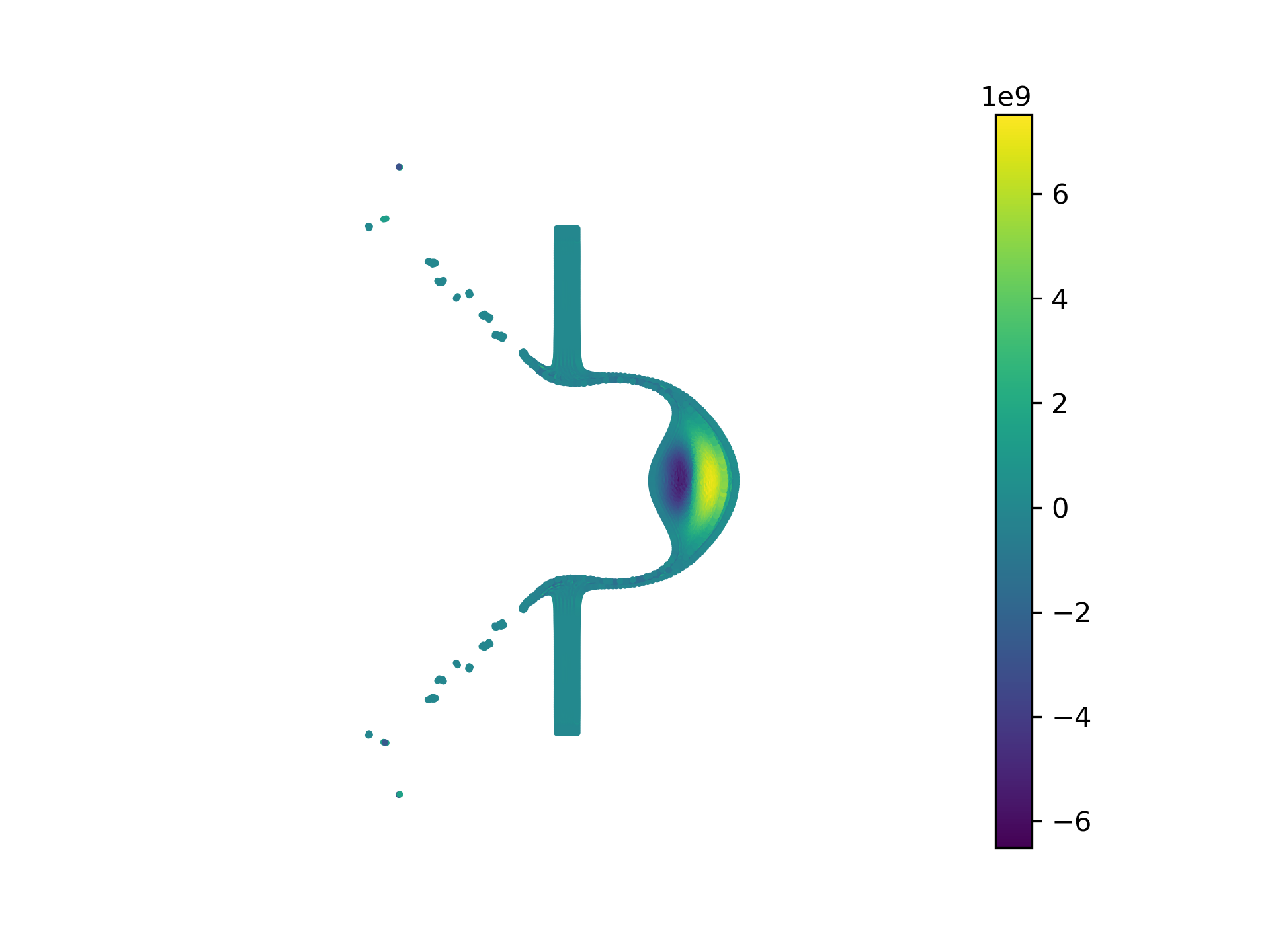}
    \subcaption{t = 1.5e-02 sec}\label{}
  \end{subfigure}
  \begin{subfigure}{0.3\textwidth}
    \centering
    \includegraphics[width=1.0\textwidth]{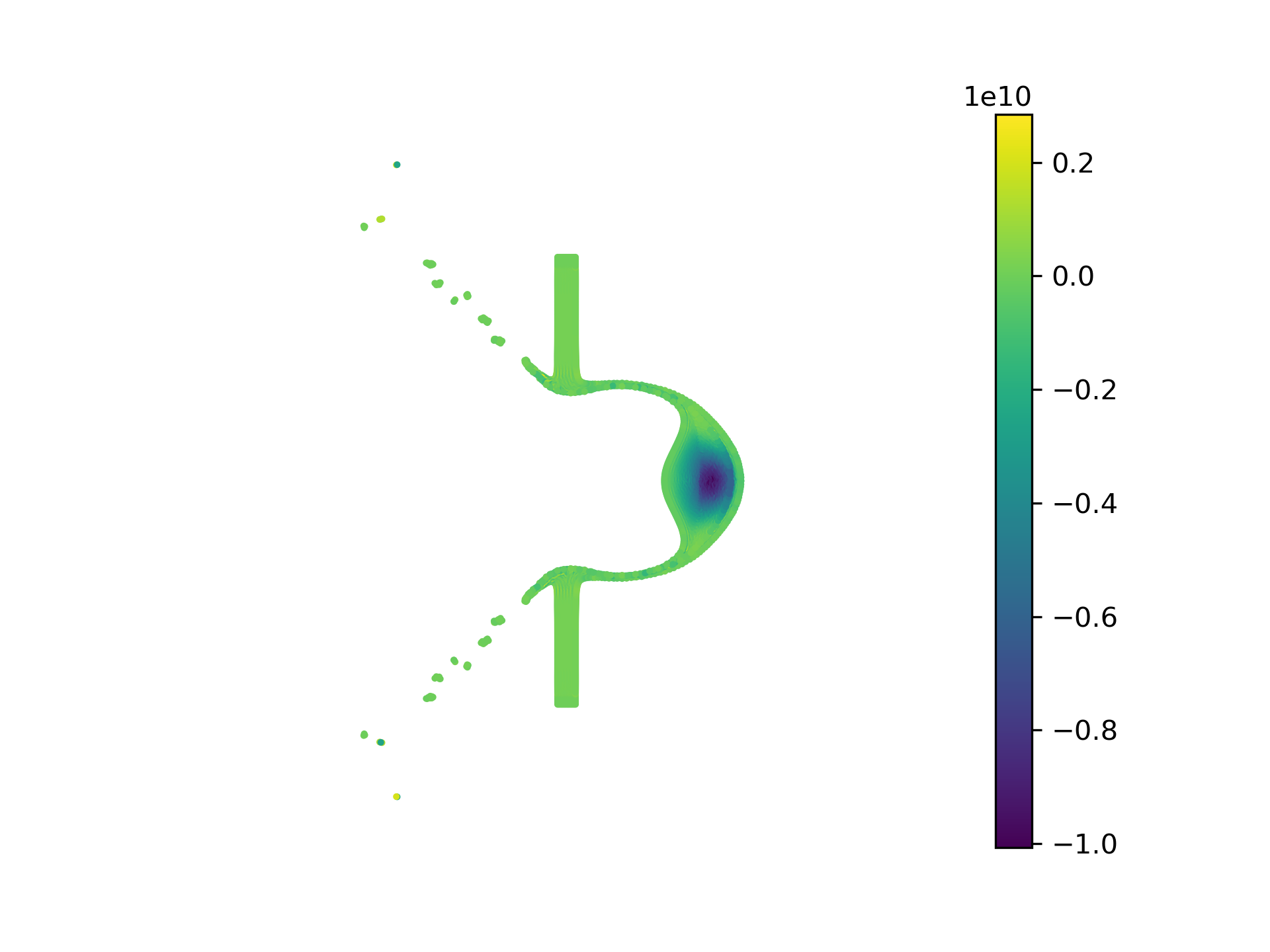}
    \subcaption{t = 1.5e-02 sec}\label{}
  \end{subfigure}
  \caption{High velocity impact of cylinder on to a structure}
\label{fig:hvi:etvf-sun2019}
\end{figure}
Here the aluminium follows an elastic-perfectly plastic constitutive model. In
elastic perfectly plastic model, the material is assumed to be elastic up to
the yield point and once the material reaches the yield point, there will be
no further increase in the stress, and is bounded by a factor
$\beta = \min\left(\frac{Y_0^2}{3J_2}, 1 \right)$, where $J_2$ is calculated
from $J_2 = \frac{1}{2} \teng{\sigma}^{'} : \teng{\sigma}^{'}$. We use an
$\alpha=1$ in \cref{eq:mom-av} in the current case.

\Cref{fig:hvi:etvf-sun2019} shows the plots of cylinder impacting the
structure at different time instants. This is computed using the particle
shifting technique of Sun~\cite{sun_consistent_2019}. The color contour
represents the pressure of the particles. The width of the hole created by the
cylinder is $19.6$ mm. When computed using the GTVF
scheme~\cite{zhang_hu_adams17} the hole has a size of $19.8$ mm. In
\citet{howell2002free}, the value cited is $19.2$ mm. We can see, that the
current scheme is closer to the one simulated by \cite{howell2002free}, which
is taken as reference in \cite{zhang_hu_adams17}.

\section{Discussion and conclusions}

The proposed CTVF scheme builds on the original TVF scheme of
\citet{Adami2013} and is as an improvement on the GTVF of
\citet{zhang_hu_adams17}. In addition it generalizes the implementation of the
EDAC-SPH method~\cite{edac-sph:cf:2019} where the TVF formulation was used for
internal flows and a separate WCSPH formulation used for fluid flows with a
free-surface. The current work proposes the addition of a few correction terms
which improve the accuracy of the method as demonstrated in the earlier
section. The addition of the terms imposes a small computational cost but
compensates through the improved accuracy. As an example, in simulating the
lid-driven cavity problem with a resolution of $50 \times 50$, the original
EDAC scheme without any of the correction terms with a one step predictor
corrector integrator takes $251$ seconds for a time of $25$ seconds, the new
scheme with a kick-drift-kick scheme takes $293$ seconds. Despite the change
of the integrator this is a small increase in the performance. For solid
mechanics problems we consider the colliding rings problem simulated for a
total time of $0.016$ seconds. This takes $98$ seconds of time to simulate
with the full CTVF scheme, and takes $73$ seconds without the corrections (run
on Intel i5-7400, quad core machine). Free-surfaces are handled carefully.
The method produces smoother pressure fields due to the use of the EDAC
scheme. Finally the method is robust to changes in the PST method used. This
has been demonstrated using both the PST of \citet{sun_consistent_2019} and
the IPST of \citet{huang_kernel_2019}.

An important feature of the proposed scheme is that it works well in the
context of both fluid mechanics and solid mechanics. For elastic dynamics we
propose correction terms that improve the accuracy and robustness of the
method. The GTVF~\cite{zhang_hu_adams17} method fails when the PST method is
changed as demonstrated in \cref{sec:oscillating-plate}, however the proposed
method is more robust. Furthermore, our method uses the true velocity in order
to compute the velocity gradient. The results of the uniaxial compression
problem in \cref{sec:uniaxial-compression} suggest that that the proposed
method is more accurate than the GTVF. The main difference between the GTVF
and the current scheme in the context of solid mechanics is the addition of
the correction terms to the continuity equation, the usage of momentum
velocity $\ten{u}$ in the computation of the velocity gradient, and the new
particle shifting technique incorporation. We have found that the additional
terms arising in the equation for the Jaumann stress rate
\cref{eq:jaumann-stress-rate} has negligible influence and can be safely
ignored. However, the computations in this work have included this term. The
additional stress term in the momentum equation is negligible and has not been
employed. We reiterate that for the fluid mechanics simulations the additional
stress terms in the momentum equation are not negligible.

We note that for solid mechanics problems the method works well with either
the traditional state equation used for the pressure evolution or the use of
the EDAC equation. This does not make a significant difference for these
problems since there is no additional damping added to the evolution equation
for the deviatoric stresses. The EDAC evolution equation does make a
significant improvement to the pressure evolution in the fluid mechanics
problems as discussed earlier in \cite{edac-sph:cf:2019}.

The newly proposed method has not been applied to three dimensional problems
or to fluid structure interaction (FSI) problems. We believe that the method
would be easier to use in the context of FSI since it can handle both fluids
and solids in the same formulation. We propose to investigate these in the
future.

\section*{Acknowledgment}

The authors would like to thank Abhinav Muta for valuable discussions and
feedback.

\section*{References}
\bibliographystyle{model6-num-names}
\bibliography{references}

\end{document}